\documentclass[acmsmall]{acmart}

\acmJournal{PACMPL}
\acmVolume{0} 
\acmNumber{POPL} 
\acmArticle{0} 
\acmYear{2020} 
\acmMonth{1} 
\acmDOI{10.1145/nnnnnnn.nnnnnnn} 
\startPage{1}

\setcopyright{none}

\usepackage{booktabs}   
\usepackage{subcaption} 
\input{packages}

\allowdisplaybreaks

\newcommand*{\adb}[1]{\ensuremath{{\llbracket #1 \rrbracket}^\sharp}}
\newcommand*{\db}[1]{\ensuremath{{\llbracket #1 \rrbracket}}}
\newcommand*{\ind}[1]{\ensuremath{{\mathds{1}_{[{#1}]}}}}

\newcommand*{\defeq}{\triangleq} 
\newcommand*{\code}[1]{\mathsf{{#1}}}

\newcommand*{\csample}{\code{sample}}
\newcommand*{\cscore}{\code{score}}
\newcommand*{\cskip}{\code{skip}}
\newcommand*{\cif}{\code{if}}

\newcommand*{\celse}{\code{else}}
\newcommand*{\cwhile}{\code{while}}

\newcommand*{\cnorm}{\code{norm}}
\newcommand*{\ctrue}{\code{true}}

\newcommand*{\cexp}{\code{exp}}

\newcommand*{\dd}{\mathrm{d}}
\newcommand*{\dom}{\mathrm{dom}}
\newcommand*{\argmin}{\mathrm{argmin}}
\newcommand*{\KL}{\mathrm{KL}}

\newcommand*{\Meas}{\mathrm{Mea}}
\newcommand*{\Spr}{\mathrm{Sp}}

\newcommand*{\smath}[1]{\mathit{{#1}}}

\newcommand*{\density}{\smath{dens}}
\newcommand*{\get}{\smath{get}}

\newcommand*{\consumed}{\smath{consumed}}

\newcommand*{\nn}{\smath{nn}}

\newcommand*{\nil}{\smath{nil}}

\newcommand*{\true}{\smath{true}}

\newcommand*{\fix}{\smath{fix}}
\newcommand*{\wfix}{\smath{wfix}}

\newcommand*{\sskip}{\smath{skip}}
\newcommand*{\cond}{\smath{cond}}
\newcommand*{\update}{\smath{update}}

\newcommand*{\sample}{\smath{sample}}
\newcommand*{\score}{\smath{score}}
\newcommand*{\widen}{\smath{widen}}

\newcommand*{\fin}{\smath{fin}}

\newcommand*{\Str}{\smath{Str}}

\newcommand*{\Var}{\smath{Var}}

\newcommand*{\Store}{\smath{Store}}
\newcommand*{\State}{\smath{State}}

\newcommand*{\Rdb}{\smath{RDB}}

\newcommand*{\aRdb}{\smath{RDB}^\sharp}
\newcommand*{\aZone}{\smath{Zone}^\sharp}
\newcommand*{\aDist}{\smath{Dist}^\sharp}

\newcommand*{\obs}{\mathit{obs}}

\newcommand*{\init}{I}

\newcommand*{\initsigma}{{\sigma_\init}}
\newcommand*{\cA}{\mathcal{A}}
\newcommand*{\cB}{\mathcal{B}}
\newcommand*{\cD}{\mathcal{D}}
\newcommand*{\cE}{\mathcal{E}}

\newcommand*{\cK}{\mathcal{K}}
\newcommand*{\cN}{\mathcal{N}}
\newcommand*{\cO}{\mathcal{O}}
\newcommand*{\cP}{\mathcal{P}}
\newcommand*{\cR}{\mathcal{R}}

\newcommand*{\cT}{\mathcal{T}}

\newcommand*{\bN}{\mathbb{N}}
\newcommand*{\bE}{\mathbb{E}}
\newcommand*{\bR}{\mathbb{R}}
\newcommand*{\EE}[2]{\bE_{#1}\left[{#2}\right]}

\newcommand*{\sinter}{\mathrel{\cap}}
\newcommand*{\sunion}{\mathrel{\cup}}

\newcommand{\abs}[1]{#1^{\sharp}}

\newcommand*{\cmt}[2]{{}}
\newcommand*{\hsy}[1]{\cmt{HSY}{#1}}
\newcommand*{\wl}[1]{\cmt{WL}{#1}}

\newcommand*{\xr}[1]{\cmt{XR}{#1}}

\newcommand*{\iftr}[1]{} 
\newcommand*{\commentout}[1]{}
\newcommand*{\mytt}[1]{{\tt\small #1}}
\newcommand*{\mytts}[1]{{\tt\footnotesize #1}}

\definecolor{lred}{rgb}{1.0,0.5,0.5}
\definecolor{lgreen}{rgb}{0.35,0.95,0.35}

\definecolor{darkgreen}{RGB}{0,64,0}
\definecolor{lime}{rgb}{0.9, 1.0, 0.6}

\begin{document}

\title[Towards Verified Stochastic Variational Inference for Probabilistic Programs]{%
Towards Verified Stochastic Variational Inference for Probabilistic Programs%
}

%


\author{Wonyeol Lee}
\affiliation{
  \institution{School of Computing, KAIST}                   
  \country{South Korea}                       
}
\email{wonyeol@kaist.ac.kr}          

\author{Hangyeol Yu}
\affiliation{
  \institution{School of Computing, KAIST}                   
  \country{South Korea}                       
}
\email{yhk1344@kaist.ac.kr}          

\author{Xavier Rival}
\affiliation{
  \institution{INRIA Paris and D\'{e}partement d'Informatique of ENS,
    CNRS/\'Ecole Normale Sup\'{e}rieure/PSL University}
  \country{France}                   
}
\email{rival@di.ens.fr}              

\author{Hongseok Yang}
\affiliation{
  \institution{School of Computing, KAIST}                   
  \country{South Korea}                       
}
\email{hongseok.yang@kaist.ac.kr}          

\begin{abstract}

Probabilistic programming is the idea of writing models from statistics and machine learning using program notations and reasoning about these models using generic inference engines. Recently its combination with deep learning has been explored intensely, which led to the development of so called deep probabilistic programming languages, such as Pyro, Edward and ProbTorch. At the core of this development lie inference engines based on stochastic variational inference algorithms.
When asked to find information about the posterior distribution of a
model written in such a language, these algorithms convert this
posterior-inference query into an optimisation problem and solve it
approximately by a form of gradient ascent or descent. In this paper, we analyse one of the most fundamental and versatile variational inference algorithms, called score estimator or REINFORCE, using tools from denotational semantics and program analysis. We formally express what this algorithm does on models denoted by programs, and expose implicit assumptions made by the algorithm on the models.
The violation of these assumptions may lead to an undefined optimisation
objective or the loss of convergence guarantee of the optimisation process.
We then describe rules for proving these assumptions, which can be automated by static program analyses.
Some of our rules use nontrivial facts from continuous mathematics, and let us replace requirements about integrals in the assumptions, such as integrability of functions defined in terms of programs' denotations, by conditions involving differentiation or boundedness, which are much easier to prove automatically (and manually).
Following our general methodology, we have developed a static program analysis for the Pyro programming language that aims at discharging the assumption about what we call model-guide support match.
Our analysis is applied to the eight representative model-guide pairs from the Pyro webpage, which include sophisticated neural network models such as AIR. It finds a bug in one of these cases, reveals a non-standard use of an inference engine in another, and shows that the assumptions are met in the remaining six cases.

\end{abstract}



\maketitle


\section{Introduction}

Probabilistic programming refers to the idea of writing models from statistics and machine learning using program notations and reasoning about these models using generic inference engines. It has been the subject of active research in  machine learning and programming languages, because of its potential for enabling scientists and engineers to design and explore sophisticated models easily; when using these languages, they no longer have to worry about developing custom inference engines for their models, a highly-nontrivial task requiring expertise in statistics and machine learning. Several practical probabilistic programming languages now exist, and are used for a wide range of applications~\cite{Stan,InferNet,Tabular,goodman_uai_2008,Mansinghka-venture14,wood-aistats-2014,TolpinMYW16,Hakaru,psi:cav:16}. 

In this paper, we consider inference engines that lie at the core of so called deep probabilistic programming languages, such as Pyro \cite{BinghamCJOPKSSH19}, Edward \cite{TranKDRLB16, TranHMSVR18} and ProbTorch \cite{siddharth2017learning}. These languages let their users freely mix deep neural networks with constructs from probabilistic programming, in particular, those for writing Bayesian probabilistic models.  In so doing, they facilitate the development of probabilistic deep-network models that may address the problem of measuring the uncertainty in current non-Bayesian deep-network models; a non-Bayesian model may predict that the price of energy goes up and that of a house goes down, but it cannot express, for instance, that the model is very confident with the first prediction but not the second. 

The primary inference engines for these deep probabilistic programming languages implement stochastic (or black-box)
variational inference%
\footnote{The term {\em stochastic} variational inference (VI)
  often refers to VI with data subsampling~\cite{HoffmanJMLR13},
  and our usage of the term is often called black-box VI~\cite{RanganathGB14}
  to stress the treatment of a model as a black-box sampler.}
algorithms.
Converting inference problems into optimisation problems is the high-level
idea of these algorithms.\footnote{The inference problems in their original forms involve solving summation/integration/counting problems, which are typically more difficult than optimisation problems. The variational-inference algorithms convert the former problems to the latter ones, by looking for approximate, not exact, answers to the former.} When asked to find information about the posterior distribution of a model written in such a language, these algorithms convert this question to an optimisation problem and solve the problem approximately by performing a gradient descent or ascent on the optimisation objective. The algorithms work smoothly with gradient-based parameter-learning algorithms for deep neural networks, which is why they form the backbone for deep probabilistic programming languages.

In this paper, we analyse one of the most fundamental and versatile variational inference algorithms, called score estimator or REINFORCE%
\footnote{REINFORCE~\cite{WilliamsMLJ1992} is an algorithm originally developed for reinforcement learning (RL), but it is commonly used as a synonym of the score-estimator algorithm. This is because REINFORCE and score estimator use a nearly identical method for estimating the gradient of an optimisation objective.}~\cite{WilliamsMLJ1992,PaisleyICML12,WingateBBVI13,RanganathGB14}, using tools from denotational semantics and program analysis~\cite{cc:popl:77, cc:popl:79, Cousot:1992vz}. We formally express what this algorithm does on models denoted by probabilistic programs, and expose implicit assumptions made by the algorithm on the models. The violation of these assumptions can lead to undefined optimisation objective or the loss of convergence guarantee of the optimisation process. We then describe rules for proving these assumptions, which can be automated by static program analyses. Some of our rules use nontrivial facts from continuous mathematics, and let us replace requirements about integrals in the assumptions, such as integrability of functions defined in terms of programs' denotations, by the conditions involving differentiation or boundedness, which are much easier to prove automatically (and manually) than the original requirements. 

Following our general methodology, we have developed a static program analysis for the Pyro programming language that can discharge one assumption of the inference algorithm about so called model-guide pairs. In Pyro and other deep probabilistic programming languages, a program denoting a model typically comes with a companion program, called guide, decoder, or inference network. This companion, which we call guide, helps the inference algorithm to find a good approximation to what the model ultimately denotes under a given dataset (i.e., the posterior distribution of the model under the dataset); the algorithm uses the guide to fix the search space of approximations, and solves an optimisation problem defined on that space. A model and a guide should satisfy an important correspondence property, which says that they should use the same sets of random variables, and for any such random variable, if the probability of the variable having a particular value is zero in the model, it should also be zero in the guide. If the property is violated, the inference algorithm may attempt to solve an optimisation problem with undefined optimisation objective and return parameter values that do not make any sense. Our static analysis checks this correspondence property for Pyro programs. When applied to eight representative model-guide pairs from the Pyro webpage, which include sophisticated neural network models such as Attend-Infer-Repeat (AIR), the analysis found a bug in one of these cases, revealed a non-standard use of the inference algorithm in another, and proved that the property holds in the remaining six cases.

Another motivation for this paper is to demonstrate an opportunity
for programming languages and verification research to have an impact
on the advances of machine learning and AI technologies.
One popular question is: what properties should we verify on
machine-learning programs?
Multiple answers have been proposed, which led to excellent research
results, such as those on robustness of neural networks~\cite{MirmanGV18}.
But most of the existing research focuses on the final outcome of
machine learning algorithms, not the process of applying these
algorithms.
One of our main objectives is to show that the process often relies
on multiple assumptions on models and finding automatic ways for
discharging these assumptions can be another way of making PL and
verification techniques contribute.
While our suggested solutions are not complete, they are intended
to show the richness of this type of problems in terms of theory
and practice.

\iftr{
Before ending this introduction, we want to comment on a reason for writing this paper. One popular question from researchers in programming languages and verification is: what properties should we verify on machine-learning programs?
There have been multiple proposals to this question, which led to excellent research outcomes, such as robustness of neural networks. But most of the existing research focuses on the final outcome of machine learning algorithms, not the process of applying these algorithms. One of our main objectives is to show that the process often relies on multiple assumptions on models and finding automatic ways for discharging these assumptions can be another way of making programming-language and verification technologies have impact on the recent advance in machine-learning and AI technologies. Our suggested solutions are not complete, but they are intended to show the richness of this type of problems in terms of theory and practice. \hsy{This paragraph is pretentious, although it expresses or attempts to express the message of this paper directly. Think about revising it or moving it to a different place in the paper.}
\xr{I like this paragraph; the conclusion should echo it
  (maybe some part of it could be moved to conclusion, but a part should remain
  in the intro)}}

We summarise the contributions of the paper:
\begin{itemize} 
\item We formally express the behaviour of the most fundamental variational inference algorithm on probabilistic programs using denotational semantics, and identify requirements on program denotations that are needed for this algorithm to work correctly.
\item We describe conditions that imply the identified requirements but
  are easier to prove.
  The sufficiency of the conditions relies on nontrivial results from
  continuous mathematics.
  We sketch a recipe for building program analyses for checking these
  conditions automatically.
\item We present a static analysis for the Pyro language that checks the correspondence requirement of model-guide pairs. The analysis is based on our recipe, but extends it significantly to address challenges for dealing with features of the real-world language. Our analysis has successfully verified 6 representative Pyro model-guide examples, and found a bug in one example.
\end{itemize} 
The proofs of lemmas and theorems of the paper are included in Appendix.


\section{Variational Inference and Verification Challenges by Examples}
\label{sec:overview}

We start by explaining informally the idea of stochastic variational
inference (in short SVI), one fundamental SVI algorithm, and the
verification challenges that arise when we use this algorithm.

\subsection{Stochastic variational inference}

In a probabilistic programming language, we specify a model by a program. The program \mytt{model()} in Figure~\ref{fig:eg:svi}(a) is an example. It describes a joint probability density $p(v,\obs)$ on two real-valued random variables $v$ and $\obs$. The value of the former is not observed, while the latter is observed to have the value $0$. Finding out the value of $v$ is the objective of writing this model. The joint density $p(v,\obs)$ is expressed in terms of \emph{prior} $p(v)$ and \emph{likelihood} $p(\obs|v)$ in the program. The prior $p(v)$ of $v$ is the normal distribution with mean $0$ and standard deviation $5$, and it expresses the belief about the possible value of $v$ before any observation. The likelihood $p(\obs|v)$ is a normal distribution whose mean and standard deviation are either $(1,1)$ or $(-2,1)$ depending on the sign of the value of $v$. The purpose of most inference algorithms is to compute exactly or approximately the \emph{posterior} density given a prior and a likelihood. In our example, the posterior $p(v|\obs{=}0)$ is:
\[
        p(v|\obs{=}0) = \frac{p(v,\obs{=}0)}{\int \dd v\,p(v,\obs{=}0)}
                      = \frac{p(v)\cdot p(\obs{=}0|v)}{p(\obs{=}0)}.
\]
Intuitively, the posterior expresses an updated belief on $v$ upon observing $\obs=0$. The dashed blue and solid orange lines in Figure~\ref{fig:eg:svi}(b) show the prior and posterior densities, respectively. Note that the density of a positive $v$ in the prior went up in the posterior. This is because when $v > 0$, the mean of $p(\obs|v)$ is $1$, a value closer to the observed value $0$ than the alternative $-2$ for the mean.

SVI algorithms approach the posterior inference problem from the optimisation angle. They consider a collection of approximating distributions to a target posterior, formulate the problem of finding a good approximation in the collection as an optimisation problem, and solve the optimisation problem. The solution becomes the result of those algorithms. In Pyro, the users specify such a collection by a single parameterised program called \emph{guide}; the collection can be generated by instantiating the parameters with different values. The program \mytt{guide()} in Figure~\ref{fig:eg:svi}(a) is an example.
It has a real-valued parameter $\theta$ (written as \mytt{theta} in the program), and states that the probability density $q_\theta(v)$ of $v$ is the normal distribution with unknown mean $\theta$ and standard deviation $1$. The lines 13--17 in the figure show how to apply a standard SVI engine of Pyro (called \mytt{Trace\_ELBO}) to find a good $\theta$. They instruct the engine to solve the following optimisation problem: 
\[
         \argmin_\theta \KL(q_{\theta}(v) || p(v|\obs{=}0)),
         \qquad \text{where }\ \KL(q_\theta(v) || p(v|\obs{=}0)) \defeq \EE{q_\theta(v)}{\log \frac{q_\theta(v)}{p(v|\obs{=}0)}}.
\]
The optimisation objective $\KL(q_\theta(v) || p(v|\obs{=}0))$ is the \emph{KL divergence} from $q_\theta(v)$ to $p(v|\obs{=}0)$, and measures the similarity between the two densities, having a small value when the densities are similar.
The KL divergence is drawn in Figure~\ref{fig:eg:svi}(c) as a function
of $\theta$, and the dotted green line in Figure~\ref{fig:eg:svi}(b) draws the density $q_\theta$ at the optimum $\theta$.
Note that the mean of this distribution is biased toward the positive side,
which reflects the fact that the property $v > 0$ has a higher probability
than its negation $v \leq 0$ in the posterior distribution.

One of the most fundamental and versatile algorithms for SVI is score estimator (also called REINFORCE). It repeatedly improves $\theta$ in two steps. First, it estimates the gradient of the optimisation objective with samples from the current $q_{\theta_n}$: 
\[
        \nabla_\theta \KL(q_\theta(v) || p(v|\obs{=}0)) \Big|_{\theta=\theta_n}
        \approx
        \frac{1}{N} \sum_{i=1}^N (\nabla_\theta \log q_{\theta_n}(v_i)) \cdot \log \frac{q_{\theta_n}(v_i)}{p(v_i,\obs{=}0)}
\]
where $v_1,\ldots,v_N$ are independent samples from the distribution $q_{\theta_n}$.
Then, the algorithm updates $\theta$ with the estimated gradient
(the specific learning rate $0.01$ is chosen to improve readability):
\[
        \theta_{n+1} \leftarrow \theta_n 
        - 0.01 \times \frac{1}{N} \sum_{i=1}^N (\nabla_\theta \log q_{\theta_n}(v_i)) \cdot \log \frac{q_{\theta_n}(v_i)}{p(v_i,\obs{=}0)}.
\]
When the learning rate $0.01$ is adjusted according to a known scheme, the algorithm is guaranteed to converge to a local optimum (in many cases) because its gradient estimate satisfies the following unbiasedness property (in those cases): 
\begin{equation}
        \label{eqn:score-unbiasedness-example}
        \nabla_\theta \KL(q_\theta(v) || p(v|\obs{=}0)) \Big|_{\theta=\theta_n}
        =
        \EE{}{\frac{1}{N} \sum_{i=1}^N (\nabla_\theta \log q_{\theta_n}(v_i)) \cdot \log \frac{q_{\theta_n}(v_i)}{p(v_i,\obs{=}0)}}
\end{equation}
where the expectation is taken over the independent samples $v_1,\ldots,v_N$ from $q_{\theta_n}$.

\begin{figure}
  \begin{subfigure}{.9\textwidth}
    \begin{minipage}[t]{.45\textwidth}
      \begin{lstlisting}
# define model and guide
def model():
  v = pyro.sample("v", Normal(0., 5.))
  if (v > 0):
    pyro.sample("obs", Normal(1., 1.), obs=0.)
  else:
    pyro.sample("obs", Normal(-2., 1.), obs=0.)
    
def guide():
  theta = pyro.param("theta", 3.)
  v = pyro.sample("v", Normal(theta, 1.))
\end{lstlisting}
    \end{minipage}
    \hfill
    \begin{minipage}[t]{.45\textwidth}
      \begin{lstlisting}[firstnumber=12]
# perform stochastic variational inference
svi = SVI(model, guide,
        Adam({"lr": 1.0e-2}),
        loss=Trace_ELBO())
for step in range(2000):
  svi.step()

# print result
print("trained theta =",
  pyro.param("theta").item())
\end{lstlisting}
    \end{minipage}
    \hrule
    \caption{Example model-guide pair for stochastic variational inference in Pyro.}
  \end{subfigure}
  \\[1ex]
  \centering
  \begin{subfigure}[t]{.45\textwidth}
    \centering
    \includegraphics[width=.9\textwidth]{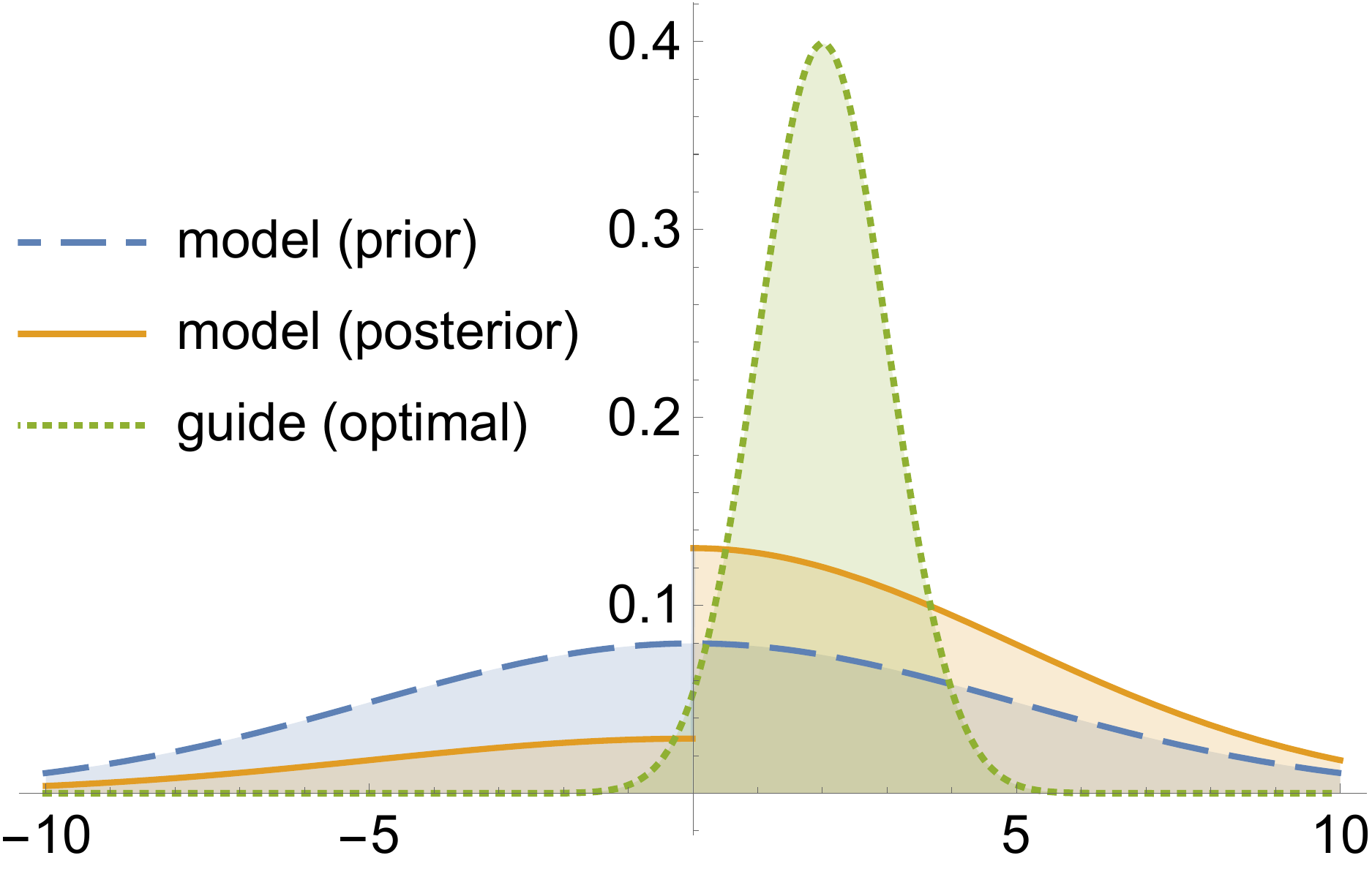}
    \caption{Probability densities of the model and the guide
      as a function of $v \in \bR$.}
  \end{subfigure}
  \qquad
  \begin{subfigure}[t]{.45\textwidth}
    \centering
    \includegraphics[width=.9\textwidth]{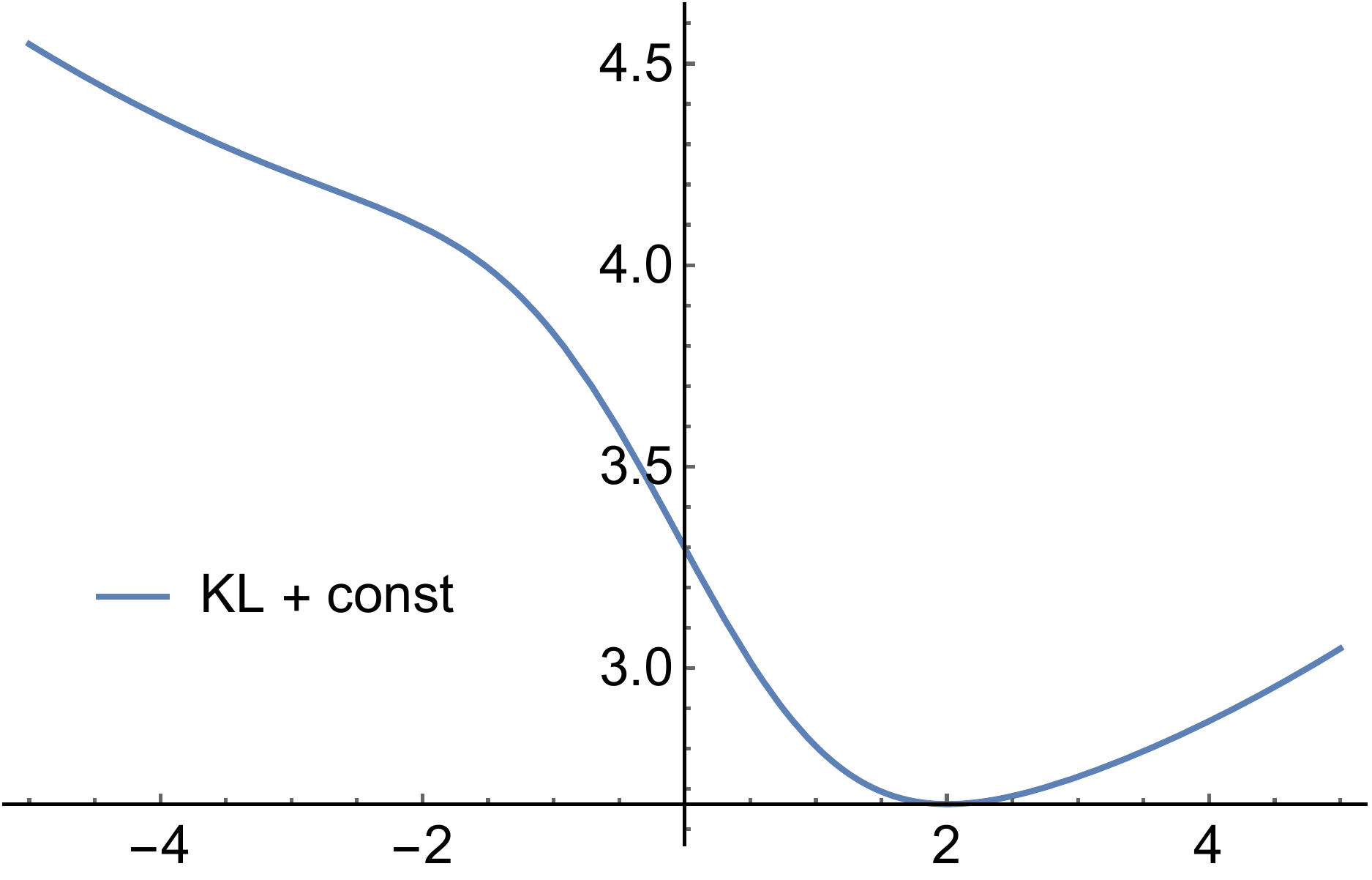}
    \caption{KL divergence from the guide to the model (plus $\log p(\obs{=}0)$)
      as a function of $\theta \in \bR$.}
  \end{subfigure}
  \caption{Example of performing stochastic variational inference.}
  \label{fig:eg:svi}
\end{figure}

\subsection{Verification challenges}

We now give two example model-guide pairs that illustrate verification challenges related to SVI.

The first example appears in Figure~\ref{fig:eg:kl:undefined}(a).
It is the Bayesian regression example from the Pyro webpage (this example
is among the benchmarks used in \S\ref{s:8:pyroai}), which solves the problem of finding a line that interpolates a given set of points in $\mathbb{R}^2$.

The problem with this example is that the KL divergence of its model-guide pair, the main optimisation objective in SVI, is undefined. The model and guide in the figure use the random variable \mytt{sigma}, but they use different non-zero-probability regions, called supports, for it. In the model, the support is $[0,10]$, while that in the guide is $\bR$. But the KL divergence from a guide to a model is defined only if for every random variable, its support in the guide is included in that in the model. We point out that this support mismatch was found by our static analyser explained in~\S\ref{s:8:pyroai}.

Figures~\ref{fig:eg:kl:undefined}(b) and~\ref{fig:eg:kl:undefined}(c) show
two attempts to resolve the undefined-KL issue. To fix the issue,
we change the distribution of \mytt{sigma} in the model in (b),
and in the guide in (c). These revisions remove the problem about the support of \mytt{sigma},
but do not eliminate that of the undefined KL. In both (b) and (c), the KL divergence is $\infty$. This happens mainly because \mytt{sigma} can be arbitrarily close to 0 in the guide in both cases,
which makes integrand in the definition of the KL divergence diverge to $\infty$.

An SVI-specific verification challenge related to this example is how to prove the well-definedness of the KL divergence and more generally the optimisation objective of an SVI algorithm. In \S\ref{subsec:condition-3}, we provide a partial answer to the question. We give a condition for ensuring the well-definedness of the KL divergence. Our condition is more automation-friendly than the definition of KL, because it does not impose the difficult-to-check integrability requirement present in the definition of KL.

\begin{figure}
  \begin{subfigure}{.9\textwidth}
    \begin{minipage}[t]{.45\textwidth}
\begin{lstlisting}[escapechar=@]
def model(...):
  ...
  sigma = pyro.sample("sigma",
            @\colorbox{lred}{Uniform}@(0., 10.))
  ...            
  pyro.sample("obs",
    Normal(..., sigma), obs=...)
\end{lstlisting}
    \end{minipage}
    \hfill
    \begin{minipage}[t]{.45\textwidth}
\begin{lstlisting}[firstnumber=8, escapechar=@]
def guide(...):
  ...
  loc = pyro.param("sigma_loc", 1.,
          constraint=constraints.positive)
  ...
  sigma = pyro.sample("sigma",
            @\colorbox{lred}{Normal}@(loc, 0.05))
\end{lstlisting}
    \end{minipage}
    \hrule
    \caption{Bayesian regression example from the Pyro webpage.}
  \end{subfigure}
  \\[0.5ex]
  \begin{subfigure}{.9\textwidth}
      \begin{subfigure}[t]{.45\textwidth}
\begin{lstlisting}[escapechar=@]
def model_r1(...):
  ...
  sigma = pyro.sample("sigma",
            @\colorbox{lgreen}{Normal}@(5., 5.))
  ...
  pyro.sample("obs",
    Normal(..., @\textcolor{black}{abs}@(sigma)), obs=...)
  
def guide_r1(...):
  # same as guide() in (a)
  ...
\end{lstlisting}
        \hrule
        \caption{The example with a revised model.}
      \end{subfigure}
      \hfill
      \begin{subfigure}[t]{.45\textwidth}
\begin{lstlisting}[escapechar=@]
def model_r2(...):
  # same as model() in (a)
  ...

def guide_r2(...):
  ...
  sigma = pyro.sample("sigma",
            @\colorbox{lgreen}{Uniform}@(0., 10.))
\end{lstlisting}~\\[1.4ex]
        \hrule
        \caption{The example with a revised guide.}
      \end{subfigure}
  \end{subfigure}
  \caption{Example model-guide pairs whose KL divergence is undefined.}
  \label{fig:eg:kl:undefined}
\end{figure}

The second example appears in Figure~\ref{fig:eg:score:biased}(a).  It uses the same model as in Figure~\ref{fig:eg:svi}(a), but has a new guide that uses a uniform distribution parameterised by $\theta \in \bR$.  For this model-guide pair, the KL divergence is well-defined for all $\theta \in \bR$, and the optimal $\theta^*$ minimising the KL is $\theta^*=1$.

However, as shown in Figure~\ref{fig:eg:score:biased}(b), the gradient of the KL divergence is undefined for $\theta \in \{-1,1\}$, because the KL divergence is not differentiable at $-1$ and $1$.  For all the other $\theta \in \bR \setminus \{-1,1\}$, the KL divergence and its gradient are both defined, but the score estimator cannot estimate this gradient in an unbiased manner (i.e., in a way satisfying \eqref{eqn:score-unbiasedness-example}), thereby losing the convergence guarantee to a local optimum. The precise calculation is not appropriate in this section, but we just point out that the expectation of the estimated gradient is always zero for all $\theta \in \bR\setminus\{-1,1\}$, but
the true gradient of the KL is always non-zero for those $\theta$, because it has the form:
$
\frac{\theta}{25}- \ind{-1 \leq \theta \leq 1} \cdot \frac{1}{2}
\log\frac{\cN(0;1,1)}{\cN(0;-2,1)}.
$
Here $\cN(v;\mu,\sigma)$ is the density of the normal distribution with mean $\mu$ and standard deviation $\sigma$
(concretely, $\cN(v;\mu,\sigma) = 1/(\sqrt{2\pi}\sigma) \cdot \exp(-(v-\mu)^2/(2\sigma^2))$).
The mismatch comes from the invalidity of one implicit assumption about interchanging integration
and gradient in the justification of the score estimator; see \S\ref{sec:inference} for detail.

To sum up, the second example shows that even if the KL divergence is defined,
its gradient is sometimes undefined,
and also that even if both the KL divergence and its gradient are defined,
the sample-based estimate of the gradient in a standard SVI algorithm may be biased---this means that
the equation similar to \eqref{eqn:score-unbiasedness-example} does not hold and an SVI algorithm
is no longer guaranteed to converge to a local optimum. Proving that these failure cases do not arise
is another SVI-specific verification challenge. In \S\ref{subsec:condition-456},
we give another example of similar flavour,
and provide an automation-friendly condition that ensures the existence of
the KL divergence and its gradient as well as the unbiasedness of the gradient estimate of the score estimator.

\begin{figure}
  \begin{subfigure}{.93\textwidth}
    \centering
    \begin{subfigure}[b]{.45\textwidth}
      \begin{lstlisting}[escapechar=@]
def model():
  v = pyro.sample("v", Normal(0., 5.))
  if (v > 0):
    pyro.sample("obs", Normal(1., 1.), obs=0.)
  else:
    pyro.sample("obs", Normal(-2., 1.), obs=0.)

def guide():
  theta = pyro.param("theta", 3.)
  v = pyro.sample("v",
        @\colorbox{lgreen}{Uniform}@(theta-1., theta+1.))
\end{lstlisting}
      \hrule
      \caption{The model from Figure~\ref{fig:eg:svi}(a),
        and a guide using a parameterised uniform distribution.}
    \end{subfigure}
    \quad\,\,
    \begin{subfigure}[b]{.47\textwidth}
      \centering
      \includegraphics[width=.88\textwidth]{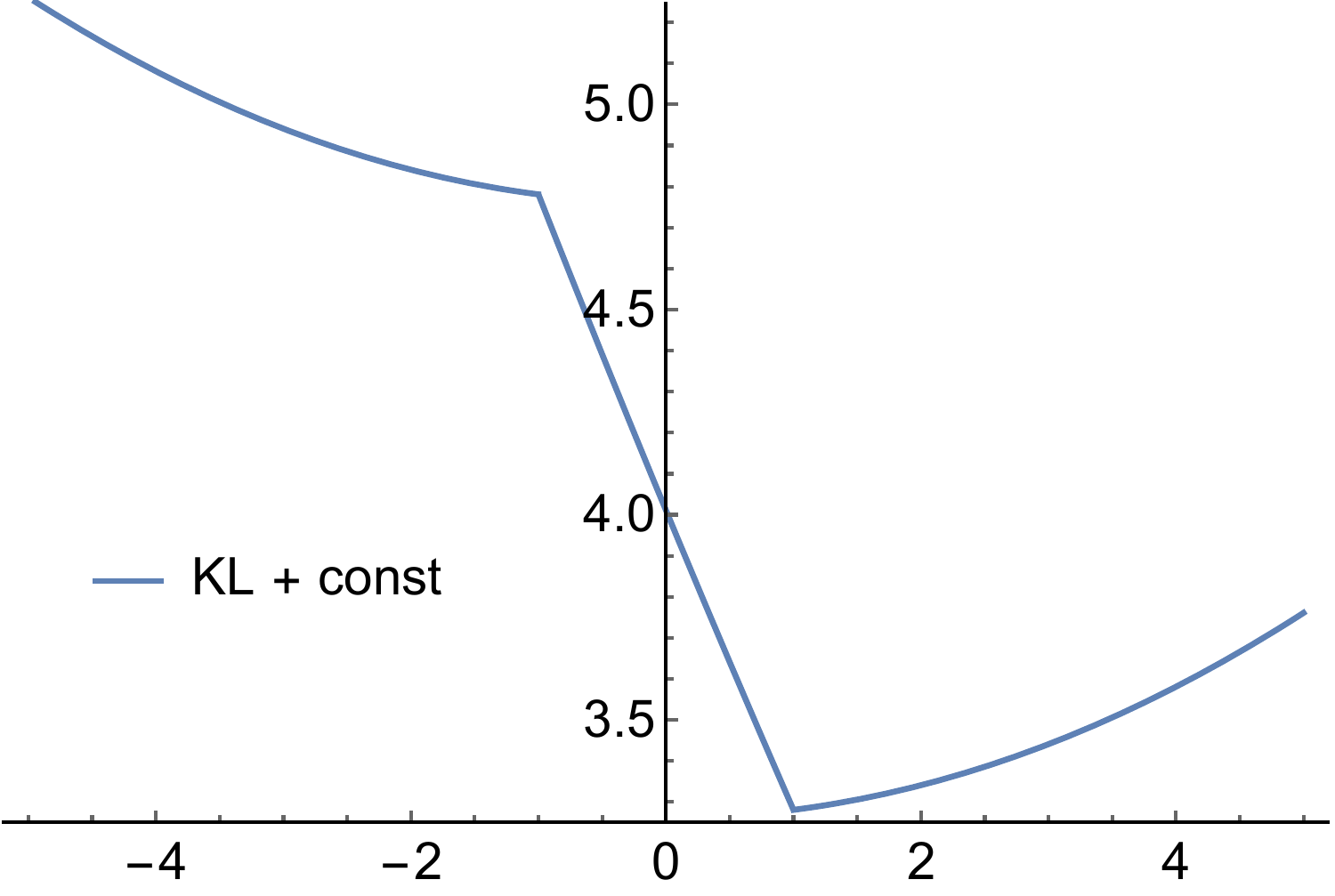}
      \caption{KL divergence from the guide to the model (plus $\log p(\obs{=}0)$)
        as a function of $\theta \in \bR$.}
    \end{subfigure}
  \end{subfigure}
  \caption{Example model-guide pair for which
    the gradient of the KL divergence is undefined,
    or the score estimator is biased.}
  \label{fig:eg:score:biased}
\end{figure}

We conclude the section by emphasising that
the aforementioned issues could have a large impact on the results of SVI.
For instance, running the Bayesian regression example in
Figure~\ref{fig:eg:kl:undefined}(a) in Pyro after a small change
(0.4 instead of 0.05 in line 14) sometimes results in a crash.
Also, running its revised version in Figure~\ref{fig:eg:kl:undefined}(c) 
leads to the complete failure of an optimiser and produces meaningless results.
These observations strengthen the importance of
resolving the verification challenges presented above.%
\footnote{%
  The aforementioned issues do not always cause problems in inference results
  (e.g., Figures~\ref{fig:eg:kl:undefined}(a) and ~\ref{fig:eg:kl:undefined}(c)
  mostly give reasonable results), because the random seeds and the initial values for SVI
  could be set well so that the probability of SVI going wrong becomes low.
  We emphasise, however, that the probability is non-zero.
}


\section{Review of Measure Theory and Notations}
\label{sec:prelim}

A \emph{$\sigma$-algebra} $\Sigma$ on a set $X$ is a collection of
subsets of $X$ such that (i) $X \in \Sigma$; (ii) $A_0 \cup A_1 \in
\Sigma$ and $X \setminus A_0 \in \Sigma$ for all $A_0,A_1 \in \Sigma$; (iii) $\bigcup_n A_n \in \Sigma$ when all subsets $A_n$ are in $\Sigma$. An equivalent but easier-to-remember characterisation is that $\Sigma$ is closed under boolean operations and countable union and intersection. We call the pair $(X,\Sigma)$ of a set and a $\sigma$-algebra \emph{measurable space}, and subsets in $\Sigma$ \emph{measurable}. A function $f$ from a measurable space $(X,\Sigma)$ to another measurable space $(X',\Sigma')$ is \emph{measurable} if $f^{-1}(B) \in \Sigma$ for all $B \in \Sigma'$.

An example of measurable space is the $n$-dimensional Euclidean space $\mathbb{R}^n$ with the
\emph{Borel $\sigma$-algebra} $\cB \defeq \sigma(\{(-\infty,r_1) \times \cdots \times (-\infty,r_n) ~\big|~ r \in \mathbb{R}^n\})$,
where $\sigma$ is the closure operator that converts a collection of subsets of $\mathbb{R}^n$ into the smallest $\sigma$-algebra containing the collection. Subsets $X$ of $\mathbb{R}^n$, such as $[0,\infty)^n$, form measurable spaces with the $\sigma$-algebra $\{X \cap A \,\mid\, A \in \cB\}$. Another example is a set $X$ with the so called \emph{discrete} $\sigma$-algebra on $X$ that consists of all subsets of $X$.

A \emph{measure} $\mu$ on a measurable space $(X,\Sigma)$ is a function from $\Sigma$ to $[0,\infty]$ such that $\mu(\emptyset) = 0$ and $\mu$ satisfies the \emph{countable additivity condition}: for a \emph{countable} family of \emph{disjoint} measurable subsets $B_n$,
\[
\mu\Big(\bigcup_{n=0}^\infty B_n\Big) = \sum_{n=0}^\infty \mu(B_n).
\]
A well-known example is the Lebesgue measure $\lambda^n$ on $\mathbb{R}^n$ which maps each measurable subset of $\mathbb{R}^n$ to its volume in the usual sense.\footnote{The Lebesgue measure $\lambda^n$ is the unique measure on $\mathbb{R}^n$ that sets the volume of the unit cube $(0,1)^n$ to $1$ and is translation invariant: for all measurable subsets $A$ and $r \in \mathbb{R}^n$, $\lambda^n(A) = \lambda^n(\{r'-r \,\mid\, r' \in A \})$.}
When $\mu(X) \leq 1$, we call $\mu$ \emph{subprobability measure}. If $\mu(X) = 1$, we may drop ``sub'', and call $\mu$ \emph{probability measure}.

The Lebesgue integral $\int$ is a partial operator that maps a measure $\mu$ on $(X,\Sigma)$ and a real-valued measurable function on the same space $(X,\Sigma)$ to a real number. It is denoted by $\int\mu(\dd x)\,f(x)$. To follow the paper, it is enough to know that this integral generalises the usual Riemann integral from calculus.\footnote{Another useful fact is that when $f$ is non-negative, $\int\mu(\dd x)\,f(x) = \sup \sum_i (\inf_{x \in A_i} f(x)) \cdot \mu(A_i)$ where the supremum is taken with respect to all finite partitions $\{A_i\}_i$ of $X$ into measurable subsets.} For a measure $\nu$ on $(X,\Sigma)$, if $\nu(A) = \int \mu(\dd x)\,(f(x) \cdot \ind{x \in A})$ for non-negative $f$, we say that $f$ is the \emph{density} of $\nu$ with respect to $\mu$ and call $\mu$ \emph{reference measure}.

In the paper, we use a few well-known methods for building measurable spaces.

The first method applies when we are given a set $X$ and a collection
of functions $\{f_i : X \to X_i \,\mid\, i \in I\}$ to measurable spaces
$(X_i,\Sigma_i)$.
The method is to equip $X$ with the smallest $\sigma$-algebra $\Sigma$
making all $f_i$'s measurable: $\Sigma \defeq \sigma(\{f^{-1}_i(B) \,\mid\, i \in I,\, B \in \Sigma_i\})$.

The second relies on two constructions for sets, i.e., product and disjoint union.
Suppose that we are given measurable spaces $(X_i,\Sigma_i)$ for all
$i \in I$.
We define a \emph{product measurable space} that has $\prod_{i \in I} X_i$
as its underlying set and the following product $\sigma$-algebra
$\bigotimes_{i \in I} \Sigma_i$ as its $\sigma$-algebra:
\[
        \bigotimes_{i \in I} \Sigma_i \defeq \sigma\Big(\Big\{ \prod_i A_i \,\Big|~\,\text{ there is a finite $I_0 \subseteq I$ such that }\ (\forall j \in I \setminus I_0.\,A_j = X_j) \wedge (\forall i \in I_0.\,A_i \in \Sigma_i)\Big\}\Big).
\]
The construction of the product $\sigma$-algebra can be viewed as a special case of the first method where we consider the smallest $\sigma$-algebra on $\prod_{i \in I} X_i$ that makes every projection map to $X_i$ measurable. When the $X_i$ are disjoint, they can be combined as disjoint union. The underlying set in this case is $\bigcup_{i \in I} X_i$,
and the $\sigma$-algebra is
\[
\bigoplus_{i \in I} \Sigma_i \defeq \{A \,\mid\, A \cap X_i \in \Sigma_i \text{ for all $i \in I$}\}.
\]
When $I = \{i,j\}$ with $i\neq j$, we denote the product measurable space by $(X_i\times X_j, \Sigma_i \otimes \Sigma_j)$. In addition, if $X_i$ and $X_j$ are disjoint, we write $(X_i \cup X_j, \Sigma_i \oplus \Sigma_j)$ for the disjoint-union measurable space.

The third method builds a measurable space out of measures or a certain type of measures, such as subprobability measures. For a measurable space $(X,\Sigma)$, we form a measurable space with measures. The underlying set $\Meas(X)$ and $\sigma$-algebra $\Sigma_{\Meas(X)}$ of the space
are defined by
\begin{align*}
\Meas(X) & \defeq \{\mu \mid \text{$\mu$ is a measure on $(X,\Sigma)$}\},
&
\Sigma_{\Meas(X)} & \defeq \sigma\big(\big\{\{\mu \mid \mu(A) \leq r\} ~\big|~ A \in \Sigma, r \in \mathbb{R}\big\}\big).
\end{align*}
The difficult part to grasp is $\Sigma_{\Meas(X)}$. Once again, a good approach for understanding it
is to realise that $\Sigma_{\Meas(X)}$ is the smallest $\sigma$-algebra that makes the function $\mu \longmapsto \mu(A)$
from $\Meas(X)$ to $\mathbb{R}$ measurable for all measurable subsets $A \in \Sigma$. This measurable space gives rise to a variety of measurable spaces, each having a subset $M$ of $\Meas(X)$ as its underlying set
and the induced $\sigma$-algebra $\Sigma_M = \{ A \cap M \,\mid\, A \in \Sigma_{\Meas(X)}\}$. In the paper, we use two such spaces, one induced by the set $\Spr(X)$ of subprobability measures on $X$ and the other by the set $\Pr(X)$ of probability measures.

A measurable function $f$ from $(X,\Sigma)$ to $(\Meas(Y),\Sigma_{\Meas(Y)})$ is called \emph{kernel}. If $f(x)$ is a subprobability measure (i.e., $f(x) \in \Spr(Y)$) for all $x$, we say that $f$ is a \emph{subprobability kernel}. In addition, if $f(x)$ is a probability measure (i.e., $f(x) \in \Pr(Y)$) for all $x$, we call $f$ \emph{probability kernel}. A good heuristic is to view a probability kernel as a random function and a subprobability kernel as a random partial function. We use well-known facts that a function $f : X \to \Meas(Y)$ is a subprobability kernel if and only if it is a measurable map from $(X,\Sigma)$ to $(\Spr(Y),\Sigma_{\Spr(Y)})$, and that similarly a function $f$ is a probability kernel if and only if it is a measurable function from $(X,\Sigma)$ to $(\Pr(Y),\Sigma_{\Pr(Y)})$.

We use a few popular operators for constructing measures throughout the paper. We say that a measure $\mu$ on a measurable space $(X,\Sigma)$ is \emph{finite} if $\mu(X) < \infty$, and \emph{$\sigma$-finite} if there is a countable partition of $X$ into measurable subsets $X_n$'s such that $\mu(X_n) < \infty$ for every $n$. Given a finite or countable family of $\sigma$-finite measures $\{\mu_i\}_{i \in I}$ on measurable spaces $(X_i,\Sigma_i)$'s, the product measure of $\mu_i$'s, denoted $\bigotimes_{i \in I} \mu_i$, is the unique measure on $(\prod_{i \in I} X_i, \bigotimes_{i \in I} \Sigma_i)$ such that for all measurable subsets $A_i$ of $X_i$, 
\[
        \Big(\bigotimes_{i \in I} \mu_i\Big)\Big(\prod_{i \in I} A_i\Big) = \prod_{i \in I} \mu_i(A_i).
\]
Given a finite or countable family of measures $\{\mu_i\}_{i \in I}$ on disjoint measurable spaces $(X_i,\Sigma_i)$'s, the sum measure of $\mu_i$'s, denoted $\bigoplus_{i \in I} \mu_i$, is the unique measure on $(\sum_{i \in I} X_i, \bigoplus_{i \in I} \Sigma_i)$ such that 
\[
        \Big(\bigoplus_{i \in I} \mu_i\Big)\Big(\bigcup_{i \in I} A_i\Big) = \sum_{i \in I} \mu_i(A_i).
\]

Throughout the paper, we take the convention that the set $\bN$ of natural numbers includes $0$. For all positive integers $n$, we write $[n]$ to mean the set $\{1,2,\ldots,n\}$.


\section{Simple Probabilistic Programming Language}
\label{s:4:lang}

In this section, we describe the syntax and semantics of a simple probabilistic programming language, which we use to present the theoretical results of the paper. 
The measure semantics in \S\ref{s:4:1:semantics}
uses results and observations from~\cite{StatonYWHK16}, but
the density semantics and the other materials in \S\ref{subsec:posterior-density}
are new in this work.

\subsection{Syntax}
\label{s:4:1:syntax}

\begin{figure}[t]
  \noindent
  \[
  \begin{array}{@{}r@{}l@{\qquad\quad}rl@{}}
    \text{\it Real constants}\quad c & \;\;\in\; \mathbb{R}
    &
    \text{\it Primitive real-valued functions}\quad f \;::=\; \ldots
    \\[.5ex]
    \text{\it String constants}\quad \alpha & \;\;\in\; \Str
    &
    \text{\it Primitive string-valued functions}\quad g \;::=\; \ldots
    \\[.5ex]
    \text{\it Real expressions}\quad E & \omit\rlap{${} \;::=\; c \,\mid\, x \,\mid\, f(E,\ldots,E)$}
    \\[.5ex]
    \text{\it Boolean expressions}\quad B & \omit\rlap{${} \;::=\; \ctrue \,\mid\, E < E \,\mid\, B \wedge B \,\mid\, \neg B$}
    \\[.5ex]
    \text{\it String expressions}\quad S & \omit\rlap{${}\;::=\; \alpha \,\mid\, g(S,\ldots,S,E,\ldots,E)$}
    \\[.5ex]
    \text{\it Commands}\quad C & \omit\rlap{${}\;::=\;
      \cskip
      \,\mid\, x := E
      \,\mid\, C;C
      \,\mid\, \cif\,B\,\{C\}\,\celse\,\{C\} \,\mid\, \cwhile\,B\,\{C\}$}
    \\[.5ex]
    & \omit\rlap{${}~\;\;\mid\;\;~ x := \csample_\cnorm(S,E,E) \,\mid\, \cscore_\cnorm(E,E,E)$}
  \end{array}\]
  \hrule \vspace{-1ex}
  \caption{Grammar of our core language.}
  \label{fig:lang:grammar}
\end{figure}
We use an extension of the standard while language with primitives for probabilistic programming. The grammar of the language is given 
in Figure~\ref{fig:lang:grammar}.
Variables in the language store real numbers, but expressions may denote reals, booleans and strings, and they are classified into $E,B,S$ based on these denoted values. The primitive functions $f$ for reals and $g$ for strings may be usual arithmetic and string operations, such as multiplication and exponentiation for $f$ and \mbox{string concatenation for $g$.}

The grammar for $C$ includes the cases for the standard constructs of the while language, such as assignment, sequencing, conditional statement and loops. In addition, it has two constructs for probabilistic programming. The first $x := \csample_\cnorm(S,E_1,E_2)$ draws a sample from the \emph{normal} distribution with mean $E_1$ and standard deviation $E_2$ and names the sample with the string $S$. The next $\cscore_\cnorm(E_0,E_1,E_2)$ expresses that a sample is drawn from the normal distribution with mean $E_1$ and standard deviation $E_2$ and the value of this sample is observed to be $E_0$. It lets the programmers express information about observed data inside programs. Operationally, this construct can be understood as an instruction for updating a global variable that stores the so called importance score of the execution. The score quantitatively records how well the random choices in the current execution match the observations, and the score statement updates this score by multiplying it with the density at $E_0$ of the appropriate normal distribution.

Consider the following program: 
\[
        x := \csample_\cnorm(``a", 0.0,5.0);\; \cscore_\cnorm(3.0,x,1.0).
\]
The program specifies a model with one random variable $``a"$. Using a relatively flat normal distribution, the program specifies a prior belief that the value of the random variable $``a"$ is likely to be close to $0.0$ and lie between $-2\times 5.0$ and $2 \times 5.0$. The next score statement refines this belief with one data point $3.0$, which is a noisy observation of the value of $``a"$ (bound to $x$). The parameters to the normal density in the statement express that the noise is relatively small, between $-2\times 1.0$ and $2 \times 1.0$. Getting the refined belief, called posterior distribution, is the reason that a data scientist writes a model like this program. It is done by an inference algorithm of the language.

Permitting only the normal distribution does not limit the type of models expressible in the language. Every distribution can be obtained by transforming the standard normal distribution.\footnote{Here we consider only Borel spaces. Although using only the normal distribution does not affect the expressiveness, it has an impact on stochastic variational inference to be discussed later, because it requires a guide to use only normal distributions. But the impact is not significant, because most well-known approaches for creating guides from the machine-learning literature (such as extensions of variational autoencoder) use normal distributions only, or can be made to do so easily.}

\subsection{Measure semantics}
\label{s:4:1:semantics}

The denotational semantics of the language just presented is mostly standard, but employs some twists to address the features for probabilistic programming~\cite{StatonYWHK16}.

Here is a short high-level overview of the measure semantics. Our semantics defines multiple measurable spaces, such as $\Store$ and $\State$, that hold mathematical counterparts to the usual actors in computation, such as program stores (i.e., mappings from variables to values) and states (which consist of a store and further components).
Then, the semantics interprets expressions $E,B,S$ and commands $C$ as measurable functions of the following types:
\begin{align*}
        \db{E} & : \Store \to \mathbb{R},
        &
        \db{B} & : \Store \to \mathbb{B},
        &
        \db{S} & : \Store \to \Str,
        &
        \db{C} & : \State \to \Spr(\State \times [0,\infty)).
\end{align*}
Here $\bR$ is the measurable space of reals with the Borel $\sigma$-algebra, and $\mathbb{B}$ and $\Str$ are discrete measurable spaces of booleans and strings.
$\Store$ and $\State$ are measurable spaces for stores (i.e., maps from
variables to values) and states which consist of a store and a part for
recording information about sampled random variables.
Note that the target measurable space of commands is built by first taking the product of measurable spaces $\State$ and $[0,\infty)$ and then forming a space out of subprobability measures on $\State \times [0,\infty)$. This construction indicates that commands denote probabilistic computations, and the result of each such computation consists of an output state and a score which expresses how well the computation matches observations expressed with the score statements in the command $C$. Some of the possible outcomes of the computation may lead to non-termination or an error, and these abnormal outcomes are not accounted for by the semantics, which is why $\db{C}(\sigma)$ for a state $\sigma \in \Sigma$ is a subprobability distribution. The semantics of expressions is much simpler. It just says that expressions do not involve any probabilistic computations, so that they denote deterministic measurable functions.

We now explain how this high-level idea gets implemented in our semantics.
Let $\Var$ be a countably infinite set of variables. Our semantics uses the following sets:
\begin{align*}
        \text{\it Stores} \qquad s
        & \in \Store \defeq [\Var \to \bR]  \quad \Big(\text{which is isomorphic to } \prod_{x \in \Var} \bR\Big)
        \\[1ex]
        \text{\it Random databases}\qquad r & \in \Rdb \defeq \bigcup_{K \subseteq_\fin \Str} [K \to \bR]
        \quad
        \Big(\text{which is isomorphic to } \bigcup_{K \subseteq_\fin \Str}  \prod_{\alpha \in K} \bR \Big)
        \\[1ex]
        \text{\it States} \qquad \sigma
        & \in \State \defeq \Store \times \Rdb,
\end{align*}
where $[X \to Y]$ denotes the set of all functions from $X$ to $Y$.
A state $\sigma$ consists of a store $s$ and a random database $r$. The former fixes the values of variables, and the latter records the name (given as a string) and the value of each sampled random variable. The domain of $r$ is the names of all the sampled random variables. By insisting that $r$ should be a map, the semantics asserts that no two random variables have the same name. For each state $\sigma$, we write $\sigma_s$ and $\sigma_r$ for its store and random database components, respectively. Also, for a variable $x$, a string $\alpha$ and a value $v$, we write $\sigma[x\mapsto v]$ and $\sigma[\alpha \mapsto v]$ to mean $(\sigma_s[x\mapsto v],\sigma_r)$ and $(\sigma_s,\sigma_r[\alpha \mapsto v])$.

We equip all of these sets with $\sigma$-algebras and turn them to measurable spaces in a standard way. Note that we constructed the sets from $\mathbb{R}$ by repeatedly applying the product and disjoint-union operators. We equip $\bR$ with the usual Borel $\sigma$-algebra. Then, we parallel each usage of the product and the disjoint-union operators on sets with that of the corresponding operators on $\sigma$-algebras.
This gives the $\sigma$-algebras for all the sets defined above.
Although absent from the above definition, the measurable spaces $\mathbb{B}$ and $\Str$ equipped with discrete $\sigma$-algebras are also used in our semantics. 

We interpret expressions $E,B,S$ as measurable functions 
$\db{E} : \Store \to \mathbb{R}$, 
$\db{B} : \Store \to \mathbb{B}$, and
$\db{S} : \Store \to \Str$, under the assumption
that the semantics of primitive real-valued $f$ of arity $n$ and string-valued $g$
of arity $(m,l)$ are given by measurable functions $\db{f} : \mathbb{R}^n \rightarrow \mathbb{R}$ and 
$\db{g} : \Str^m \times \mathbb{R}^l \rightarrow \Str$. It is standard, and we describe it
only for some sample cases of $E$ and $S$:
\begin{align*}
        \db{x}s & \defeq s(x),
        &
        \db{f(E_0,\ldots,E_{n-1})}s & \defeq \db{f}(\db{E_0}s,\ldots,\db{E_{n-1}}s),
        \\
        \db{\alpha}s & \defeq \alpha,
        & 
        \db{g(S_0,\ldots,S_{m-1},E_0,\ldots,E_{l-1})}s & \defeq \db{g}(\db{S_0}s,\ldots,\db{S_{m-1}}s,\db{E_0}s,\ldots,\db{E_{l-1}}s).
\end{align*}
\begin{lemma}\label{lemma:expression-welldefined}
        For all expressions $E$, $B$, and $S$, their semantics $\db{E}$, $\db{B}$ and $\db{S}$ are measurable functions 
        from $\Store$ to $\mathbb{R}$, $\mathbb{B}$ and $\Str$, respectively.
\end{lemma}

We interpret commands $C$ as measurable functions from $\State$ to $\Spr(\State \times [0,\infty))$, i.e., subprobability kernels
from $\State$ to $\State \times [0,\infty)$. Let $\cK$ be the set of subprobability kernels from $\State$ to $\State \times [0,\infty)$,
and $\Sigma_{\State \times [0,\infty)}$ be the $\sigma$-algebra of the product space ${\State \times [0,\infty)}$. We equip $\cK$ with the partial order $\sqsubseteq$: for all $\kappa,\kappa' \in \cK$, $\kappa \sqsubseteq \kappa'$ 
if and only if 
$\kappa(\sigma)(A) \leq \kappa'(\sigma)(A)$ for all $\sigma \in \State$ and $A \in \Sigma_{\State \times [0,\infty)}$.
The next lemma is a minor adaptation of a known result~\cite{kozen:81}.
\begin{lemma}\label{lemma:subprob-kernels-cpo}
        $(\cK,\sqsubseteq)$ is an $\omega$-complete partial order with the least element $\bot \defeq (\lambda \sigma.\,\lambda A.\,0)$.
\end{lemma}

The measure semantics of a command $C$ is defined in Figure~\ref{f:sampling:sem}.
\begin{figure}[t]
  \noindent
  \[
  \begin{array}{@{}r@{\;}c@{\;}l@{}}
    \db{\cskip}(\sigma)(A)
    & \defeq
    & \ind{(\sigma,1) \in A}
    \\[1ex]
    \db{x:=E}(\sigma)(A)
    & \defeq
    & \ind{(\sigma[x\mapsto \db{E}\sigma_s],1) \in A}
    \\[1ex]
    \db{C_0;C_1}(\sigma)(A)
    & \defeq
    & \int \db{C_0}(\sigma)(\dd(\sigma',w'))\int
    \db{C_1}(\sigma')(\dd(\sigma'',w''))\,\ind{(\sigma'',w'\cdot w'') \in A}
    \\[1ex]
    \db{\cif\,B\,\{C_0\}\,\celse\,\{C_1\}}(\sigma)(A)
    & \defeq
    & \ind{\db{B}\sigma_s = \true} \cdot \db{C_0}(\sigma)(A)
    + \ind{\db{B}\sigma_s \neq \true} \cdot \db{C_1}(\sigma)(A)
    \\[1ex]
    \db{\cwhile\,B\,\{C\}}(\sigma)(A)
    & \defeq
    & (\fix\,F)(\sigma)(A)
    \\
    \multicolumn{3}{@{}l@{}}{
      \qquad\big(\text{where } F(\kappa)(\sigma)(A) \defeq
      \ind{\db{B}\sigma_s \neq \true} \cdot \ind{(\sigma,1) \in A}
    }
    \\
    \multicolumn{3}{@{}l@{}}{
      \phantom{\qquad\big(\text{where } F(\kappa)(\sigma)(A) \defeq}
      {} + \ind{\db{B}\sigma_s = \true} \cdot \int \db{C}(\sigma)(\dd
      (\sigma',w'))\int \kappa(\sigma')(\dd (\sigma'',w''))\,\ind{(\sigma'',
        w'\cdot w'') \in A}
      \big)
    }
    \\[1ex]
    \db{x:=\csample_\cnorm(S,E_1,E_2)}(\sigma)(A)
    & \defeq & {}
    \ind{\db{S}\sigma_s \not\in \dom(\sigma_r)} \cdot \ind{\db{E_2}\sigma_s \in (0,\infty)}
    \\[0.5ex]
    &&
          \cdot 
          \int \dd v\,\big(
                \cN(v;\db{E_1}\sigma_s,\db{E_2}\sigma_s)
                \cdot
                \ind{((\sigma_s[x\mapsto v], \sigma_r[\db{S} \sigma_s \mapsto v]), 1) \in A}\big)
    \\[1ex]
    \db{\cscore_\cnorm(E_0,E_1,E_2)}(\sigma)(A)
          & \defeq & \ind{\db{E_2}\sigma_s \in (0,\infty)} \cdot \ind{(\sigma, \cN(\db{E_0}\sigma_s;\db{E_1}\sigma_s,\db{E_2}\sigma_s)) \in A}
  \end{array}
  \]
  \hrule \vspace{-1ex}
        \caption{Measure semantics \( \db{C} \in \cK \) of commands \( C \). Here $\cN(v;\mu,\sigma)$ is the density of the normal distribution with mean $\mu$ and standard deviation $\sigma$.}
  \label{f:sampling:sem}
\end{figure}
The interpretation of the loop is the least fixed point of the function $F$ on the $\omega$-complete partial order $\cK$. The function $F$ is continuous, so the least fixed point is obtained by the $\omega$-limit of the sequence $\{F^n(\bot)\}_n$. In the definition
of $F$, the argument $\kappa$ is a sub-probability kernel, and it represents the computation after the first iteration of the loop.
The semantics of the sample statement uses an indicator function to
exclude erroneous executions where the argument $S$ denotes a name
already used by some previous random variable, or the standard deviation
$\db{E_2}\sigma_s$ is not positive.
When this check passes, it distils $A$ to a property on the value of $x$ and computes the probability of the property using the normal distribution with mean $\db{E_1}\sigma_s$ and standard deviation $\db{E_2}\sigma_s$.

\begin{theorem}\label{thm:command-welldefined}
        For every command $C$, its interpretation $\db{C}$ is well-defined and belongs to $\cK$.
\end{theorem}


\subsection{Posterior inference and density semantics}
\label{subsec:posterior-density}

We write a probabilistic program to answer queries about the model and
data that it describes.
Among such queries, posterior inference is one of the most important and popular. Let $\initsigma = (s_I, [])$ be the initial state that consists of some fixed store and the empty random database. In our setting, posterior inference amounts to finding information about the following probability measure $\Pr(C,\cdot)$ for a command $C$. For a measurable $A \subseteq \Rdb$,
\begin{align}
        \Meas(C,A) & \defeq \int \db{C}(\initsigma)(\dd(\sigma,w)) (w\cdot\ind{\sigma \in \Store \times A}),
        & Z_C & \defeq \Meas(C,\Rdb),
        & \Pr(C,A) & \defeq \frac{\Meas(C,A)}{Z_C}.
        \label{eqn:posterior-definition}
\end{align}
The probability measure $\Pr(C,\cdot)$ is called the
\emph{posterior distribution of $C$}, and $\Meas(C,\cdot)$ the
\emph{unnormalised posterior distribution of $C$}
(in $\Meas(C,\cdot)$ and $\Pr(C,\cdot)$, we elide the dependency on $s_I$
to avoid clutter).
Finding information about the former is the goal of most inference engines of existing probabilistic programming languages. Of course, $\Pr(C,\cdot)$ is not defined when the normalising constant $Z_C$ is infinite or zero.
The inference engines regard such a case as an error that a programmer
should avoid, and consider only $C$ without those errors.

Most algorithms for posterior inference use the density semantics of commands. They implicitly pick measures on some measurable spaces used in the semantics.
These measures are called \emph{reference measures}, and constructed out of Lebesgue and counting measures~\cite{BhatAVG12,BhatBGR13,HurNRS15}. Then, the algorithms interpret commands as density functions with respect to these measures. One outcome of this density semantics is that the unnormalised posterior distribution $\Meas(C,\cdot)$ of a command $C$ has a measurable function $f : \Rdb \to [0,\infty)$ such that $\Meas(C,A) = \int \rho(\dd r)\,(\ind{r \in A} \cdot f(r))$,
where $\rho$ is a reference measure on $\Rdb$.
Function $f$ is called \emph{density} of $\Meas(C,\cdot)$ with respect
to $\rho$. 

In the rest of this subsection, we reformulate the semantics of commands
using density functions.
To do this, we need to set up some preliminary definitions.

First, we look at a predicate and an operator for random databases, which are about the possibility and the very act of merging two databases. For $r,r' \in \Rdb$, define the predicate $r \# r'$ by: 
\[
        r \# r' \iff \dom(r) \cap \dom(r') = \emptyset.
\]
When $r\#r'$, let $r \uplus r'$ be the random database obtained by merging $r$ and $r'$:
\begin{align*}
        \dom(r \uplus r') \defeq \dom(r) \cup \dom(r');
        \qquad
        (r \uplus r')(\alpha) \defeq
                \text{if } \alpha \in \dom(r) \text{ then } r(\alpha) \text{ else } r'(\alpha).
\end{align*}

\begin{lemma}
  \label{lemma:rdb-disj-measurable}
  For every measurable $h : \Rdb \times \Rdb \times \Rdb \to \mathbb{R}$,
  the function $(r,r') \longmapsto \ind{r \# r'} \times h(r,r',r\uplus r')$
  from $\Rdb \times \Rdb$ to $\mathbb{R}$ is measurable.
\end{lemma}

Second, we define a reference measure $\rho$ on $\Rdb$:
\[
        \rho(R) \defeq
        \sum_{\substack{K \subseteq_\fin \Str}} \Big(\bigotimes_{\alpha \in K} \lambda\Big)(R \cap [K\to\bR]),
\]
where $\lambda$ is the Lebesgue measure on $\bR$. 
As explained in the preliminary section, the symbol $\otimes$ here represents the operator for constructing a product measure. In particular, $\bigotimes_{\alpha \in K} \lambda$ refers to the product of the $|K|$ copies of the Lebesgue measure $\lambda$ on $\bR$. 
In the above definition, we view functions in $[K\to\bR]$ as tuples with $|K|$ real components
and measure sets of such functions using the product measure $\bigotimes_{\alpha \in K} \lambda$. When $K$ is the empty set, $\bigotimes_{\alpha \in K} \lambda$ is the nullary-product measure on $\{[]\}$, which assigns $1$ to $\{[]\}$ and $0$ to the empty set.

The measure $\rho$ computes the size of each measurable subset $R \subseteq \Rdb$ in three steps. It splits a given $R$ into groups based on the domains of elements in $R$. Then, it computes the size of each group separately, using the product of the Lebesgue measure. Finally, it adds the computed sizes. The measure $\rho$ is not finite, but it satisfies
the \emph{$\sigma$-finiteness} condition,\footnote{The condition lets us use Fubini theorem when showing the well-formedness of the density semantics in this subsection and relating this semantics with the measure semantics in the previous subsection.} 
the next best property.

Third, we define a partially-ordered set $\cD$ with certain measurable functions. We say that
a function $g :  \Store \times \Rdb \to \{\bot\} \cup (\Store \times \Rdb \times [0,\infty) \times [0,\infty))$
\emph{uses random databases locally} or \emph{is local} if for all $s, s' \in \Store$, $r, r' \in \Rdb$, and $w',p' \in [0,\infty)$, 
\begin{align*}
        g(s,r) = (s',r',w',p') \implies  {}
        &
        (\exists r''.\, r'\# r'' \wedge r = r' \uplus r'' \wedge g(s,r'') = (s',[],w',p'))
        \\
        &
        {}
        \wedge
        (\forall r'''.\, r\#r''' \implies g(s,r\uplus r''') = (s',r' \uplus r''',w',p')).
\end{align*}
The condition describes the way that $g$ uses a given random database $r$,
which plays the role of a bank of random seeds (that \( g \) may partially
consume as it needs random values).
Some part $r''$ of $r$ may be consumed by $g$, but the unconsumed part $r'$ of $r$ does not change and is returned in the output. Also, the behaviour of $g$ does not depend on the unconsumed $r'$. We define the set $\cD$ by
\begin{align}
        \label{eqn:definition-D}
        \cD \defeq
        &
        \Big\{ g : \Store \times \Rdb \to \{\bot\} \cup (\Store \times \Rdb \times [0,\infty) \times [0,\infty))
        ~\Big|~ \text{ $g$ is measurable and local}\Big\}.
\end{align}
Here we view $\{\bot\}$ and $[0,\infty)$ as measurable spaces equipped with discrete and Borel $\sigma$-algebras. Also, we regard $\Store \times \Rdb$ and $\{\bot\} \cup (\Store \times \Rdb \times [0,\infty) \times [0,\infty))$ as measurable spaces constructed by the product and disjoint-union operators on measurable spaces, as explained in \S\ref{sec:prelim}. 

The locality in the definition of $\cD$ formalises expected behaviours of commands. In fact, as we will show shortly, it is satisfied by all commands in our density semantics. This property plays an important role when we establish the connection between the density semantics in this subsection and the standard measure semantics in \S\ref{s:4:1:semantics}. 

The functions in $\cD$ are ordered pointwise: for all $g,g' \in \cD$, 
\[
        g \sqsubseteq g'
        \iff 
        \forall (s,r) \in \Store \times \Rdb.\, (g(s,r) = \bot \vee g(s,r) = g'(s,r)).
\]
\begin{lemma}
  \label{lem:4:6:cpo}
  $(\cD,\sqsubseteq)$ is an $\omega$-complete partial order and has
  the least element $a \longmapsto \bot$.
  Thus, every continuous function $G$ on $\cD$ has a least fixed point
  (and this least fixed point is unique).
\end{lemma}

For each $g \in \cD$, let  $g^\ddagger$ be the following lifting to a function on
$\{\bot\} \cup (\Store \times \Rdb \times [0,\infty) \times [0,\infty))$: 
\begin{align*}
        g^\ddagger(\bot) \defeq \bot,
        \quad\;
        g^\ddagger(s,r,w,p)  
        \defeq 
        \left\{\begin{array}{ll}
                \bot & \text{ if } g(s,r) = \bot,
                \\[0.5ex]
                (s',r',w \times w',p \times p') & \text{ if } g(s,r) = (s',r',w',p').
        \end{array}\right.
\end{align*}
This lifting lets us compose two functions in $\cD$.

Using these preliminary definitions, we define a density semantics in Figure~\ref{f:density:sem}, where a command $C$ means a function \( \db{C}_d \in \cD \).  The notation $r \setminus v$ in the figure means the removal of the entry $v$ from the finite map $r$ if $v \in \dom(r)$; otherwise, $r \setminus v$ is just $r$. The set membership $\db{C}_d \in \cD$ says that $\db{C}_d$ is a local measurable function from $\Store \times \Rdb$ to $\{\bot\} \cup \Store \times \Rdb \times [0,\infty) \times [0,\infty)$. Thus, the function $\db{C}_d$ takes a store $s$ and a random database $r$ as inputs, where the former fixes the values of variables at the start of $C$ and the latter specifies random seeds some of which $C$ may consume to sample random variables. Given such inputs, the function outputs an updated store $s'$, the part $r'$ of $r$ not consumed by $C$, the total score $w'$ expressing how well the execution of $C$ matches observations, and the probability density $p'$ of $C$ at the consumed part of $r$. If $r$ does not contain enough random seeds, or the execution of $C$ encounters some runtime error, or it falls into an infinite loop, then the function returns $\bot$.

\begin{figure}[t]
  \noindent
  \[
  \begin{array}{@{}r@{\;}c@{\;}l@{}}
    \db{\cskip}_d(s,r)
    & \defeq & (s,r,1,1)
    \\[1ex]
    \db{x{:=}E}_d(s,r)
    & \defeq 
    & (s[x\mapsto \db{E}s],r,1,1)
    \\[1ex]
    \db{C_0;C_1}_d(s,r)
    & \defeq 
    & (\db{C_1}_d^\ddagger \circ \db{C_0}_d)(s,r)
    \\[1ex]
    \db{\cif\,B\,\{C_0\}\,\celse\,\{C_1\}}_d(s,r)
    & \defeq 
    & \text{if}\ (\db{B}s = \true)\ 
    \text{then}\ \db{C_0}_d(s,r)\ \text{else}\ \db{C_1}_d(s,r)
    \\[1ex]
    \db{\cwhile\,B\,\{C\}}_d(s,r)
    & \defeq
    & (\fix\, G)(s,r)
    \\
    \multicolumn{3}{@{}l@{}}{
    \qquad\qquad\qquad\qquad\qquad
     (\text{where }
      G(g)(s,r) =  
      \text{if}\;(\db{B}s \neq \true)\;
      \text{then}\;(s,r,1,1)\;
          \text{else}\;(g^\ddagger \circ \db{C}_d)(s,r))
    }
    \\[1ex]
    \db{x{:=}\csample_\cnorm(S,E_1,E_2)}_d(s,r)
    & \defeq
    & \text{if}\;(\db{S}s \not\in \dom(r) \vee \db{E_2}s \not\in (0,\infty))\; \text{then}\;\bot
    \\
    &
    &
    \text{else}\;
    (s[x\mapsto r(\db{S}s)],\,
    r \setminus \db{S}s,\,
    1,\,
    \cN(r(\db{S}s);\db{E_1}s,\db{E_2}s))
    \\[1ex]
    \db{\cscore_\cnorm(E_0,E_1,E_2)}_d(s,r)
    & \defeq 
    & \text{if}\;(\db{E_2}s \not\in (0,\infty))\; \text{then}\;\bot\;
    \text{else}\; (s,r,\cN(\db{E_0}s;\db{E_1}s,\db{E_2}s),1)
  \end{array}
  \]
  \hrule \vspace{-1ex}
  \caption{Density semantics \( \db{C}_d \in \cD \) of commands \( C \)}
  \label{f:density:sem}
\end{figure}

\begin{lemma}
  \label{lem:4:7:dsemwd}
  For every command $C$, its semantics $\db{C}_d$ is well-defined and belongs to $\cD$.
\end{lemma}

The density semantics $\db{C}_d$ is closely related to the measure semantics $\db{C}$ defined in \S\ref{s:4:1:semantics}. Both interpretations of $C$ describe the computation of $C$ but from
slightly different perspectives. To state this relationship formally, we need a few notations. For $g \in \cD$ and $s \in \Store$, define
\begin{align*}
        \density(g,s) & : \Rdb \to [0,\infty),
        &
        \density(g,s)(r) & \defeq
        \left\{\begin{array}{ll}
                w' \times p' & \text{if } \exists s',w',p'.\, (g(s,r) = (s',[],w',p')),
                \\[0.5ex]
                0 & \text{otherwise},
        \end{array}\right.
        \\[1ex]
        \get(g,s) & : \Rdb \to \Store \cup \{\bot\},
        &
        \get(g,s)(r) & \defeq
        \left\{\begin{array}{ll}
                s' & \text{if } \exists s',w',p'.\, (g(s,r) = (s',[],w',p')),
                \\[0.5ex]
                \bot & \text{otherwise}.
        \end{array}\right.
\end{align*}
Both $\density(g,s)$ and $\get(g,s)$ are concerned with random databases $r$ that precisely describe the randomness needed by the execution of $g$ from $s$. This is formalised by the use of $[]$ in the definitions. The function $\density(g,s)$ assigns a score to such an $r$, and in so doing, it defines a probability density on $\Rdb$ with respect to the reference measure $\rho$.
The function $\get(g,s)$ computes a store that the computation of $g$ would result in when started with such an $r$.
We often write $\density(C,s)$ and $\get(C,s)$ to mean
$\density(\db{C}_d,s)$ and $\get(\db{C}_d,s)$, respectively.

\begin{lemma}
        \label{lemma:density-get-measurability}
        For all $g \in \cD$, the following functions from $\Store \times \Rdb$ to $\mathbb{R}$ and $\{\bot\} \cup \Store$ are measurable: $(s,r) \longmapsto \density(g,s)(r)$
        and $(s,r) \longmapsto \get(g,s)(r)$.
\end{lemma}

The next lemma is the main reason that we considered the locality property. It plays a crucial role in proving Theorem~\ref{thm:correspondence:measure-density}, the key result of this subsection.
\begin{lemma}
        \label{lemma:dens-get-sequencing}
        For all non-negative bounded measurable functions $h : (\{\bot\} \cup \Store) \times \Rdb \to \mathbb{R}$, stores $s$,
        and functions $g_1,g_2 \in \cD$, we have that
        \begin{align*}
                & \int \rho(\dd r)\, \Big(\density(g_2^\ddagger \circ g_1,s)(r) \cdot h\Big(\get(g_2^\ddagger \circ g_1, s)(r), r\Big)\Big)
                \\
                & {}
                =
                \int \rho(\dd r_1)\,
                \Big(\density(g_1,s)(r_1)
                \cdot \ind{\get(g_1,s)(r_1) \neq \bot}
                \\
                &
                \qquad
                \qquad
                \cdot \int \rho(\dd r_2)\,
                \Big(
                \density(g_2,\get(g_1,s)(r_1))(r_2) \cdot \ind{r_1 \# r_2} \cdot h\Big(\get(g_2,\get(g_1,s)(r_1))(r_2), r_1 \uplus r_2\Big)\Big)
                \Big).
        \end{align*}
\end{lemma}

Assume that $(\{\bot\} \cup \Store) \times \Rdb$ is ordered as follows:
for all $(a,r), (a',r') \in (\{\bot\} \cup \Store) \times \Rdb$,
\[
(a,r) \sqsubseteq (a',r') 
\iff
(a = \bot \vee a = a') \wedge r = r'.
\]
\begin{theorem}
        \label{thm:correspondence:measure-density}
        For all non-negative bounded measurable monotone functions $h : (\{\bot\} \cup \Store) \times \Rdb \to \mathbb{R}$ and states $\sigma$,
        \[
                \int \db{C}(\sigma)(\dd (\sigma',w'))\, (w' \cdot h(\sigma'_s,\sigma'_r))
                = \int \rho (\dd r')\, \Big(\density(C,\sigma_s)(r') \cdot \ind{r' \# \sigma_r} \cdot h(\get(C,\sigma_s)(r'), r' \uplus \sigma_r)\Big).
        \]
\end{theorem}

\begin{corollary}
        \label{coro:measure-density-correspondence}
        $\Meas(C,A) = \int \rho(\dd r)\,(\ind{r \in A} \cdot \density(C,s_I)(r))$
        for all $C$ and all measurable $A$.
\end{corollary}
\begin{proof}
        We instantiate Theorem~\ref{thm:correspondence:measure-density} with $h(a,r) = \ind{r \in A}$ and $\sigma = \sigma_I$.
        Recall that $(\sigma_I)_s = s_I$ and $(\sigma_I)_r = []$. Thus, the conclusion of Theorem~\ref{thm:correspondence:measure-density} in this case says that
        \begin{align*}
                \int \db{C}(\sigma_I,\dd (\sigma',w'))\, (w' \cdot \ind{\sigma'_r \in A})
                & = \int \rho (\dd r')\, \Big(\density(C,s_I)(r') \cdot \ind{r' \# (\sigma_I)_r} \cdot \ind{r' \uplus (\sigma_I)_r \in A}\Big)
                \\
                & = \int \rho (\dd r')\, \Big(\density(C,s_I)(r') \cdot \ind{r' \in A}\Big).
        \end{align*}
        This gives the equation claimed by the corollary.
\end{proof}
\noindent
This corollary says that $\density(C,s_I)$ is the density of the measure $\Meas(C,\cdot)$ with respect to $\rho$, and supports our informal claim that $\db{C}_d$ computes the density of the measure $\db{C}$.
%
%


\section{Stochastic variational inference}
\label{sec:inference}
In this section, we explain stochastic variational inference (SVI) algorithms using the semantics that we have developed so far. In particular, we describe the requirements made implicitly by one fundamental SVI algorithm, which is regarded most permissive by the ML researchers because the algorithm does not require the differentiability of the density of a given probabilistic model.

We call a command $C$ \emph{model} if it has a finite nonzero normalising
constant: 
\[
        Z_C = \Meas(C,\Rdb) = \Big(\int \db{C}(\initsigma)(\dd(\sigma,w))\, w\Big) \in (0,\infty),
\]
where $\initsigma$ is the initial state. Given a model $C$, the SVI algorithms attempt to infer a good approximation of $C$'s posterior distribution $\Pr(C,\cdot)$ defined in \eqref{eqn:posterior-definition}. They tackle this posterior-inference problem in two steps. 

First, the SVI algorithms fix a collection of approximate distributions. They usually do so by asking the developer of $C$ to provide a command $D_\theta$ parameterised by $\theta \in \mathbb{R}^p$, which can serve as a template for approximation distributions. The command $D_\theta$ typically has a control-flow structure similar to that of $C$, but it is simpler than $C$: it does not use any score statements, and may replace complex computation steps of $C$ by simpler ones. In fact, $D_\theta$ should satisfy two formal requirements, which enforce this simplicity. The first is 
\[
        \Meas(D_\theta,\Rdb) = 1 \quad \text{ for all $\theta \in \mathbb{R}^p$},
\]
which means that the normalising constant of $D_\theta$ is $1$. The second is that $D_\theta$ should keep the score (i.e., the $w$ component) to be $1$, i.e., 
\[
        \db{D_\theta}(\initsigma)(\State \times ([0,\infty)\setminus\{1\})) = 0.
\]
Meeting these requirements is often not too difficult. A common technique is
to ensure that $D_\theta$ does not use the score statement and always terminates. Figure~\ref{fig:model-guide-example1} gives an example of $(C,D_\theta)$ for simple Bayesian linear regression with three data points. Note that in this case, $D_\theta$ is obtained from $C$ by deleting the score statements and replacing the arguments $0.0$ and $5.0$ of normal distributions by parameter $\theta = (\theta_1,\theta_2,\theta_3,\theta_4)$. Following the terminology of Pyro, we call a parameterised command $D_\theta$ \emph{guide} if it satisfies the two requirements just mentioned.

Second, the SVI algorithms search for a good parameter $\theta$ that makes the distribution described by $D_\theta$ close to the posterior of $C$. Concretely, they formulate an optimisation problem where the optimisation objective expresses that some form of distance from $D_\theta$'s distribution to $C$'s posterior should be minimised. Then, they solve the problem by a version of gradient descent. 

\begin{figure}[t]
  \noindent
  \[
  \begin{array}{@{}r@{\;\;}c@{\;\;}c@{}l@{}}
    \text{model } C
    & \equiv & \big( &
    s := \csample_\cnorm({\rm ``slope"},0.0, 5.0 ); \;
    i := \csample_\cnorm({\rm ``intercept"}, 0.0, 5.0 );
    \\ & & &
    x_1 := 1.0; \;
    y_1 := 2.3; \;
    x_2 := 2.0; \;
    y_2 := 4.2; \;
    x_3 := 3.0; \;
    y_3 := 6.9;
    \\ & & &
    \cscore_\cnorm( y_1, s \cdot x_1 + i, 1.0 ); \;
    \cscore_\cnorm( y_2, s \cdot x_2 + i, 1.0 ); \;
    \cscore_\cnorm( y_3, s \cdot x_3 + i, 1.0 )
    \big)
    \\[1ex]
    \text{guide } D_{\theta}
    & \equiv & \big( &
    s := \csample_\cnorm({\rm ``slope"}, \theta_1, \cexp(\theta_2) ); \;
    i := \csample_\cnorm({\rm ``intercept"}, \theta_3, \cexp(\theta_4) )
    \big)
  \end{array}
  \]
  \hrule \vspace{-1ex}
  \caption{Example model-guide pair for simple Bayesian linear regression.\label{fig:model-guide-example1}}
\end{figure}

The KL divergence is a standard choice for distance. Let $\mu,\mu'$ be measures on $\Rdb$ that have densities $g$ and $g'$ with respect to the measure $\rho$. The KL divergence from $g$ to $g'$ is defined by
\begin{equation}
        \label{eqn:KL:definition}
        \KL(g{||}g') \defeq 
        \int \rho(\dd r)\, \left(g(r) \cdot \log \frac{g(r)}{g'(r)}\right).
\end{equation}
In words, it is the ratio of densities $g$ and $g'$ averaged according to $g$. If $g = g'$, the ratio is always $1$, so that the KL becomes $0$.
The KL divergence is defined only if the following conditions are met: 
\begin{itemize} 
        \item \emph{Absolute continuity}: $g'(r) = 0 \implies g(r) = 0$ for all $r \in \Rdb$,\footnote{This condition can be relaxed in a more general formulation of the KL divergence stated in terms of the so called Radon-Nikodym derivative. We do not use the relaxed condition to reduce the amount of materials on \mbox{measure theory in the paper.}} which ensures that the integrand in \eqref{eqn:KL:definition} is well-defined even when the denominator $g'(r)$ in \eqref{eqn:KL:definition} takes the value $0$; 
        \item \emph{Integrability}: the integral in \eqref{eqn:KL:definition} has a finite value. 
\end{itemize} 

Using our semantics, we can express the KL objective as follows:
\begin{equation}
        \label{eqn:SVI:objective}
        \argmin_{\theta \in \mathbb{R}^p} \KL\left(\density(D_\theta,s_I) {\Big|\Big|} \frac{\density(C,s_I)}{Z_C}\right).
\end{equation}
Recall that $\density(D_\theta,s_I)$ and $\density(C,s_I)/Z_C$ are densities of the probability measures of the command $D_\theta$ and the posterior of $C$ (Corollary~\ref{coro:measure-density-correspondence}), and they are defined by means of our density semantics in \S\ref{subsec:posterior-density}. Most SVI engines solve this optimisation problem by a version of gradient descent.

In the paper, we consider one of the most fundamental and versatile SVI algorithms. The algorithm is called score estimator or REINFORCE, and it works by estimating the gradient of the objective in \eqref{eqn:SVI:objective} using samples and performing the gradient descent with this estimated gradient. More concretely, the algorithm starts by initialising $\theta$ with some value (usually chosen randomly) and updating it repeatedly by the following procedure:
\begin{align*}
        \text{(i)}\,\, & \text{Sample $r_1,\ldots,r_N$ independently from $\density(D_\theta,s_I)$}
        \\
        \text{(ii)}\,\, & \theta \leftarrow \theta - \eta \times \left(\frac{1}{N} \sum_{i = 1}^N \Big(\nabla_\theta \log \density(D_\theta,s_I)(r_i)\Big) \cdot \log \frac{\density(D_\theta,s_I)(r_i)}{\density(C,s_I)(r_i)}\right)
\end{align*}
Here $N$ and $\eta$ are hyperparameters to this algorithm, the former determining the number of samples used to estimate the gradient and the latter, called learning rate, deciding how much the algorithm should follow the direction of the estimated gradient. Although we do not explain here, sampling $r_1,\ldots,r_N$ and computing all of $\density(D_\theta,s_I)(r_i)$, $\density(C,s_I)(r_i)$ and $\nabla_\theta (\log \density(D_\theta,s_I)(r_i))$ can be done by executing $D_\theta$ and $C$ multiple times under slightly unusual operational semantics~\cite{YangPPCourseNote19}. The SVI engines of Pyro and Anglican implement such operational semantics. 

The average over the $N$ terms in the $\theta$-update step is the core of the algorithm. It approximates the gradient of the optimisation objective in \eqref{eqn:SVI:objective}:
\begin{align*}
        \nabla_\theta
        \KL\left(\density(D_\theta,s_I) {\Big|\Big|} \frac{\density(C,s_I)}{Z_C}\right)
        \approx
        \frac{1}{N} \sum_{i = 1}^N \Big(\nabla_\theta \log \density(D_\theta,s_I)(r_i)\Big) \cdot \log \frac{\density(D_\theta,s_I)(r_i)}{\density(C,s_I)(r_i)}.
\end{align*}
The average satisfies an important property called \emph{unbiasedness},
summarised by Theorem~\ref{thm:score-unbiasedness}.
\begin{theorem}
\label{thm:score-unbiasedness}
Let $C$ be a model, $D_\theta$ be a guide, and $N \neq 0 \in \bN$. 
Define $\KL_{(-)} : \bR^p \to \bR_{\geq 0}$ as
$\KL_\theta$ $\defeq$ $\KL(\density(D_\theta,s_I) {||} \density(C,s_I)/Z_C)$.
Then, $\KL_{(-)}$ is well-defined and continuously differentiable with
\begin{align}
  \label{eqn:score-estimator}
        &
        \nabla_\theta \KL_\theta
        =
        \mathbb{E}_{\prod_i\density(D_\theta,s_I)(r_i)}
        \left[ 
        \frac{1}{N}\sum_{i=1}^N
                \Big(\nabla_\theta \log \density(D_\theta,s_I)(r_i)\Big)
                \log \frac{\density(D_\theta,s_I)(r_i)}{\density(C,s_I)(r_i)}
        \right]
\end{align}
if
\begin{enumerate}[label=(R\arabic*), ref=R\arabic*]
\item\label{req:1}
  $\density(C,s_I)(r) = 0 \implies \density(D_\theta,s_I)(r) = 0$,
        for all $r \in \Rdb$ and $\theta \in \mathbb{R}^p$;
\item\label{req:2}
  for all $(r,\theta,j) \in \Rdb \times \mathbb{R}^p \times [p]$, the function $v \longmapsto \density(D_{\theta[j:v]},s_I)(r)$ on $\bR$ is differentiable;
\item\label{req:3}
  for all $\theta \in \mathbb{R}^p$,
        \[
                \int \rho(\dd r)\,
                \left(\density(D_\theta,s_I)(r) \cdot
                \log\frac{\density(D_\theta,s_I)(r)}{\density(C,s_I)(r)}\right)
                < \infty;
        \]
\item\label{req:4}
  for all $(\theta,j) \in \mathbb{R}^p \times [p]$, the function 
        \[
                v \longmapsto 
        \int \rho(\dd r)\,
                \left(\density(D_{\theta[j:v]},s_I)(r) \cdot
                \log\frac{\density(D_{\theta[j:v]},s_I)(r)}{\density(C,s_I)(r)}\right)
        \]
        on $\mathbb{R}$ is continuously differentiable;
\item\label{req:5}
  for all $\theta \in \mathbb{R}^p$, 
\begin{multline*}
  \!\!\!\!\!\!\!\!\!
        \nabla_\theta
        \int \rho(\dd r)
                \left(\density(D_\theta,s_I)(r) \cdot
                \log \frac{\density(D_\theta,s_I)(r)}{\density(C,s_I)(r)}\right)
        = 
        \int \rho(\dd r)\,
        \nabla_\theta
        \left(\density(D_\theta,s_I)(r) \cdot
                \log \frac{\density(D_\theta,s_I)(r)}{\density(C,s_I)(r)}\right);
\end{multline*}%
\item\label{req:6}
  for all $\theta \in \mathbb{R}^p$,
        \[
        \int \rho(\dd r)\,
        \nabla_\theta
        \density(D_\theta,s_I)(r)
        =
        \nabla_\theta
        \int \rho(\dd r)\,
        \density(D_\theta,s_I)(r).  
        \]
\end{enumerate}
Here $\theta[j : v]$ denotes a vector in $\bR^p$
that is the same as $\theta$ except that its $j$-th component is $v$.
\end{theorem}
\noindent
The conclusion of this theorem (Equation~\eqref{eqn:score-estimator})
and its proof are well-known~\cite{RanganathGB14},
but the requirements in the theorem
(and the continuous differentiability of $\KL_\theta$ in the conclusion)
are rarely stated explicitly in the literature.

The correctness of the algorithm crucially relies on the unbiasedness property in Theorem~\ref{thm:score-unbiasedness}. The property ensures that the algorithm converges to a local minimum with probability $1$. Thus, it is important that the requirements in the theorem are met. In fact, some of the requirements there are needed even to state the optimisation objective in \eqref{eqn:SVI:objective}, because without them, the objective does not exist. In the next section, we describe conditions that imply those requirements and can serve as target properties of program analysis for probabilistic programs. The latter point is worked out in detail in \S\ref{sec:analysis} and \S\ref{s:8:pyroai} where we discuss program analysis for probabilistic programs and SVI.


\section{Conditions for stochastic variational inference}
\label{sec:conditions}

Ideally we want to have program analysers that discharge the six requirements \ref{req:1}-\ref{req:6} in Theorem~\ref{thm:score-unbiasedness}.
However, except \ref{req:1} and \ref{req:2},
the requirements are not ready for serving as the targets of static analysis algorithms. Automatically discharging them based on the first principles (such as the definition of integrability with respect to a measure) may be possible, but seems less immediate than doing so using powerful theorems from continuous mathematics.

In this section, we explain conditions that imply the requirements \ref{req:3}-\ref{req:6} 
and are more friendly to program analysis than the requirements themselves. The conditions are given in two boxes \eqref{eqn:condition-2b} and \eqref{eqn:condition-456}.
Throughout the section, we fix a model $C$ and a guide $D_\theta$.

%
%

\subsection{Assumption}
\label{subsec:requirement-assumption}

Throughout the section, we assume that the densities of $D_\theta$ and $C$ have the following form:
\begin{align}
  \label{eqn:guide-factorisation}
  & \density(D_\theta,s_I)(r) = 
        \sum_{i=1}^M 
        \ind{r \in A_i} 
        \prod_{\alpha \in K_i}
        \cN\big(r(\alpha); \mu_{(i,\alpha)}(\theta),\sigma_{(i,\alpha)}(\theta)\big),
  \\
  \nonumber 
  & \density(C,s_I)(r) = 
        \sum_{i=1}^M 
        \ind{r \in A_i} 
        \left(\prod_{\alpha \in K_i}
        \cN\big(r(\alpha); \mu'_{(i,\alpha)}(r),\sigma'_{(i,\alpha)}(r)\big)\right)
        \left(\prod_{j \in [N_i]}
        \cN\big(c_{(i,j)}; \mu''_{(i,j)}(r),\sigma''_{(i,j)}(r)\big)\right),
\end{align}
where 
\begin{itemize} 
        \item $M, N_i \in \bN \setminus \{0\}$; 
        \item $A_1,\ldots,A_M$ are disjoint measurable subsets of $\Rdb$; 
        \item $K_i$'s are finite sets of strings such that $\dom(r) = K_i$ for all $r \in A_i$; 
        \item $\mu_{(i,\alpha)}$ and $\sigma_{(i,\alpha)}$ are measurable functions from $\bR^p$ to $\bR$ and $(0,\infty)$, respectively; 
        \item $\mu'_{(i,\alpha)}$ and $\mu''_{(i,j)}$ are measurable functions from $[K_i \to \bR]$ to $\bR$; 
        \item $\sigma'_{(i,\alpha)}$ and $\sigma''_{(i,j)}$ are measurable functions from $[K_i \to \bR]$ to $(0,\infty)$; 
        \item $c_{(i,j)}$ is a real number.
\end{itemize} 
In programming terms, our assumption first implies that both $C$ and $D_\theta$ use at most a fixed number of random variables. That is, the number of random variables they generate must be finite not only within a single execution but also across all possible executions 
since the names of all random variables are found in a finite set $\bigcup_{i=1}^M K_i$. This property is met if the number of steps in every execution of $C$ and $D_\theta$ from $\sigma_I$ is bounded by some $T$ and the executions over the same program path use the same set of random variables. Note that the bound may depend on $s_I$. Such a bound exists for most probabilistic programs.\footnote{Notable exceptions are models using probabilistic grammars or those from Bayesian nonparametrics.} Also, the assumption says that the parameters of normal distributions in sample statements in $D_\theta$ may depend only on $\theta$, but not on other sampled random variables. This is closely related to a common approach for designing approximate distributions in variational inference, called mean-field approximation, where the approximate distribution consists of independent normal random variables.

We use the term ``assumption'' here, instead of ``condition'' in the following subsections because the assumed properties are rather conventional and they are not directly related to the requirements in Theorem~\ref{thm:score-unbiasedness}, at least not as much as the conditions that we will describe next. 

\subsection{Condition for the requirement~\ref{req:3}}
\label{subsec:condition-3}

Note that the integral in the requirement~\ref{req:3} can be written as the sum of two expectations:
\begin{align}
        &
        \int \rho(\dd r)\,
                \left(\density(D_\theta,s_I)(r) \cdot
                \log\frac{\density(D_\theta,s_I)(r)}{\density(C,s_I)(r)}\right)
                \nonumber
        \\
        & \quad {} =
        \EE{\density(D_\theta,s_I)(r)}{\log \density(D_\theta,s_I)(r)}
        - 
        \EE{\density(D_\theta,s_I)(r)}{\log \density(C,s_I)(r)}.
        \label{eqn:KL-decompose}
\end{align}


The minus of the first term (i.e., $-\EE{\density(D_\theta,s_I)(r)}{\log \density(D_\theta,s_I)(r)}$) is called the \emph{differential entropy} of the density $\density(D_\theta,s_I)$. Intuitively, it is large when the density on $\bR^n$ is close to the Lebesgue measure, which is regarded to represent the absence of information. The differential entropy is sometimes undefined~\cite{GhourchianCL17}.
Fortunately, a large class of probability densities (containing many commonly used probability distributions) have well-defined entropies~\cite{NairArxiv06, GhourchianCL17}. Our $\density(D_\theta,s_I)$ is one of such fortunate cases.

\begin{theorem}
        \label{thm:condition-2a}
        $\EE{\density(D_\theta,s_I)(r)}{|\log \density(D_\theta,s_I)(r)|} < \infty$
	under our assumption in \S\ref{subsec:requirement-assumption}.
\end{theorem}

We remark that a violation of the assumption~\eqref{eqn:guide-factorisation}
for $D_\theta$ can result in an undefined entropy, as illustrated by the following examples.
\begin{example}
  \label{eg:condition-2a}
  Consider guides $D_{(i, \theta)}$ defined as follows ($i =1,2$):
  \[
  D_{(i,\theta)} \equiv
  (x_1 := \csample_\cnorm(``a_1", \theta_1,1);\; x_2 := \csample_\cnorm(``a_2", \theta_2,E_i[x_1]))
  \]
  where for some $n \geq 1$ and $c \neq 0 \in \mathbb{R}$,\footnote{Formally, we should
  implement $E_1[x_1]$ as an application of a primitive function $f_1$ to $x_1$ that has
  the semantics described by the if-then-else statement here.}
  \begin{align*}
    E_1[x_1] & \equiv \code{if}\,(x_1{=}0)\,\code{then}\, 1\,\code{else}\,{\code{exp}(-1/|x_1|^n)},
    &
    E_2[x_1] & \equiv \code{exp}(\code{exp}(c \cdot x_1^3)).
  \end{align*}
  None of $\density(D_{(i,\theta)},s_I)$'s
  satisfies the assumption~\eqref{eqn:guide-factorisation}
  because the standard deviation of the normal distribution for
  $x_2$ depends on the value of $x_1$. The entropies of $\density(D_{(i,\theta)},s_I)$'s
  are all undefined: $\bE_{\density(D_{(i,\theta)},s_I)(r)}{[|\log \density(D_{(i,\theta)},s_I)(r)|]} = \infty$
  for all $i = 1,2$. 
  \qed
\end{example}

Since the first term of~\eqref{eqn:KL-decompose} is always finite
by Theorem~\ref{thm:condition-2a}, it is enough to ensure the finiteness of the second term of~\eqref{eqn:KL-decompose}.
For $i \in [M]$, define the set of (absolute) affine functions on $[K_i \to \bR]$ as:
\[
        \cA_i \defeq \Big\{ f \in [[K_i \to \bR] \to \bR] \,~\Big|~\,
        f(r) = c + \sum_{\alpha \in K_i} (c_\alpha \cdot |r(\alpha)|)\
        \text{ for some $c, c_\alpha \in \bR$}\Big\}.
\]
Our condition for ensuring the finiteness of the second term
is as follows:
\begin{align}
  \label{eqn:condition-2b}
  \Aboxed{
    \begin{array}{l}
      \text{For all $i\in [M]$, there are $f',f'',l',u',l'',u'' \in \cA_i$ 
        such that}
      \\[.5ex]
      \qquad
      |\mu'_{(i,\alpha)}(r)| \leq \exp(f'(r)),
      \qquad\
      \exp(l'(r)) \leq \sigma'_{(i,\alpha)}(r) \leq \exp(u'(r)),
      \\[.5ex]
      \qquad
      |\mu''_{(i,j)}(r)| \leq \exp(f''(r)),
      \qquad
      \exp(l''(r)) \leq \sigma''_{(i,j)}(r) \leq \exp(u''(r)),
      \\[.5ex]
      \text{all four hold for every $(\alpha, j, r) \in K_i \times [N_i] \times A_i$.}
    \end{array}
  }
\end{align}
%
\begin{theorem}
        \label{thm:condition-2}
	The condition \eqref{eqn:condition-2b} implies
        $\EE{\density(D_\theta,s_I)(r)}{|\log \density(C,s_I)(r)|} < \infty$
        under our assumption in \S\ref{subsec:requirement-assumption}. 
        Thus, in that case, it entails the requirement~\ref{req:3} (i.e., the objective in \eqref{eqn:SVI:objective} is well-defined).
\end{theorem}

Our condition in \eqref{eqn:condition-2b} is sufficient but not necessary for the objective in \eqref{eqn:SVI:objective} to be defined. However, its violation is a good warning, as illustrated by our next examples.
\begin{example}
  \label{eg:condition-2b}
  Consider models $C_1, \ldots, C_4$ and a guide $D_\theta$ defined
  as follows:
  \[\begin{array}{rll}
          C_i \equiv {} \!\!\!\!\!
          & (x_1 := \csample_\cnorm(``a_1", 0,1);\; x_2 := \csample_\cnorm(``a_2", E_i[x_1],1)) 
          & \text{for $i=1,2$}
          \\[.5ex]
          C_i \equiv {} \!\!\!\!\!
          & (x_1 := \csample_\cnorm(``a_1", 0,1);\; x_2 := \csample_\cnorm(``a_2",  0, E_i[x_1]))
          &\text{for $i=3,4$}
          \\[.5ex]
          D_\theta \equiv {} \!\!\!\!\!
          & (x_1 := \csample_\cnorm(``a_1", \theta_1, 1);\; x_2 := \csample_\cnorm(``a_2", \theta_2, 1))
  \end{array}\]
  where for some $n \geq 1$ and $c \neq 0 \in \mathbb{R}$,
  \begin{align*} 
          E_1[x_1] & \,{\equiv}\, \code{if}\,(x_1{=}0)\,\code{then}\,0\,\code{else}\,\frac{1}{x_1^n},
          &
          E_2[x_1] & \,{\equiv}\, E_4[x_1] \,{\equiv}\, \code{exp}(c {\cdot} x_1^3),
          &
          E_3[x_1] & \,{\equiv}\, \code{if}\,(x_1{=}0)\,\code{then}\, 1\,\code{else}\,{|x_1|^n}.
  \end{align*}
  Let $A = [\{``a_1",``a_2"\} \to \bR]$. For $r \in A$, define
  \[\begin{array}{rlrl}
          \mu'_1(r) \!\!\!\!
          & \defeq \text{if } r(``a_1") = 0 \text{ then } 0 \text{ else } 1/r(``a_1")^{n},
          &
          \qquad\mu'_2(r) \!\!\!\!
          & \defeq \exp({c \cdot r(``a_1")^3}),
          \\[.5ex]
          \sigma'_3(r) \!\!\!\!
          & \defeq \text{if } r(``a_1") = 0\text{ then } 1 \text{ else } |r(``a_1")|^n,
          &
          \qquad\sigma'_4(r) \!\!\!\!
          & \defeq \exp({c \cdot r(``a_1")^3}).
  \end{array}\]
  Then, we have that
  \[\begin{array}{rll}
          \density(C_i,s_I)(r) \!\!\!\!
          & = 
          \ind{r \in A} 
          \cdot \cN(r(``a_1");0,1) 
          \cdot \cN(r(``a_2");\mu'_i(r),1)
          & \text{ for $i = 1,2$}
          \\[.5ex]
          \density(C_i,s_I)(r) \!\!\!\!
          & = 
          \ind{r \in A} 
          \cdot \cN(r(``a_1");0,1) 
          \cdot \cN(r(``a_2");0,\sigma'_i(r))
          & \text{ for $i = 3,4$}
          \\[.5ex]
          \density(D_\theta,s_I)(r) \!\!\!\!
          & =
          \ind{r \in A}
          \cdot
          \cN(r(``a_1");\theta_1,1)
          \cdot
          \cN(r(``a_2");\theta_2,1).
  \end{array}\]
        None of $\mu'_1$, $\mu'_2$, $\sigma'_3$, and $\sigma'_4$ satisfies the condition in \eqref{eqn:condition-2b}. The function $\mu'_1$ is not bounded in $\{ r \in A \,\mid\, -1 \leq r(``a_1") \leq 1 \wedge -1 \leq r(``a_2") \leq 1\}$, but every $\mu'$ satisfying the condition in \eqref{eqn:condition-2b} should be bounded. Also, the cubic exponential growth of $\mu'_2$ cannot be bounded by any linear exponential function on $r$. The violation of the condition by $\sigma'_3$ and $\sigma'_4$ can be shown similarly. 
        
       In fact, the objective in \eqref{eqn:SVI:objective} is not defined for all of the four cases. This is because 
       for all $i = 1, \ldots, 4$,
       $\EE{\density(D_\theta,s_I)(r)}{|\log \density(C_i,s_I)(r)|} = \infty$.
\qed
\end{example}

We now show that the condition in \eqref{eqn:condition-2b} is satisfied by a large class of functions in machine learning applications,
including functions using neural networks.
We call a function $\mathit{nn} :\bR^n \to \bR$ an \emph{affine-bounded neural network}
if there exist functions
$f_j : \bR^{n_j} \to \bR^{n_{j+1}}$
and affine functions $l_j : \bR^{n_j} \to \bR$ for all $1 \leq j \leq d$
such that (i) $n_1 = n$ and $n_{d+1} = 1$; (ii) $\nn = f_d \circ \cdots \circ f_1$; and (iii)
$\|f_j(v)\|_1 \leq l_j(|v_1|,\ldots,|v_{n_j}|)$
for all $1 \leq j \leq d$ and $v \in \bR^{n_j}$,
where $\|\cdot\|_1$ denotes the $\ell_1$-norm.
Note that each component of $f_j$ can be, for instance, an affine function, one of commonly used activation functions (e.g., relu, tanh, sigmoid, softplus), or one of min/max functions ($\min$ and $\max$).
Therefore, most of neural networks used in machine learning applications
are indeed affine-bounded.
Lemma~\ref{lem:condition-2b-eg} indicates that
a wide range of functions satisfy the condition~\eqref{eqn:condition-2b}.
\begin{lemma}
  \label{lem:condition-2b-eg}
  Pick $i \in [M]$ and $\alpha \in K_i$.
  Let $(\alpha_1,\ldots,\alpha_J)$ be an enumeration of the elements in $K_i$,
  and $\overline{r} \defeq (r(\alpha_1),\ldots,r(\alpha_J)) \in \bR^J$
  be an enumeration of the values of $r \in A_i$.
  Consider any affine-bounded neural network $\mathit{nn}: \bR^J \to \bR$,
  polynomial $\mathit{poly}: \bR^J \to \bR$,
  and $c\in (0,\infty)$.
  Then, the below list of functions $\mu'_{(i,\alpha)}$ and $\sigma'_{(i,\alpha)}$ on $A_i$
  satisfy the condition in~\eqref{eqn:condition-2b}:
  \[\begin{array}{r@{~}lr@{~}lr@{~}l}
    \mu'_{(i,\alpha)}(r) &= \mathit{nn}(\overline{r}),\quad
    & \mu'_{(i,\alpha)}(r) &= \mathit{poly}(\overline{r}),\quad
    & \mu'_{(i,\alpha)}(r) &= \mathrm{exp}(\mathit{nn}(\overline{r})),
    \\[.5ex]
    \sigma'_{(i,\alpha)}(r) &= |\mathit{nn}(\overline{r})|+c,\quad
    & \sigma'_{(i,\alpha)}(r) &= |\mathit{poly}(\overline{r})|+c,\quad
    & \sigma'_{(i,\alpha)}(r) &= \mathrm{exp}(\mathit{nn}(\overline{r})),
    \\[.5ex]
    \sigma'_{(i,\alpha)}(r) &= (|\mathit{nn}(\overline{r})|+c)^{-1},\quad
    & \sigma'_{(i,\alpha)}(r) &= (|\mathit{poly}(\overline{r})|+c)^{-1},\quad
    & \sigma'_{(i,\alpha)}(r) &= \mathrm{softplus}(\mathit{nn}(\overline{r})),
  \end{array}\]
  where $\mathrm{softplus}(v) \defeq \log(1+\exp(v))$.
  Moreover, the same holds for $\mu''_{(i,j)}$ and $\sigma''_{(i,j)}$ as well.
\end{lemma}

Lemma~\ref{lem:condition-2b-eg} follows from properties about functions having affine-exponential functions as their upper or lower bounds, summarised by the next lemma:
\begin{lemma}
  \label{lem:lin-exp}
  Assume the setting of Lemma~\ref{lem:condition-2b-eg}.
  Then, we have the following.
  \begin{itemize}
  \item
    $\EE{q(\overline{r})}{\exp(l(r))} < \infty$ for
    every $l \in \cA_i$ and every $J$-dimensional normal distribution $q$.
  \item
    For some $l,l' \in \cA_i$,
    $|\mathit{poly}(\overline{r})| \leq \exp(l(r))$
    and
    $|\nn(\overline{r})| \leq l'(r)$
    for all $r \in A_i$.
  \item
    For some affine functions $l,l' : \bR \to \bR$, 
    $\exp(l(|v|)) \leq \mathrm{softplus}(v) \leq \exp(l'(|v|))$
    for all $v \in \bR$.
  \item
    For every $l_1, l_2 \in \cA_i$ and every $\odot \in \{ +, -, \times, /, \max\}$,
    there exists $l \in \cA_i$ such that
    $\exp(l_1(r)) \odot \exp(l_2(r)) \leq \exp(l(r))$
    for all $r \in A_i$.
  \end{itemize}
\end{lemma}


\subsection{Condition for the requirements~\ref{req:4}-\ref{req:6}}
\label{subsec:condition-456}

Assume that the model $C$ and the guide $D_\theta$ satisfy our assumption and condition in the previous two subsections.
Our condition for the requirements~\ref{req:4}-\ref{req:6} is given below:
\begin{align}
  \label{eqn:condition-456}
  \Aboxed{
    \begin{array}{@{\,}l@{\,}}
            \text{For all $i \in [M]$, $\alpha \in K_i$,
              and $(\theta, j) \in \mathbb{R}^p \times [p]$,}
            \\[.5ex]
            \quad \text{the function $v\in\bR \longmapsto \mu_{(i,\alpha)}(\theta[j : v])$
            is continuously differentiable;}
            \\[.5ex]
            \quad \text{the function $v\in\bR \longmapsto \sigma_{(i,\alpha)}(\theta[j : v])$
            is continuously differentiable.}
    \end{array}
    }
\end{align}

\begin{theorem}
        \label{thm:condition-456}
        If both our assumption in \S\ref{subsec:requirement-assumption}
        and the condition \eqref{eqn:condition-2b} hold, then
        the condition \eqref{eqn:condition-456} implies the requirements~\ref{req:4}-\ref{req:6}.
\end{theorem}

The proof of the theorem uses the following nontrivial result~\cite[Theorem~6.28]{klenkeBook14} about exchanging differentiation and integration, a consequence of the dominated convergence theorem.
\begin{theorem}
  \label{thm:diff-under-int}
  Let $V \subset \bR$ be an open interval, and $(X,\Sigma,\mu)$ be a measure space.
  Suppose that a measurable function $f: V \times X \to \bR$ satisfies the following conditions:
  (a) for all $v \in V$, the integral $\int \mu(\dd x)\, f_v(x)$ is well-defined;
  (b) for almost all $x \in X$ (w.r.t. $\mu$) and all $v \in V$, the partial derivative $\nabla_v f_v(x)$
       with respect to $v$ is well-defined;\footnote{A more popular notation is $(\partial f_v(x))/(\partial v)$, but we opt for $\nabla_v f_v(x)$ to avoid clutter.}
  (c) there is a measurable function $h : X \to \bR$ such that $\int \mu(\dd x)\, h(x)$ is well-defined
          and $\left|\nabla_v f_v(x)\right| \leq h(x)$ for all $v \in V$ and almost all $x \in X$ (w.r.t. $\mu$).
  Then, for all $v \in V$,
  both sides of the below equation are well-defined,
  and the equality holds:
  \[
          \nabla_v \int \mu(\dd x)\, f_v(x) 
          = 
          \int \mu(\dd x)\,\nabla_v f_v(x).
  \]
\end{theorem}
\noindent
Note that the theorem ensures
not only the validity of interchanging differentiation and integration,
but also the differentiability of $\int \mu(\dd x)\, f_v(x)$ (w.r.t. $v$)
and the integrability of $\nabla_v f_v(x)$ over $x \in X$.

Our condition in~\eqref{eqn:condition-456} is sufficient but not necessary
for the requirements~\ref{req:4}-\ref{req:6} to hold,
in particular for the objective in~\eqref{eqn:SVI:objective}
to have well-defined partial derivatives in $\theta$.
However, its violation is a good indication of a potential problem. The following example illustrates this point.

\begin{example}
  Consider a model $C$ and a guide $D_\theta$ defined as follows:
  \begin{align*}
    C & \equiv x := \csample_\cnorm(``a", 0,1)
    &
    D_\theta & \equiv x := \csample_\cnorm(``a", 0,E[\theta])
  \end{align*}
  where $E[\theta] \equiv \code{relu}(\theta)+2$
  and ${\rm relu}(v) \defeq \ind{v \geq 0} \cdot v$.
  Such $E[\theta]$ can definitely appear in machine learning applications,
  once a guide starts to use neural networks with parameters $\theta$.
  Let $A = [\{``a"\} \to \bR]$ and $\sigma(\theta) \defeq {\rm relu}(\theta)+2$.
  Then,
  $\density(C,s_I)(r) = \ind{r \in A} \cdot \cN(r(``a"); 0, 1)$
  and
  $\density(D_\theta, s_I)(r) = \ind{r \in A} \cdot \cN(r(``a"); 0, \sigma(\theta)).$
  Note~\eqref{eqn:condition-456} is violated:
  $\sigma$ is non-differentiable at $\theta=0$.
  A simple calculation shows:
  \begin{align*}
    &\nabla_\theta
    \int \rho(\dd r) \left( \density(D_\theta, s_I)(r) 
      \cdot \log \frac{\density(D_\theta, s_I)(r)}{ \density(C, s_I)(r)} \right)
      = \left\{
    \begin{array}{ll}
      0 & \text{if $\theta \in (-\infty,0)$} \\
      ((2+\theta)^2-1)/(2+\theta) & \text{if $\theta \in (0,\infty)$} \\
      {\rm undefined} & \text{if $\theta = 0$}.
    \end{array}\right.
  \end{align*}
  Hence, the objective in~\eqref{eqn:SVI:objective}
  does not have a well-defined partial derivative in $\theta$ at $\theta = 0$.
  \qed
\end{example}


\section{Analysis}
\label{sec:analysis}
In this section, we describe a recipe for building a static analysis
that automatically discharges some of the assumptions and conditions
given in \S\ref{sec:conditions}.
The recipe ensures that the constructed static analyses are sound with
respect to the density semantics in \S\ref{subsec:posterior-density}.
We illustrate it by describing four static analyses for verifying
the model-guide support match
(the requirement~\ref{req:1}), 
the guide-parameter differentiability
(the requirement~\ref{req:2}), 
the condition~\eqref{eqn:condition-2b},
and the condition~\eqref{eqn:condition-456}.
The analysis for the model-guide support match has been developed
significantly more for the Pyro programming language, and applied
to analyse realistic examples of the language.
This fully-blown analysis and our experiments will be described in
\S\ref{s:8:pyroai}.

Throughout this section, we assume that the parameters $\theta$ of a
guide $D_\theta$ are included in $\Var$, and are only read by $D_\theta$
and not accessed by a model $C$.
What we used to call $s_I$ will be the part of the store for variables
in $\Var \setminus \theta$, and what we used to write $\theta$ will
correspond to the other part for $\theta$. 

\subsection{A generic program analysis framework}
\label{s:7:1:fwk}
Our recipe is for building a static analysis that infers
information about the state transformation of a given command.
It is similar to the conventional methodology for building a so called \emph{relational} static analysis,
which also attempts to find information about the relationship between input and 
output states of a given command. However, our recipe diverges from the convention
in one important point: while the abstract states of conventional relational analyses
represent relations on states, we let abstract states directly express sets 
of concrete state transformers. This departure from the convention is due to 
the difficulty of using relations for expressing properties of state transformers 
that we desire. For instance, we could not express a set of functions with a certain type
of differentiability using relations.

Recall the domain $\cD$ in \eqref{eqn:definition-D}, and the notion of
\emph{admissible} subset from domain theory: $D_0 \subseteq \cD$ is
\emph{admissible} if it contains $\bot$ and is closed under taking the
limits of $\omega$-chains in $D_0$.

Our recipe assumes an abstraction instance defined by the following items:
\begin{itemize} 
\item An \emph{abstract domain}, i.e., a set \( \cT^\sharp \) with a
        designated element \(\bot^\sharp\). 
\item A \emph{concretisation function}, i.e., a function
  \( \gamma : \cT^\sharp \to \cP(\cD) \)
  such that for every $t \in \cT^\sharp$, $\gamma(t)$ is an admissible
  subset of $\cD$.
  Note that the concretisation interprets each abstract element $t$ as a
  set of concrete transformers in $\cD$. The admissibility is imposed to
  enable the sound analysis of loops.
\item A \emph{widening operator}
  \(
  \widen : \cT^\sharp \times \cT^\sharp \to \cT^\sharp,
  \)
  such that for all $t_1,t_2 \in \cT^\sharp$ and $i \in [2]$,
  \(
  \gamma(t_i) \subseteq \gamma(\widen(t_1, t_2))
  \)
  and for every sequence $\{t_n\}_{n\geq 1}$ in $\cT^\sharp$, its widened
  sequence $\{t'_n\}_{n \geq 1}$, defined by
  $t'_1 \defeq t_1$
  and
  $t'_{n+1} \defeq \widen(t'_n, t_{n+1})$
  for $n \geq 1$,
  has an index $m$ such that $t'_m = t'_{m+1}$.
\item An \emph{abstract conditional operator} for every expression $E$,
  that is, a function
  \(
  \cond(E)^\sharp : \cT^\sharp \times \cT^\sharp \to \cT^\sharp 
  \)
  such that for all $t_1,t_2 \in \cT^\sharp$ and  $g_1,g_2 \in \cD$,
  if $g_1 \in \gamma(t_1)$ and $g_2 \in \gamma(t_2)$, then
  \(
  \big(\lambda (s,r). (\text{if}\ (\db{E}s {=} \true)
  \ \text{then}\ g_1(s,r)\ \text{else}\ g_2(s,r))\big)
  \in \gamma(\cond(E)^\sharp(t_1,t_2)).
  \)
\item An \emph{abstract composition operator}
  \(
  \circ^\sharp : \cT^\sharp \times \cT^\sharp \to \cT^\sharp 
  \)
  such that for all $t_1,t_2 \in \cT^\sharp$ and $g_1,g_2 \in \cD$,
  if $g_1 \in \gamma(t_1)$ and $g_2 \in \gamma(t_2)$, then
  \(
  g_2^\ddagger \circ g_1 \in \gamma(t_2 \circ^\sharp t_1).
  \)
\item For all expressions $E_0,E_1,E_2$ and for all variables $x$, the
  abstract elements
  \( \sskip^\sharp \),
  \( \update(x,E_0)^\sharp \),
  \( \sample(x,S,E_1,E_2)^\sharp \), and
  \( \score(E_0,E_1,E_2)^\sharp \,\in\, \cT^\sharp \)
  such that 
  \begin{align*}
    \db{\cskip}_d
    & \in \gamma(\sskip^\sharp),
    &
    \db{x:=\csample_\cnorm(S,E_1,E_2)}_d
    & \in \gamma(\sample(x,S,E_1,E_2)^\sharp),
    \\
    \hspace{-1.3em}
    \db{x:=E_0}_d
    & \in \gamma(\update(x,E_0)^\sharp),
    &
    \db{\cscore_\cnorm(E_0,E_1,E_2)}_d
    & \in \gamma(\score(E_0,E_1,E_2)^\sharp).
  \end{align*}
\end{itemize} 

Given these data, we define the static analysis \( \adb{C} \in \cT^\sharp \)
of a command $C$ in Figure~\ref{f:asem}.
\begin{figure}[t]
  \noindent
  \[
  \hspace{-1.3em}
  \begin{array}{@{}r@{\;}c@{\;}lr@{\;}c@{\;}l@{}}
    \adb{\cskip} & \defeq & \sskip^\sharp
    &
    \adb{\cif\ E\ \{C_0\}\ \celse\ \{C_1\}}
    & \defeq & \cond(E)^\sharp(\adb{C_0}, \adb{C_1})
    \\[1ex]
    \adb{x:=E}
    & \defeq & \update(x,E)^\sharp
    &
    \adb{x := \csample_\cnorm(S,E_1,E_2)}
    & \defeq & \sample(x,S,E_1,E_2)^\sharp
    \\[1ex]
    \adb{C_0;C_1}
    & \defeq & \adb{C_1} \circ^\sharp \adb{C_0}
    &
    \adb{\cscore_\cnorm(E_0,E_1,E_2)}
    & \defeq & \score(E_0,E_1,E_2)^\sharp
    \\[1ex]
    \adb{\cwhile\ E\ \{C\}}
    & \omit\rlap{$
      \defeq (\wfix\, T)
      \,\,
      (\text{where $T(t') \defeq \cond(E)^\sharp(t' \circ^\sharp
        \adb{C},\, \sskip^\sharp)$})
      $}
  \end{array}
  \]
  \hrule \vspace{-1ex}
  \caption{Abstract semantics \( \adb{C} \in \cT^\sharp \) of commands \( C \)}
  \label{f:asem}
\end{figure}
Here the $(\wfix\ T)$ is the usual widened fixed point $t_\mathit{fix}$
of $T$, which is defined as the first element $t_m$ with $t_m = t_{m+1}$
in the widened sequence \( (t_n)_{n \geq 1} \) where \( t_1 \defeq
\bot^\sharp \) and \( t_{n+1} \defeq \widen(t_n, T(t_n)) \).

\begin{theorem}[Soundness]
  \label{thm:analysis-soundness}
  For all commands $C$, we have $\db{C}_d \in \gamma(\adb{C})$.
\end{theorem}
In the rest of this section, we instantiate this framework into four
static analysis instances.
In each case, we describe the abstract domain, the abstract bottom
element, and the concretisation function.
Moreover, in the first two cases, we detail the transfer functions.
In the following, for a tuple $(s',r',w',p') \in \Store \times \Rdb
\times [0,\infty) \times [0,\infty)$, we use the subscripts  $-_s$,
$-_r$, $-_w$, and $-_p$ to denote its components.
For instance, $(s',r',w',p')_s = s'$ and $(s',r',w',p')_r = r'$.

\subsection{Analysis for the model-guide match}
\label{s:7:2:mgm}
The first instance analysis finds information about the names of
sampled random variables.
Such information can be used for discharging the requirement~\ref{req:1},
the correspondence between the
support of a model and that of a guide.
The analysis is based on the below abstraction:
\[
\begin{array}{c}
  \cT^\sharp \defeq \{\bot^\sharp,\top^\sharp\} \cup \cP(\Str),
  \qquad
  \bot^\sharp \defeq \bot^\sharp,
  \qquad
  \gamma(\bot^\sharp) \defeq \{\lambda (s,r).\,\bot\},
  \\[0.5ex]
  \gamma(\top^\sharp) \defeq \cD,
  \qquad
  \gamma(K) \defeq
  \big\{ g \in \cD \,~\big|~\, \forall s,r.\,\, g(s,r) \neq \bot \wedge
  (g(s,r))_r = [] \implies \dom(r) = K \big\},
\end{array}
\]
where $[]$ denotes the empty random database.
A typical abstract element in $\cT^\sharp$ is a set of names $K$,
which represents concrete commands sampling random variables in $K$.
The domain $\cT^\sharp$ contains $\bot^\sharp$ and $\top^\sharp$ to
express two extreme cases, the set containing only one command that
always returns $\bot$, and the set of all commands.

Sound abstract operations can be derived from the density semantics
and from the abstraction following the abstract interpretation
methodology~\cite{cc:popl:77}:
\newcommand{\absS}{S}
\[\begin{array}{lll}
  \begin{array}{@{}r@{\;}c@{\;}l}
    \widen(\bot^{\sharp}, \absS) & = &
    \widen(\absS, \bot^{\sharp}) = \absS; \\
    \widen(\top^{\sharp}, \absS) & = &
    \widen(\absS, \top^{\sharp}) = \top^{\sharp}; \\
    \widen(\absS_0, \absS_1)    & = &
    \omit\rlap{$
      \absS_0 \text{ if } \absS_0 = \absS_1,
      \,
      \top^\sharp \text{ otherwise }
      \,\,\, \text{for } \absS_0, \absS_1 \in \cP(\Str)$;}
  \end{array}
  &\,\,
  \begin{array}{@{}r@{\;}c@{\;}l}
    \cond(E)^\sharp
    & = & \widen;
    \\
    \circ^{\sharp}
    & = & \widen;
    \\
    &&
  \end{array}
  &\,\,
  \begin{array}{@{}r@{\;}c@{\;}l}
    \sskip^{\sharp} = 
    \update(x,E_0)^{\sharp}
    & = & \emptyset;
    \\
    \sample(x,S,E_1,E_2)^{\sharp}
    & = & \{ x \};
    \\
    \score(E_0,E_1,E_2)^{\sharp}
    & = & \emptyset.
  \end{array}
\end{array}\]

We can use the resulting analysis to discharge the requirement~\ref{req:1}.
We just need to run it on both $C$ and $D_\theta$, and check
whether $\adb{C} = \adb{D_\theta} = K$ for some $K \in \cP(\Str)$.
The positive answer implies the requirement~\ref{req:1}, because all the random
variables are drawn from the normal distribution.
Our extension of this analysis for Pyro (\S\ref{s:8:pyroai}) does not rely on this exclusive
use of the normal distribution, and tracks information about the type
of distribution of each random variable and state properties, so as to
prove the model-guide support match for realistic Pyro programs.

\subsection{Analysis for the guide parameter differentiability}
\label{s:7:3:gpd}
The second instance analysis aims at proving the differentiability of the
density of a guide $D_\theta$ with respect to its parameters $\theta$.
It infers the continuous partial differentiability of multiple functions
with respect to variables in the input state.
The analysis is defined by the below abstraction:
\begin{align*}
  \cT^\sharp & \defeq \cP(\Var) \times \cP(\Var) \times \cP(\Var \times \Var),
  \qquad\qquad
  \bot^\sharp \defeq (\Var,\, \Var,\, \Var \times \Var),
  \qquad\quad
  \\
  \gamma(X,Y,R) & \defeq {} \big\{g \in \cD \,~\big|~\,
  \big(\forall x \in X.\, \forall s,r.\,\, g(s,r) \neq \bot
  \implies \,s(x)=(g(s,r))_s(x)\big) 
  \\[-0.5ex]
  \omit\rlap{$
    \qquad \wedge {}
    \big(\forall y \in Y.\, \forall s,r.\,\, g(s,r)\neq \bot
    \implies
    \text{$\lambda v \in \mathbb{R}.\,
      \density(g,s[y\mapsto v])(r)$ is $C^1$}
    \big)
    $}
  \\[-0.5ex]
  \omit\rlap{$
    \qquad \wedge {}
    \big(\forall (z,u) \in R.\,\forall s,r.\,\, g(s,r)\neq \bot
    \implies \text{$\lambda v \in \mathbb{R}.\, (g(s[z\mapsto v],r))_s(u)$
      is $\mathbb{R}$-valued and $C^1$} \big)\big\}.
    $}
\end{align*}
By ``$C^1$'', we mean that the relevant function is continuously
differentiable.
Being a $\bR$-valued function in the last part requires that
$g(s[z\mapsto v],r)$ be never $\bot$.
An $(X,Y,R)$ in $\cT^\sharp$ expresses a property of a transformer
$g \in \cD$ (which can be viewed as semantic command) that $g$ does
not change variables in $X$, its density is $C^1$ with respect to each
variable in $Y$ in the input state, and for each $(z,u) \in R$, it
assigns a real value to $u$ in the output state in a $C^1$ manner with
respect to $z$ in the input state.

\newcommand{\varsof}[1]{\mathcal{V}(#1)}
\newcommand{\varscd}[1]{\mathcal{C}^1(#1)}
We now define the abstract operations induced by this abstraction.
Given an expression \( E \), we let \( \varsof{E} \) denote the set
of variables that occur in \( E \), and we write \( \varscd{E} \) for
the set of variables with respect to which \( \db{E} \) is $C^1$
(based on classical differentiability rules).
The definitions below follow from general principles such as the
multivariate chain rule and account for discontinuities induced by
conditions which break differentiability.
\[
\begin{array}{@{}r@{\;}c@{\;}l}
  \widen((X_0, Y_0, R_0), (X_1, Y_1, R_1))
  & = & (X_0 \sinter X_1, Y_0 \sinter Y_1, R_0 \sinter R_1)
  \\
  \cond(E)^\sharp((X_0, Y_0, R_0), (X_1, Y_1, R_1))
  & = & (X_0 \sinter X_1, (Y_0 \sinter Y_1) \setminus \varsof{E},
  \\
  & & \quad
  \{ (z,u) \in R_0 \sinter R_1 \mid z \not\in \varsof{E} \vee
  u \in X_0 \sinter X_1 \})
  \\
  \circ^{\sharp}((X_0, Y_0, R_0), (X_1, Y_1, R_1))
  & = & (X_0 \sinter X_1, \{ x \in Y_1 \mid \forall y \in \Var. \;
  (x,y) \in R_1 \wedge y \in Y_0 \},
  \\
  & & \quad
  \{ (z,v) \mid \forall u \in \Var. \; (z,u) \in R_1 \wedge (u,v) \in R_0 \})
  \\
  \sskip^{\sharp}
  & = & (\Var, \Var, \Var \times \Var)
  \\
  \update(x,E)^{\sharp}
  & = & (\Var \setminus \{ x \}, \Var,
  \Var \times (\Var \setminus \{ x \}) \sunion
  \{ (y,x) \mid y \in \varscd{E} \} )
  \\
  \sample(x,S,E_1,E_2)^{\sharp}
  & = &
  (\Var \setminus \{ x \},
  \varscd{E_1} \setminus (\varsof{S} \sunion \varsof{E_2}),
  \\
  & & \quad
  (\Var \setminus (\varsof{S} \sunion \varsof{E_2})) \times \Var)
  \\
  \score(E_0,E_1,E_2)^{\sharp}
  & = &
  (\Var, (\varscd{E_0} \cap \varscd{E_1}) \setminus \varsof{E_2},
  (\Var \setminus \varsof{E_2}) \times \Var)
\end{array}
\]

To discharge the differentiability requirement~\ref{req:2},
we need to run this analysis
on a guide $D_\theta$.
If the $Y$ component of the analysis result contains all the parameters
$\theta$ (i.e., there exists $(X,Y,R)$ such that $\adb{D_\theta} = (X,Y,R)$
and $\theta \subseteq Y$), then the requirement~\ref{req:2} is met.

\subsection{Analysis for condition~\eqref{eqn:condition-456}}
The third analysis extends the second by tracking and checking more
properties.
Its aim is to prove the condition~\eqref{eqn:condition-456}.
Just like the second analysis, it infers information about the continuous
partial differentiability of multiple functions involving the output state
and the density.
Also, it checks whether the density of a given command $C_0$ has the form
\begin{equation}
  \label{eqn:condition-456-analysis:g-form}
  \density(\db{C_0}_d,s)(r) =
  \density(g,s)(r) =
  \sum_{i=1}^M
  \ind{r \in A_i}
  \prod_{\alpha \in K_i}
  \cN\big(r(\alpha); \mu_{(i,\alpha)}(s),\sigma_{(i,\alpha)}(s)\big)
\end{equation}
for some \emph{finite} $M$, and some $A_i$, $K_i$, $\mu_{(i,\alpha)}$ and
$\sigma_{(i,\alpha)}$, and if so, it tracks properties of the
$\mu_{(i,\alpha)}$ and $\sigma_{(i,\alpha)}$. Here is the abstraction for
the analysis:
\begin{align*}
  \cT^\sharp
  &\defeq \{\top^\sharp\} \cup \big(\cP(\Var) \times \cP(\Var) \times \cP(\Var \times \Var)\big),
  \qquad
  \bot^\sharp \defeq (\Var,\, \Var,\, \Var \times \Var),
  \\
  \gamma(\top^\sharp)
  &\defeq {} \cD,
  \\
  \gamma(X,Y,R)
  &\defeq {} \big\{g \in \cD \,~\big|~\, 
  \text{$g$ has the form \eqref{eqn:condition-456-analysis:g-form}} \wedge {}
  \big(\forall x \in X.\, \forall s,r.\,\, g(s,r) \neq \bot
  \implies \,s(x)=(g(s,r))_s(x)\big)
  \\[-0.5ex]
  \omit\rlap{$\qquad \wedge {}
    \big(\forall s,r.\,\, g(s,r)\neq \bot 
    \implies \text{$\lambda v \in \mathbb{R}.\, \mu_{(i,\alpha)}(s[\theta_j \mapsto v])$
      is $C^1$ for all $i,j,\alpha$}\big) $}
  \\[-0.5ex]
  \omit\rlap{$\qquad \wedge {}
    \big(\forall s,r.\,\, g(s,r)\neq \bot 
    \implies \text{$\lambda v \in \mathbb{R}.\, \sigma_{(i,\alpha)}(s[\theta_j \mapsto v])$
      is $C^1$ for all $i,j,\alpha$}\big) $}
  \\[-0.5ex]
  \omit\rlap{$\qquad \wedge {}
    \big(\forall y \in Y.\, \forall s,r.\,\, g(s,r)\neq \bot 
    \implies \text{$\lambda v \in \mathbb{R}.\, \mu_{(i,\alpha)}(s[y\mapsto v])$
      is $C^1$ for all $i,\alpha$}\big) $}
  \\[-0.5ex]
  \omit\rlap{$\qquad \wedge {}
    \big(\forall y \in Y.\, \forall s,r.\,\, g(s,r)\neq \bot 
    \implies \text{$\lambda v \in \mathbb{R}.\, \sigma_{(i,\alpha)}(s[y\mapsto v])$
      is $C^1$ for all $i,\alpha$}\big) $}
  \\[-0.5ex]
  \omit\rlap{$\qquad \wedge {}
    \big(\forall (z,u) \in R.\,\forall s,r.\,\, g(s,r)\neq \bot 
    \implies \text{$\lambda v \in \mathbb{R}.\, (g(s[z \mapsto v],r))_s(u)$ is $\mathbb{R}$-valued and $C^1$} \big)\big\}.$}
\end{align*}
The abstract operations are similar to those for the differentiability
analysis. Thus, we omit their definitions.
We can use the analysis to prove the condition~\eqref{eqn:condition-456}.
We just need to run the analysis on a guide $D_\theta$ and check whether
$\adb{D_\theta}$ is not $\top^\sharp$.
If so, the condition holds.

\subsection{Analysis for condition~\eqref{eqn:condition-2b}}
\label{s:7:5:condition-2b}
The last instance is a static analysis that aims at proving the
condition~\eqref{eqn:condition-2b}.
The analysis checks whether a given command $C_0$ has a density of the
following form:
\begin{multline}
  \density(C_0,s)(r) =
  \density(g,s)(r) =
  \\
  \label{eqn:condition-3-analysis:g-form}
  \sum_{i=1}^M
  \ind{r \in A_i}
  \left(\prod_{\alpha \in K_i}
    \cN\big(r(\alpha); \mu'_{(i,\alpha)}(s,r),\sigma'_{(i,\alpha)}(s,r)\big)\right)
  \left(\prod_{j \in [N_i]}
    \cN\big(c_{(i,j)}; \mu''_{(i,j)}(s,r),\sigma''_{(i,j)}(s,r)\big)\right)
\end{multline}
for some \emph{finite} $M$, and some $A_i$, $K_i$, $N_i$, $\mu'_{(i,\alpha)}$,
$\sigma'_{(i,\alpha)}$, $\mu''_{(i,j)}$, and $\sigma''_{(i,j)}$.
If so, it tracks whether the $\mu'_{(i,\alpha)}$, $\sigma'_{(i,\alpha)}$,
$\mu''_{(i,j)}$, and $\sigma''_{(i,j)}$ and some other functions can be
bounded by affine exponentials on the input variables.
The abstraction for the analysis is as follows:
\begin{align*}
  \cT^\sharp
  & \defeq \{\top^\sharp\} \cup \cP(\Var) \times \cP(\Var) \times (\Var \rightharpoonup \cP(\Var)),
  \qquad
  \bot^\sharp \defeq (\Var,\Var,\lambda x{\in}\Var.\,\emptyset),
  \\
  \gamma(\top^\sharp)
  & \defeq \cD,
  \\
  \gamma(X,Y,R)
  & \defeq 
  \big\{g \in \cD \,~\big|~\, 
  \text{$g$ has the form \eqref{eqn:condition-3-analysis:g-form}} \wedge {}
  \big(\forall x \in X.\, \forall s,r.\,\, g(s,r) \neq \bot
  {\implies} s(x)=(g(s,r))_s(x)\big) 
  \\[-0.5ex]
    \omit\rlap{${}\,\, \wedge {}
    \big(\forall i \in [M],\alpha \in K_i.\, \exists \text{ an affine function $l$ from $[Y \cup K_i \to \bR]$ to $\bR$ such that for all $s,r$,}
    $}
  \\[-0.5ex]
  \omit\rlap{$\qquad
    g(s,r) \neq \bot {\implies}
    \max(|\mu'_{(i,\alpha)}(s,r)|, \sigma'_{(i,\alpha)}(s,r), \sigma'_{(i,\alpha)}(s,r)^{-1})
    \leq \exp(l(\{|s(x)|\}_{x \in Y},\{|r(\beta)|\}_{\beta \in K_i}))\big) $}
  \\[-0.5ex]
    \omit\rlap{${}\,\, \wedge {}
    \big(\forall i \in [M], j \in [N_i].\, 
        \exists \text{ an affine function $l$ from $[Y \cup K_i \to \bR]$ to $\bR$ such that 
        for all $s,r$,}
    $}
  \\[-0.5ex]
  \omit\rlap{$\qquad
    g(s,r) \neq \bot {\implies}
    \max(|\mu''_{(i,j)}(s,r)|, \sigma''_{(i,j)}(s,r), \sigma''_{(i,j)}(s,r)^{-1})
    \leq \exp(l(\{|s(x)|\}_{x \in Y},\{|r(\beta)|\}_{\beta \in K_i}))\big) $}
  \\[-0.5ex]
    \omit\rlap{${}\,\, \wedge {}
    \big(\forall y \in \dom(R), i \in [M].\,
        \exists \text{ an affine function $l$ from $[R(y) \cup K_i \to \bR]$ to $\bR$ such that 
        for all $s,r$,}
    $}
  \\[-0.5ex]
  \omit\rlap{$\qquad
    g(s,r) \neq \bot {\implies} (g(s,r))_s(y) \leq \exp(l(\{|s(x)|\}_{x \in R(y)}, \{|r(\beta)|\}_{\beta \in K_i}))\big)\big\}.$}
\end{align*}
Here we use the notation $\{|s(x)|\}_{x \in Y}$ to mean a partial map
from variables $x$ in $Y$ to values $|s(x)|$.
The abstract operations that derive from this abstraction are quite
similar to those of the differentiability analysis, therefore we do
not detail them.
To verify the condition~\eqref{eqn:condition-2b}, we run the analysis
on a model $C$ and check whether $\adb{C} \neq \top^\sharp$.
If the check succeeds, the condition holds.


\section{A Static Analysis for Model-Guide Pairs in Pyro and its Evaluation}
\label{s:8:pyroai}
\newcommand{\sref}[1]{\S\ref{s:#1}}
We present a static analysis that can verify the support correspondence
for Pyro model-guide pairs. The analysis extends the first instance of the framework
presented in \S\ref{sec:analysis}.
Our presentation focuses on the most salient aspects of this analysis and experimental results.


\subsection{Some features of Pyro programs}
\label{s:8:1:pyrolang}
Pyro is a probabilistic programming framework based on Python and PyTorch.
It supports a wide spectrum of distributions and neural networks, and
features commands for sampling and scoring. It comes with multiple inference
engines, including SVI.
We chose Pyro over other probabilistic programming languages (e.g., Anglican)
because unlike most of other languages, in Pyro,
SVI algorithms are considered as main inference engines
and neural networks can be used together with probabilistic programming,
which leads to more interesting examples.

In Pyro programs, random-variable names are often created at runtime, for
instance by concatenating a fixed string with a sequence of
dynamically-generated indices.

\begin{example}[Use of indexed random-variable names]
  \label{e:tensor:code}
  The code excerpt (of a model or guide program in Pyro) below
  samples \( N \times M \) instances of independent random variables,
  and names them with $``x\_1\_1", \ldots, ``x\_N\_M"$.
\begin{lstlisting}[numbers=none, basicstyle=\footnotesize\ttfamily, escapechar=@]
      for i in range(1,N+1):
        for j in range(1,M+1):
          val = pyro.sample("x_{}_{}".format(i,j), Normal(m,d)) @\hfill{\normalsize\qed}@
\end{lstlisting}
\end{example}

Since Pyro is based on PyTorch and is adapted to implement
data-science applications, Pyro programs heavily use 
multidimensional arrays, called tensors, and operations over them
from PyTorch, in particular, element-wise operations and broadcasting.
More precisely, when a binary operation (such as addition) is applied
to two tensors of identical size, it outputs a new tensor of the same
size where each output element is computed separately from the corresponding
input elements. This principle often makes it possible to parallelise computations.
Broadcasting occurs when a binary operation is applied to two tensors
of different dimensions that can somehow be unified.
Intuitively, it introduces and duplicates dimensions to produce tensors
of identical dimensions, so that binary operations can be performed
element-wise.

Tensors are also heavily used for sampling, which has several consequences.
First, it means that an analysis targeted at Pyro programs should be able
to track information about at least the dimensions of sampled tensors.
Second, the dimensions of sampled tensors are grouped such that
each of these groups has a different property with respect to probabilistic 
independence of tensor components. Pyro inference engines exploit this property 
for optimisation, for instance, via subsampling, but for this optimisation to be valid,
the grouping of dimensions in a model should match that of a guide.
Our analysis
tracks information about dimension grouping of sampled tensors and the
use of the Pyro construct called plate, which enables the optimisation
just mentioned.

\subsection{Abstract domain}
\label{s:8:2:abs}

We extend our analysis from \S\ref{sec:analysis} so that it tracks
not just how each part of a given program transforms states, but
also which states can reach at each program point.
We need the latter to get information about random variables with
dynamically generated names that is precise enough for our
verification task. 
To describe states, we rely on an abstract domain that consists of
multiple subdomains (combined by product).
Among them, the key part is $\aRdb \defeq [\Str \rightharpoonup_\fin
\{\top\} \cup (\cP_\fin(\aZone) \times \aDist)]$,
where $\rightharpoonup_\fin$ denotes a finite partial map.
$\aDist$ is an abstract domain whose element represents a set
of elementary probability distributions, such as the standard normal
distribution.
The other $\aZone$ is our custom domain for zones, which express
higher-dimensional rectangles in $\bN^n$.

\newcommand{\gammar}{\gamma_{r}}
\newcommand{\cTst}{\cT^{\sharp}_{s}}
\newcommand{\gammast}{\gamma_{s}}
An element $\abs{r} \in \aRdb$ means a set of concrete random databases
$r$ for each concrete store $s$. The domain of $r$ consists of names that
are obtained by concatenating baseline strings $\alpha$ in $\dom(\abs{r})$
with index sequences.
If $\abs{r}$ maps a baseline string $\alpha$ to $\top$, it does not give
any information about index sequence and a probability distribution used
for the random variable.
Otherwise, the second component of $\abs{r}$ specifies the information
about the distribution, and the first component a region of $\bN^n$
that contains the index sequences used in the names.
The region is described by a finite subset $\{\abs{Z_1},\ldots,\abs{Z_n}\}$
of $\aZone$, which means the disjoint union of the rectangles represented
by the $\abs{Z_i}$.
To emphasise this union interpretation, we write $\abs{Z_1} \cup \ldots
\cup \abs{Z_n}$ for the subset.
The following table summarises the definitions of our abstract domain, based
on $\aZone$:
\[
\begin{array}{@{}r@{\;}c@{\;}l@{}}
  \aRdb & \defeq
  & [\Str \rightharpoonup_\fin \{\top\} \cup (\cP_\fin(\aZone) \times \aDist)];
  \\[0.5ex]
  \abs{Z} & \defeq & \abs{I}_1 \times ... \times \abs{I}_{m},
  \ \, \text{product of intervals in $\bN$};
  \\[0.5ex]
  \abs{I} & \defeq & [\abs{B}_l, \abs{B}_r],
  \ \, \text{closed interval specified by bounds};
  \\[0.5ex]
  \abs{B} & \defeq & c
  \,\mid\, x + c
  \,\mid\, x + c = c',
  \ \, \text{equality to constant, variable plus constant, or both}.
\end{array}
\]
A higher-dimensional rectangular zone $\abs{Z}$ is described by a finite
sequence of intervals $\abs{I}$, each of which is made of two bounds.
A bound $\abs{B}$ may be defined as one or two constraints which express that
this bound is equal to a constant, or to a variable plus a constant, or both.
This intuitive denotation defines a concretisation function $\gammar : \aRdb \to \cP(\State)$.

\begin{example}
  \label{e:tensor:abs}
  Consider the code of Example~\ref{e:tensor:code}.
  After \( i-1 \) iterations in the main loop and \( j \) iterations in
  the last execution of the inner loop, we expect the analysis to come
  up with the invariant, $[``x" \mapsto ([1,i{-}1] \times [1, M] \cup [i,i]
  \times [1,j],\, \mathrm{normal}(m,d))]$.
  In turn, at the exit of the main loop, we 
  expect the analysis to infer the invariant, $[``x" \mapsto ([1,i{=}N]
  \times [1, j{=}M],\, \mathrm{normal}(m,d))]$. \qed
\end{example}

In addition to these constraints over random databases, our analyser
also uses an abstraction that maintains typing information and
numerical constraints over variables.
We do not fully formalise these constraints as the overall structure
of the domain relies on a classical reduced product.

Finally, we describe the combination of the above state abstraction with
the abstraction of \S\ref{s:7:2:mgm}.
More precisely, we start with the abstract domain exposed in \S\ref{s:7:2:mgm}
(which we denote by \( \cT^\sharp \)) and build a novel abstract domain \( \cTst \) that
also satisfies the requirements of \S\ref{s:7:1:fwk}.
We let the set of abstract elements be
\( \cTst \defeq [\abs{RDB} \rightarrow \cT^\sharp \times \abs{RDB}] \).
Intuitively, such an element maps an input abstract random database into a
set of functions together with an over-approximation of their output, when
applied to this abstract input.
The concretisation below formalises this: for all $t_s \in \cTst$,
\[
\gammast( t_s ) \defeq
        \begin{array}[t]{@{}l@{}} 
        \Big\{ g \in \cD ~\Big|~ \forall \abs{r}_i \in \abs{RDB}. \; 
                t_s( \abs{r}_i ) = (t, \abs{r}_o) \Longrightarrow {} 
        \\[0.5ex] 
        \qquad
	\exists g' \in \gamma(t).\; \forall (s,r) \in \gammar( \abs{r}_i ).\;
                g(s,r) = g'(s,r) \wedge \big(g(s,r) = \bot \vee (g(s,r)_s,g(s,r)_r) \in \gammar( \abs{r}_o )\big) \Big\}.
        \end{array}
\]

\subsection{Computation of loop invariants}
\label{s:8:3:aops}
Although our static analysis for Pyro requires a state abstraction, its
principles and structure are similar to those of the general analysis
shown in \S\ref{sec:analysis}.
In the following, we first describe the integration of the state
abstraction in the analysis of \S\ref{s:7:2:mgm}.

\newcommand{\compst}{\circ^{\sharp}_{s}}
The abstract operations in \( \cTst \) can all be derived by lifting
those in \( \cT^\sharp \) into functions. We illustrate this for the abstract
composition \( \compst \) for \( \cTst \). Recall the 
operator \( \circ^\sharp \) for \( \cT^\sharp \) in \S\ref{s:7:2:mgm}.
Given \( t_s, t'_s \in \cTst \), 
\[
t_s \compst t'_s \defeq
\lambda \abs{r}_0.\,  (t \circ^{\sharp} t', \abs{r}_2)
\quad \text{where }
(t', \abs{r}_1) = t'_s( \abs{r}_0 )
\text{ and }
(t, \abs{r}_2) = t_s( \abs{r}_1 )
\]
The other abstract operators over \( \cTst \) are defined in a similar
manner.

As in most static analysis problems, 
the computation of precise loop invariants
requires a carefully designed widening operator.
In the case of zones, the analysis needs to generalise information
about the bounds.
We assume \( \abs{r}_0, \abs{r}_1 \in \aRdb \) are abstract random
databases, and present the main steps in the computation of the widening
\( \widen(\abs{r}_0, \abs{r}_1 ) \).
\begin{enumerate}[label=(\roman*)]
\item For each $\abs{r}_i$, we fuse zones $\abs{Z}_0,\abs{Z}_1$ in
  $\abs{r}_i$ into a single zone when $\abs{Z}_0$ and $\abs{Z}_1$ are
  the same except for one component and they represent adjacent
  high-dimensional rectangles.
  The following rewriting illustrates this step.
  \[
  \begin{array}{@{}l@{}}
    \big(``x" \mapsto \big([1,i{-}1] \times [1, j{=}M] \sunion [i,i] \times [1,j{=}M],\, \mathrm{normal}(m, d)\big)\big)
    \\[0.5ex]
    \qquad\qquad\qquad\qquad\qquad
    {} \leadsto
    \big(``x" \mapsto \big([1,i] \times [1,j{=}M],\, \mathrm{normal}(m, d)\big)\big)
  \end{array}
  \]
\item We generalise the intervals of corresponding zones in $\abs{r}_0$ and $\abs{r}_1$ 
      using a weak version of unification on their bounds. The bounds of corresponding intervals 
      in two such zones survive this step in a weakened form if they share at least 
        one syntactically equal expression. Otherwise, the bounds are dropped, which causes the introduction of
        $\top$ into abstract random databases.
        The following rewriting instances illustrate this bound generalisation.
  \[
  \begin{array}{@{}l@{}}
          \widen\big(
                \big(``x" \mapsto \big([1,i] \times [1, j{-}1{=}M],\, \mathrm{normal}(m,d)\big)\big),\,\,
                \big(``x" \mapsto \big([1,i] \times [1, j{=}M],\, \mathrm{normal}(m,d)\big)\big) 
                \big) 
          \\[0.5ex] 
          \qquad\qquad
          {} = \big(``x" \mapsto \big([1,i] \times [1, M],\, \mathrm{normal}(m,d)\big)\big) 
          \\[1ex] 
          \widen\big(
                \big(``x" \mapsto \big([1,i] \times [1, j{-}1],\, \mathrm{normal}(m,d)\big)\big),\,\,
                \big(``x" \mapsto \big([1,i] \times [1, j{=}M],\, \mathrm{normal}(m,d)\big)\big)
                \big) 
          \\[0.5ex] 
          \qquad\qquad
          {} = \big(``x" \mapsto \top\big)
  \end{array}
  \]
\end{enumerate}
The analysis also applies standard widening techniques to typing information and
numerical constraints mentioned in \sref{8:2:abs}.
 Finally, a model-guide pair 
can be verified if and only if their analyses return equivalent abstract 
random databases, without any name mapped into \( \top \).

\subsection{Experimental evaluation}
\label{s:8:4:exp}
\label{sec:experiments}
\newcommand{\eg}{e.g.\xspace}
We have implemented a prototype analyser
and carried out experiments
so as to assess the effectiveness of our analysis to verify the support
correspondence for Pyro model-guide pairs.%
\footnote{Code is available at \url{https://github.com/wonyeol/static-analysis-for-support-match}.}
More precisely, we evaluated the following three research questions:
\begin{enumerate}[label=(\roman*)] 
\item
  Can our analysis discover incorrect model-guide pairs
  in realistic probabilistic programs?
\item
  Can the analysis verify correct model-guide pairs 
  despite the complexity of the definition?
\item
  Is the analysis efficient enough so that it can cope with realistic
  probabilistic programs?
\end{enumerate} 

\paragraph{Benchmark programs.} We took example programs from the Pyro webpage that can be handled by standard SVI engines or can be naturally rewritten to be so. Those engines do not use Pyro's recent vectorised enumeration-based optimisation
(enabled by the option \mytt{TraceEnum\_ELBO}),
and are run with the option \mytt{Trace\_ELBO}.
We made this choice because the optimisation is newly introduced
and whether it is used or not changes the set of random variables used in a Pyro program. 

We have applied our analysis to two groups of Pyro programs. The first is 
the \emph{Pyro regression test suite}~\cite{pyro:regr:test},
which comprises 66 examples exercising the basic programming patterns expected in Pyro programs.  While each of these is small, they cover many standard patterns of both correct and incorrect model-guide pairs.
Among these, 33 test cases use only \mytt{TraceEnum\_ELBO} and fall
outside the scope of our analysis, and 6 of the remaining 33 test cases come with two subsampling options. Hence, we can consider 39 experiments based on this regression suite.

The  second is a collection of \emph{examples} from the Pyro webpage~\cite{pyro:examples}.
The webpage features 18 \emph{Pyro examples}, among which 
9 involve model-guide pairs (other examples use automatic guide
generation, or do not perform SVI at all).
Out of these 9 examples, 5 use \mytt{Trace\_ELBO}, and three can 
be converted naturally into \mytt{Trace\_ELBO} versions although they do not use \mytt{Trace\_ELBO}.
Thus, 8 Pyro examples fall within the scope of our analysis.
These examples correspond to advanced probabilistic models%
\footnote{%
  The models include
  variational autoencoder (VAE)~\cite{KingmaICLR14},
  semi-supervised VAE~\cite{KingmaNIPS14},
  deep exponential family~\cite{RanganathAISTATS15}, 
  attend-infer-repeat~\cite{EslamiNIPS16},
  deep Markov model~\cite{KrishnanAAAI17},
  inference compilation~\cite{LeAISTATS17},
  and
  amortised latent Dirichlet allocation~\cite{SrivastavaICLR17}.
}
from the machine-learning literature, 
most of which use sophisticated neural networks and probabilistic modelling.
Table~\ref{tab:pyro:example:features} describes the structure of these 8 examples in detail.


\begin{table}[]
  \aboverulesep=0.3mm
  \belowrulesep=0.3mm
  \small\addtolength{\tabcolsep}{-3pt}
  \begin{tabular}{l@{\hskip8pt}lrrrrrrrr}
    \toprule
    & & 
    & \multicolumn{4}{c}{Total \#}
    & \multicolumn{3}{c}{Total dimension}
    \\ \cmidrule(lr){4-7} \cmidrule(lr){8-10}
    \multicolumn{1}{c}{Name}
    & \multicolumn{1}{c}{Corresponding probabilistic model}
    & \multicolumn{1}{c}{LoC}
    & \mytts{for} & \mytts{plate} 
    & \mytts{sample}& \mytts{score}
    & \multicolumn{1}{c}{\mytts{sample}}
    & \multicolumn{1}{c}{\mytts{score}}
    & \multicolumn{1}{c}{$\theta$} 
    \\ \midrule
    \mytts{br} & Bayesian regression
    & 27
    & 0 & 1 &  10 & 1 &  10 & 170 & 9
    \\
    \mytts{csis}
    &
    Inference compilation 
    & 31
    & 0 & 0 &   2 & 2 &  2 & 2 & 480
    \\
    \mytts{lda} &
    Amortised latent Dirichlet allocation
    & 76
    & 0 & 5 &   8 & 1 &  21008 & 64000 & 121400
    \\
    \mytts{vae} &
    Variational autoencoder (VAE)
    & 91
    & 0 & 2 &   2 & 1 &  25600 & 200704 & 353600
    \\
    \mytts{sgdef} &
    Deep exponential family 
    & 94
    & 0 & 8 &  12 & 1 &  231280 & 1310720 & 231280
    \\
    \mytts{dmm} &
    Deep Markov model
    & 246
    & 3 & 2 &   2 & 1 &  640000 & 281600 & 594000
    \\
    \mytts{ssvae} &
    Semi-supervised VAE
    & 349
    & 0 & 2 &   4 & 1 &  24000 & 156800 & 844000
    \\
    \mytts{air} &
    Attend-infer-repeat
    & 410
    & 2 & 2 &   6 & 1 &  20736 & 160000 & 6040859
    \\
    \bottomrule
  \end{tabular}
  \\[1.5ex]
  \caption{Key features of the model-guide pairs from Pyro examples.
    LoC denotes the lines of code of model and guide.
    The columns ``Total \#'' show the number of objects/commands of each type
    used in model and guide,
    and the columns ``Total dimension'' show the total dimension of tensors in model and guide,
    either sampled from \mytts{sample} or used inside \mytts{score},
    as well as the dimension of $\theta$ in guide.}
  \label{tab:pyro:example:features}
\end{table}

\paragraph{Prototype analyser and results.}
Our analyser is implemented in OCaml, and supports the main data-structures and operations defined
by Python, PyTorch, and Pyro.
In particular, it precisely abstracts
the shape of PyTorch tensor objects,
the shape transformation information of PyTorch neural-network-related objects,
the automatic broadcasting information of Pyro plate objects, and
the shape of allocated indices for sample names, using the zone abstraction described above. 
It also supports common Pyro probability distributions,
and can precisely cope with standard Pyro and PyTorch operations manipulating
the Pyro distribution objects and PyTorch tensor objects.
While our prototype supports a wide range of Python, PyTorch, and Pyro features,
we point out that we did not implement a static analysis
for the full Python (plus PyTorch and Pyro) language
(e.g., no support for classes, dictionaries, named tuples, and user-defined functions).

\iftr{
  The front-end utilises \mytt{ast} and \mytt{pyml} in order to parse
  and convert Python programs into OCaml values.
  The numerical abstract domain is based on the \mytt{apron} library,
  and the analysis uses the Octagon domain~\cite{Mine:2006hh} by
  default.}

\begin{table}[]
  \aboverulesep=0.3mm
  \belowrulesep=0.3mm
  \small\addtolength{\tabcolsep}{-2pt}
  \begin{subtable}[t]{.6\textwidth}
    \centering
    \begin{tabular}[t]{lcccc}
      \toprule
      \multicolumn{1}{c}{Category}
      & \#Same & \#Diff & \#Crash & Time
      \\ \midrule
      No plates
      & 9 & 0 & 0
      & 0.001
      \\
      Single \mytts{for}-plate
      & 4 & 0 & \colorbox{lgreen}{3}
      & 0.004
      \\
      Nested \mytts{for}-plates
      & 2 & 0 & \colorbox{lgreen}{2}
      & 0.026
      \\
      Single \mytts{with}-plate
      & 5 & 0 & 0
      & 0.001
      \\
      Nested \mytts{with}-plates
      & 7 & \colorbox{lgreen}{2} & 0
      & 0.002
      \\
      Non-nested \mytts{with}-plates
      & 2 & 0 & 0
      & 0.002
      \\
      Nested \mytts{for}-plate \& \mytts{with}-plate
      & 0 & 0 & \colorbox{lgreen}{3}
      & N/A
      \\ \midrule
      Total & 29 & 2 & 8
      & 0.003
      \\
      \bottomrule
    \end{tabular}
    \\[1.5ex]
    \caption{Results for Pyro test suite.
      39 model-guide pairs are grouped into 7 categories,
      based on which type of plate objects are used.
      \#Same (or \#Diff) denotes the number of model-guide pairs for which
      the output of our analyser, valid or invalid,
      is the same as (or different from) the documented output.
      \#Crash denotes the number of pairs for which our analyser crashes.
    }
  \end{subtable}
  \quad\,
  \begin{subtable}[t]{.34\textwidth}
    \centering
    \begin{tabular}[t]{lcr}
      \toprule
      Name & Valid? & Time
      \\ \midrule
      \mytts{br} & \colorbox{lgreen}{\sf x}
      & 0.006
      \\
      \mytts{csis} & o
      & 0.007
      \\
      \mytts{lda} & \colorbox{lgreen}{\sf x}
      & 0.014
      \\
      \mytts{vae} & o
      & 0.005
      \\
      \mytts{sgdef} & o
      & 0.070
      \\
      \mytts{dmm} & o
      & 0.536
      \\
      \mytts{ssvae} & o
      & 0.013
      \\
      \mytts{air} & o
      & 4.093
      \\
      \bottomrule
    \end{tabular}
    \\[2.9ex]
    \caption{Results for Pyro examples.
    The column ``Valid?'' shows the output of our analysis, valid or invalid.}
  \end{subtable}
  \caption{Analysis results on two benchmark sets.
    The column ``Time'' shows the analysis time in seconds;
    in~(a), it is averaged over those model-guide pairs (in each category)
    for which our analyser does not crash.}
  \label{tab:analysis:res}
\end{table}

The analysis results are summarised in Table~\ref{tab:analysis:res}
and are discussed in detail in the following.
Run-times were measured on an Intel Core i7-7700 machine running
Linux Ubuntu 16.04.

\paragraph{Discovery of invalid model-guide pairs.}
The analysis rejected two Pyro examples, \mytt{br} and \mytt{lda},
as incorrect due to an invalid model-guide pair.
 \mytt{br} is the Bayesian regression example
discussed in \S\ref{sec:overview}.

For \mytt{br}, the analysis discovers that a random variable \mytt{sigma}
is sampled from \mytt{Uniform(0.,10.)} in the model $C$, but from
\mytt{Normal(..)} in the guide $D_\theta$
(Figure~\ref{fig:eg:kl:undefined}(a)).
Since the support of \mytt{sigma} in $D_\theta$ is not a subset of that in $C$
(i.e., $\bR \not\subseteq [0,10]$),
the requirement~\ref{req:1} is violated.
Thus, the SVI objective, $\KL(D_\theta \| C)$, is undefined,
and \mytt{br} has an invalid model-guide pair.

For \mytt{lda}, the analysis discovers that a random variable
\mytt{doc\_topics} is sampled from \mytt{Dirichlet(..)} in the model $C$,
but from \mytt{Delta(..)} in the guide $D_\theta$.
Since the reference measures of \mytt{doc\_topics} in
$C$ and $D_\theta$ are different
(the Lebesgue measure vs. the counting measure),
$\KL(D_\theta \| C)$ cannot be computed by~\eqref{eqn:KL:definition}.
For this reason, \mytt{lda} has an invalid model-guide pair.
Our analyser tracks the reference measure implicitly
by regarding the support of any distribution with Lebesgue measure,
as disjoint from that of any distribution with counting measure
(which is sound due to the aforementioned reason),
and this allowed us to discover the issue of \mytt{lda}.

In both cases, the found correctness issues are unknown before and subtle.
In particular, it turned out that \mytt{lda}, though incorrect when viewed as an example of SVI,
is a valid implementation, because it performs not SVI but variational expectation-maximisation (EM)~\cite{Neal99}.%
\footnote{We thank an anonymous reviewer and Eli Bingham
  for pointing this out and explaining it in detail.}
The \mytt{lda} uses an SVI engine just to solve an optimisation problem in variational EM
(not to do SVI), and
uses the \mytt{Delta} distribution 
to perform a particular realisation of the M-step in variational EM.

\paragraph{Verification of probabilistic programs relying on model-guide pairs.}
Among the Pyro test suite, the analysis successfully verifies 31 examples
among 39.
Interestingly, two of these 31 successful validations,
highlighted in Table~\ref{tab:analysis:res}(a),
correspond to cases
that were flagged as ``invalid model-guide pairs'' in the Pyro git repository.
Upon inspection, these two examples turn out to be correct.

On the other hand, 8 examples from the Pyro test suite could not be verified
due to the crashes of the analyser.
One of these failures is due to the need to reason more precisely about
the content of a \mytt{for} loop (\eg, using some partitioning techniques),
and seven are due to the use of plates with subsampling, as ranges for
\mytt{for} loops.
Therefore, these failures could be resolved using existing static analysis
techniques and a more precise handling of the semantics of Python
constructions.

Moreover, 6 Pyro examples (among the 8 that we considered) were verified
successfully, which means all correct Pyro examples were verified.
Finally, we corrected the two examples that were rejected due to invalid
model-guide pairs, and these two examples were also successfully
verified.


\paragraph{Analysis efficiency.}
The analysis returned within a second on each program in the Pyro test
suite, and on most of the Pyro examples.
The slowest analysis was observed on \mytt{air}, which was
analysed within 5 seconds.
Most of the Pyro examples sample from and score with 
distributions of very high dimension arranged in complex tensors,
using nested \mytt{for} and \mytt{plate}'s.
While they are not large, they present a high degree of
logical complexity, that is representative of realistic probabilistic
programs.
The fact that such programs get analysed within seconds shows that
the analysis and the underlying abstract domain to describe 
zones, sampled dimensions, and distributions can
generalise predicates quickly so that precise loop invariants
can be computed. 

\section{Related Work, Conclusion and Limitation}

\paragraph{Related work.}
As far as we know, the idea of using SVI 
for probabilistic programs first appeared in \cite{WingateBBVI13}. When further insights into how to create generic (sometimes also called black-box) SVI engines were found~\cite{RanganathGB14,KucukelbirRGB15,KucukelbirTRGB17}, the idea was tried for realistic probabilistic programming languages, such as Stan~\cite{KucukelbirRGB15} and Anglican~\cite{MeentPTW16}, resulting in impressive case studies~\cite{KucukelbirRGB15}. The major inference engines for deep probabilistic programming languages~\cite{BinghamCJOPKSSH19,TranKDRLB16,TranHMSVR18,siddharth2017learning} are all based on SVI nowadays. However, we do not know of any prior work that attempts to reveal implicit assumptions made by these SVI engines and to discharge these assumptions manually or automatically, as we did in this paper.

The correctness of a different type of inference engines based on Monte-Carlo methods has been the subject of study in the PL community. Such engines have clearly formulated correctness conditions from Markov chain theory~\cite{metropolis:53,hastings:70,10.1093/biomet/82.4.711,geyer:intro-mcmc}, such as ergodicity and correct stationarity. Tools from formal semantics have been employed to show that the inference engines satisfy these conditions~\cite{HurNRS15,Kiselyov16,gordon:icfp:16,abs-1711-03219}. While looking at different algorithms, some of these work consider more expressive languages than what we used in the paper, in particular, those supporting higher-order functions. One interesting direction is to extend our results to such expressive languages using the ideas from these works, especially the operational technique in \cite{gordon:icfp:16} and the denotational monad-based technique in \cite{abs-1711-03219}.

The consequence of having random choice in a programming language has been actively investigated by the semantics researchers from the early days~\cite{kozen:81,JonesP89,EhrhardTP14,TorontoMH15,gordon:icfp:16,StatonYWHK16,Staton17,SmolkaKFK017,HeunenKSY17,VakarKS19}.
Our work uses the technique developed in this endeavour, such as Giry monad and denotational formulation of idealised importance sampling~\cite{StatonYWHK16}. Also, just as we connected the measure semantics with the density semantics, \cite{kozen:81} and \cite{gordon:icfp:16} related two semantics with similar flavours, although the considered languages (with or without continuous distribution and score statements) and the style of semantics (operational vs denotational) are different. Our density semantics uses a reference measure built out of Lebesgue measure as in \cite{BhatAVG12,BhatBGR13,HurNRS15}. This choice is found to cause an issue when the score statement is treated not as a scoring mechanism but in terms of conditional expectation~\cite{WuSHDR18}. How to resolve this matter is still open.

Static analyses for probabilistic programming languages or languages with probabilistic choice typically attempt to check probabilistic properties~\cite{m:sas:00,cm:esop:12,cs:cav:13,reps:pldi:18}, or perform posterior inference~\cite{ChagantyNR13,reps:pldi:18}, or find information useful for posterior inference~\cite{NoriHRS14}. For instance, \cite{m:sas:00} defines a probabilistic abstract interpretation framework, which is applied to estimate the failure probability \cite{m:esop:01}. More recently, \cite{cm:esop:12} sets up a general framework to design probabilistic abstract interpretations, which lift conventional static analysis tools to a probabilistic setup, and \cite{reps:pldi:18} proposes a framework for analysing probabilistic programs based on hyper-graphs with probabilistic actions on edges. Our program analyses aim at a different type of verification tasks, namely, proving the safety requirements imposed by the SVI engines. Static analyses for checking the continuity properties of programs are proposed in \cite{ChaudhuriGL10}. Some techniques used in those analyses may help track the kind of smoothness properties considered in this paper precisely.

\paragraph{Conclusion.} In this paper, we have set up a semantic model of probabilistic programming
languages that allows to reason about the correctness of subtle optimisation
algorithms such as SVI.
In particular, we have identified a set of conditions that guarantee that
these algorithms will produce meaningful results.
These conditions simplify the verification of correctness, and open the
way towards the automation of this verification.
To evaluate the effectiveness of our framework, we have completed the
design of a static analysis to validate the correspondence assumptions
for the model-guide pairs.
Our analysis could verify nontrivial implementations as well as
uncover nontrivial bugs.
Such bugs would lead to incorrect results, which would be very difficult
to diagnose. 


\paragraph{Limitation.} We described four static analyses (\S\ref{s:7:2:mgm}-\S\ref{s:7:5:condition-2b})
for discharging the requirements~\ref{req:1}-\ref{req:6},
and developed one of the four analyses further (\S\ref{s:8:2:abs}-\S\ref{s:8:4:exp})
to build a prototype analyser for actual Pyro programs.
Here we clarify the underlying assumptions and limitations of these analyses.

The static analysis for model-guide support match (\S\ref{s:7:2:mgm})
was implemented into our static analyser for Pyro (\S\ref{s:8:4:exp})
but with additional extensions (\S\ref{s:8:2:abs}-\S\ref{s:8:3:aops}),
so that it does not make the assumption in~\S\ref{subsec:requirement-assumption};
the assumption was introduced mainly to develop analyses
for the requirements~\ref{req:3}-\ref{req:6}.
Hence, our static analyser handles both continuous and discrete random variables.

On the other hand, other analyses (\S\ref{s:7:3:gpd}-\S\ref{s:7:5:condition-2b})
are unimplemented and make the assumption in \S\ref{subsec:requirement-assumption}.
We expect that the assumption can be relaxed, without much difficulty,
to permit continuous or discrete distributions
having finite entropy and moments of all degrees,
because our proofs of theorems and lemmas bound various expectations
mostly by entropies and moments.
It would be more challenging, however, to relax the assumption
to allow distributions not having finite entropy and moments of all degrees,
or models having unboundedly many random variables (of any kind).
In particular, addressing the later limitation might require techniques
developed for proving that probabilistic systems has finite expected execution time.

This paper focuses on a particular optimisation objective~\eqref{eqn:SVI:objective} for SVI,
but we point out that several other optimisation objectives have been proposed
for different inference or learning algorithms,
such as variational EM (e.g., \mytt{lda} in \S\ref{s:8:4:exp})
and importance weighted autoencoder (IWAE)~\cite{BurdaICLR16}. One interesting research
direction is to develop techniques for verifying the well-definedness of these optimisation objectives.

\begin{acks}                            
        We thank Fritz Obermeyer and Eli Bingham for explaining the subtleties of Pyro and suggesting us to try the Pyro regression test suite. Sam Staton, Ohad Kammar and Matthijs V{\'a}k{\'a}r helped us to understand the semantics of recursion in the probabilistic programming languages better. Lee, Yang and Yu were supported by the Engineering Research Center Program through the National Research Foundation of Korea (NRF) funded by the Korean Government MSIT (NRF-2018R1A5A1059921), and also by Next-Generation Information Computing Development Program through the National Research Foundation of Korea (NRF) funded by the Ministry of Science, ICT (2017M3C4A7068177).

\end{acks}

\bibliography{refs}

\clearpage
\appendix



\section{Proofs of Lemmas and Theorems in \S\ref{s:4:lang}}
$\,$

{\sc Lemma~\ref{lemma:expression-welldefined}.\ }{\it For all expressions $E$, $B$, and $S$, their semantics $\db{E}$, $\db{B}$ and $\db{S}$ are measurable functions from $\Store$ to $\mathbb{R}$, $\mathbb{B}$ and $\Str$, respectively.}

\begin{proof}
        We prove the lemma by structural induction. We present the proof for $E$ only, because those for $B$ and $S$ are similar. When $E$ is a constant, $\db{E}$ is a constant function and so it is measurable. When $E$ is a variable, $\db{x}$ is a projection of the $x$ component of the given store $s$. Since $\Store$ is given the product $\sigma$-algebra, the projection $\db{x}$ is measurable. Finally, assume that $E$ is $f(E_0,\ldots,E_{n-1})$. Then, by induction hypothesis, the $\db{E_i}$ are measurable functions from $\Store$ to $\bR$. Thus,
        \[
                \lambda s.\,(\db{E_0}s,\ldots,\db{E_{n-1}}s) : \Store \to \bR^n
        \]
        is measurable. Furthermore, $\db{f} : \bR^n \to \bR$ is measurable. Thus, their composition is measurable as well. This implies the measurability of $\db{E}$ because
        \[
                \db{E} = \db{f} \circ (\lambda s.\,(\db{E_0}s,\ldots,\db{E_{n-1}}s)).
        \]
\end{proof}

{\sc Lemma~\ref{lemma:subprob-kernels-cpo}.\ }{\it $(\cK,\sqsubseteq)$ is an $\omega$-complete partial order with the least element $\bot$.}

\begin{proof}
The fact that $\sqsubseteq$ is a partial order immediately follows from its definition. The least element is
the function $\lambda \sigma.\,\lambda A.\,0$. It remains to show the $\omega$-completeness property. Consider an $\omega$-chain
\[
        \kappa_0 \sqsubseteq \kappa_1 \sqsubseteq \ldots
\]
Let $\kappa = \sup_{n} \kappa_n$. We will show that $\kappa$ is the desired upper bound.

Pick a state $\sigma$. Then, $\kappa(\sigma)(A)$ is non-negative for all measurable subsets $A$, and $\kappa(\sigma)(\emptyset) = 0$. Furthermore, for every countable collection $\{A_j\}_{j}$ of disjoint measurable subsets of $\State \times [0,\infty)$,
\begin{align*}
        \kappa(\sigma)\Big(\bigcup_j A_j \Big)
        & = \sup_n \kappa_n(\sigma)\Big(\bigcup_j A_j \Big)
        = \lim_n \kappa_n(\sigma)\Big(\bigcup_j A_j \Big)
        \\
        & = \lim_n \sum_j \kappa_n(\sigma)(A_j) = \sum_j \lim_n \kappa_n(\sigma)(A_j)
        \\
        & = \sum_j \sup_n \kappa_n(\sigma)(A_j) = \sum_j \kappa(\sigma)(A_j).
\end{align*}
The exchange between $\sup$ and $\lim$ in the derivation uses the fact that the sequence involved is increasing. The third equality follows from the countable additivity of $\kappa_n(\sigma)$, and the fourth from the monotone convergence theorem applied to the counting measure on the set of natural numbers. Finally,
\[
        \kappa(\sigma)(\State \times [0,\infty)) = \sup_n \kappa_n(\sigma)(\State \times [0,\infty))  \leq 1.
\]
We have just shown that $\kappa(\sigma)$ is a subprobability measure.

To show the measurability of $\kappa$, we use the fact that a map $f$ from a measurable space $(X,\Sigma)$ to $\Spr(\State \times [0,\infty))$ is measurable if and only if for all measurable subsets $A$ of $\State \times [0,\infty)$ and reals $r \in \mathbb{R}$,
\[
        f^{-1}\Big(\{\mu \,\mid\,\mu(A) \leq r\}\Big) \in \Sigma.
\]
Pick a measurable subset $A$ of $\State \times [0,\infty)$ and a real $r \in \mathbb{R}$. Then,
\begin{align*}
        \kappa^{-1}(\{\mu \,\mid\,\mu(A) \leq r\})
        & = \{\sigma \,\mid\, \kappa(\sigma)(A) \leq r\}
        \\
        & = \{\sigma \,\mid\, \sup_n\kappa_n(\sigma)(A) \leq r\}
        \\
        & = \{\sigma \,\mid\, \kappa_n(\sigma)(A) \leq r \text{ for all $n$}\}
        \\
        & = \bigcap_n \{\sigma \,\mid\, \kappa_n(\sigma)(A)\}.
\end{align*}
But the intersection at the end of this derivation gives a measurable set because each
$\{\sigma \,\mid\, \kappa_n(\sigma)(A)\}$ is measurable and the countable intersection preserves measurability.
\end{proof}

\begin{lemma}\label{lemma:integration-continuity-measure} 
        Let $\{\kappa_n\}_n$ be an $\omega$-chain in $\cK$. For all non-negative measurable functions $h : \State \times [0,\infty) \to [0,\infty)$ and all states $\sigma$,
        \[
                \int \Big(\bigsqcup_n \kappa_n\Big)(\sigma)(\dd(\sigma',w'))\, h(\sigma',w')
                =
                \lim_n \int \kappa_n(\sigma)(\dd(\sigma',w'))\, h(\sigma',w').
        \]
\end{lemma}
%
%
        
\begin{proof}
We prove it by a simpler version of the so called monotone class theorem. This version says that in order to show the claimed equality for all non-negative measurable $h$, it suffices to show the following properties:
\begin{itemize}
        \item The claimed equality holds if $h$ is the indicator function $\lambda (\sigma,w).\,\ind{(\sigma,w) \in A}$ for a measurable subset $A$ of $\State \times [0,\infty)$.
        \item If non-negative measurable $h_1$ and $h_2$ satisfy the claimed equality and $r_1,r_2$ are non-negative real numbers, the linear combination $r_1 \cdot h_1 + r_2 \cdot h_2$ also satisfies the equality.
        \item If $\{h_m\}_m$ is a countable family of non-negative measurable functions such that each $h_m$ satisfies the claimed equality and $h_m \leq h_{m+1}$ for every $m$, then $\sup_m h_m$ also satisfies the claimed equality.
\end{itemize}
We will show these properties one by one.

Let us start with the proof of the first property. Pick a measurable subset $A$ of $\State \times [0,\infty)$. Then,
\begin{align*}
        \int \Big(\bigsqcup_n \kappa_n\Big)(\sigma)(\dd(\sigma',w'))\, \ind{(\sigma',w') \in A}
        &  =
        \Big(\bigsqcup_n \kappa_n\Big)(\sigma)(A)
        = \sup_n (\kappa_n(\sigma)(A)) = \lim_n (\kappa_n(\sigma)(A))
        \\
        & = \lim_n \int \kappa_n(\sigma)(\dd(\sigma',w'))\, \ind{(\sigma',w') \in A}.
\end{align*}
The first and the last equalities use the definition of integration, the second the characterisation of the least upper bound in $\cK$, and the third the fact that $\{\kappa_n(\sigma)(A)\}_n$ is increasing.

For the second property, assume that the claimed equality holds for non-negative measurable functions $h_1$ and $h_2$. Let $r_1$ and $r_2$ be non-negative real numbers. Then,
\begin{align*}
        & \int \Big(\bigsqcup_n \kappa_n\Big)(\sigma)(\dd(\sigma',w'))\, (r_1 h_1(\sigma',w') + r_2 h_2(\sigma',w'))
        \\
        & = r_1 \int \Big(\bigsqcup_n \kappa_n\Big)(\sigma)(\dd(\sigma',w'))\, h_1(\sigma',w')
            + r_2 \int \Big(\bigsqcup_n \kappa_n\Big)(\sigma)(\dd(\sigma',w'))\, h_2(\sigma',w')
        \\
        & = r_1 \lim_n \int \kappa_n(\sigma)(\dd(\sigma',w'))\, h_1(\sigma',w')
            + r_2 \lim_n \int \kappa_n(\sigma)(\dd(\sigma',w'))\, h_2(\sigma',w')
        \\
        & = \lim_n \Big(r_1 \int \kappa_n(\sigma)(\dd(\sigma',w'))\, h_1(\sigma',w')
            + r_2 \int \kappa_n(\sigma)(\dd(\sigma',w'))\, h_2(\sigma',w')\Big)
        \\
        & = \lim_n \int \kappa_n(\sigma)(\dd(\sigma',w'))\, (r_1 h_1(\sigma',w')  + r_2 h_2(\sigma',w')).
\end{align*}
The first and the last equalities use the linearity of integration, the second the assumption on $h_1$ and $h_2$, and the third the continuity of constant multiplication and addition.

For the final third property, consider an increasing sequence $\{h_m\}_m$ in the property. Then,
\begin{align*}
        \int \Big(\bigsqcup_n \kappa_n\Big)(\sigma)(\dd(\sigma',w'))\, \Big((\sup_m h_m)(\sigma',w')\Big)
        & = \int \Big(\bigsqcup_n \kappa_n\Big)(\sigma)(\dd(\sigma',w'))\, \sup_m (h_m(\sigma',w'))
        \\
        & = \int \Big(\bigsqcup_n \kappa_n\Big)(\sigma)(\dd(\sigma',w'))\, \lim_m (h_m(\sigma',w'))
        \\
        & = \lim_m \int \Big(\bigsqcup_n \kappa_n\Big)(\sigma)(\dd(\sigma',w'))\, h_m(\sigma',w')
        \\
        & = \lim_m \lim_n \int \kappa_n(\sigma)(\dd(\sigma',w'))\, h_m(\sigma',w')
        \\
        & = \lim_n \lim_m \int \kappa_n(\sigma)(\dd(\sigma',w'))\, h_m(\sigma',w').
\end{align*}
The third equality uses the monotone convergence theorem, and the fourth the assumption on the $h_m$.
\end{proof}

{\sc Theorem~\ref{thm:command-welldefined}.\ }{\it For every command $C$, its interpretation $\db{C}$ is well-defined and it belongs to $\cK$.}

\begin{proof}
We prove the theorem by induction on the structure of $C$. In the proof, we use concepts and results from the measure theory that go beyond what we reviewed in \S\ref{sec:prelim}.

We start with the case $C \equiv (x:=E)$. For every $\sigma \in \State$, $\db{x:=E}(\sigma)$
is the Dirac measure at $(\sigma[x\mapsto \db{E}\sigma_s],1)$. Thus, it suffices to show the
measurability of the following function $F$ for every measurable subset $A$:
\[
        F(\sigma) = \ind{(\sigma[x\mapsto \db{E}\sigma_s],1) \in A}.
\]
This holds because the functions
\[
        (\sigma,w) \longmapsto \ind{(\sigma,w) \in A} : \State \times [0,\infty) \to [0,1],
        \quad
        \sigma \longmapsto (\sigma[x\mapsto \db{E}\sigma_s],1) : \State \to \State \times [0,\infty)
\]
are both measurable.

The case $C \equiv \cskip$ follows from our argument for the previous case $C \equiv (x:=E)$, because $\db{\cskip} = \db{x:=x}$.

Next we consider sequential composition. Assume that $C \equiv (C_0;C_1)$. By induction hypothesis, we know that $\db{C_0}$ and $\db{C_1}$ are subprobability kernels. Then, for all $\sigma$ and $A$,
\begin{align*}
        \db{C_0;C_1}(\sigma)(A)
        & = \int \db{C_0}(\sigma)(\dd (\sigma',w')) \int \db{C_1}(\sigma')(\dd (\sigma'',w''))\,\ind{(\sigma'',w'\cdot w'') \in A}
        \\
        & \leq \int \db{C_0}(\sigma)(\dd (\sigma',w'))
          \leq 1.
\end{align*}
Furthermore, it is not difficult to show that $\db{C_0;C_1}(\sigma)$ is a measure. The measurability of $\db{C_0;C_1}$ follows from three facts. First, it suffices to show that for all measurable subsets $A$, the function
\[
        \sigma \longmapsto \db{C_0;C_1}(\sigma)(\dd(\sigma',w'))(A)
\]
is measurable. Second, for every measurable $f$ and every subprobability kernel $\kappa$, the function $x\longmapsto \int \kappa(x)(\dd y)f(y)$ is measurable. Second, the functions $(\sigma,w) \longmapsto \db{C_1}(\sigma) \otimes \delta_w$ and $((\sigma'',w''),w) \longmapsto (\sigma'',w'\cdot w'')$ are measurable. Here $\delta_w$ is the Dirac measure at $w$ and $\otimes$ is the operator for product measure. 
%
%
%

We move on to the case of $C \equiv (\code{if}\ B\ \{C_0\}\ \code{else}\ \{C_1\})$. We first show that the function $\db{C}$ is a well-defined function from $\State$ to $\Spr(\State \times [0,\infty))$, that is, $\db{C}(\sigma)$ is a subprobability measure for every $\sigma \in \State$. The non-negativity and countable additivity of $\db{C}(\sigma)$ follow from the same properties of $\db{C_0}(\sigma)$ and $\db{C_1}(\sigma)$. Also,
\begin{align*}
        \db{C}(\sigma)(A)
        & =
        \db{\cif\,B\,\{C_0\}\,\celse\,\{C_1\}}(\sigma)(A)
        \\
        & =
        \ind{\db{B}\sigma_s = \true} \cdot \db{C_0}(\sigma)(A)
          + \ind{\db{B}\sigma_s \neq \true} \cdot \db{C_1}(\sigma)(A)
        \\
        & \leq
        \ind{\db{B}\sigma_s = \true} \cdot 1
          + \ind{\db{B}\sigma_s \neq \true} \cdot 1
          = 1.
\end{align*}
It remains to show that $\db{C}$ is measurable. This boils down to showing that
\[
        \{ \sigma \,\mid\, \ind{\db{B}\sigma_s = \true} \cdot \db{C_0}(\sigma)(A) + \ind{\db{B}\sigma_s \neq \true} \cdot \db{C_1}(\sigma)(A) \leq r\}
\]
is measurable for all $r \in \mathbb{R}$ and measurable subset $A$ of $\State \times [0,\infty)$. But this follows easily from the induction hypothesis and the measurability of $\db{B}$.

        The next case is $C \equiv (\cwhile\,B\,\{C_0\})$. Recall Lemma~\ref{lemma:subprob-kernels-cpo}, which says that the set of subprobability kernels from $\State$ to $\State \times [0,\infty)$, denoted by $\cK$, is an $\omega$-cpo when given the pointwise order $\sqsubseteq$.  Let $F$ be the function on $\cK$ used in the semantics of $C$. That is,  for all $\kappa \in \cK$, $\sigma \in \State$ and measurable $A \subseteq \Store \times [0,\infty)$,
\begin{align*}
        F(\kappa)(\sigma)(A) \defeq {}
        &
        \ind{\db{B}\sigma_s \neq \true} \cdot
        \ind{(\sigma,1) \in A}
        \\
        &
        {} +
        \ind{\db{B}\sigma_s = \true} \cdot
        \int \db{C_0}(\sigma)(\dd (\sigma',w')) \int \kappa(\sigma')(\dd(\sigma'',w''))\,\ind{(\sigma'',w'\cdot w'') \in A}.
\end{align*}
It suffices to show that $F$ is a continuous function on $\cK$. The arguments used to handle the if and sequencing cases show that $F(\kappa)$ is a subprobability kernel. The monotonicity of $F$ follows from the fact that the integration is monotone with respect to the measure and its integrand. It remains to show that $F$ preserves the limits of $\omega$-chains. Consider an $\omega$-chain $\{\kappa_n\}_n$ of subprobability kernels in $\cK$. Then, the least upper bound of this $\omega$-chain is $\kappa = \bigsqcup_n \kappa_n$. The following calculation shows that $F$ preserves this limit.
\begin{align*}
        & F(\kappa)(\sigma)(A)
        \\
        & {} =
        \ind{\db{B}\sigma_s \neq \true} \cdot
        \ind{(\sigma,1) \in A}
        \\
        & \qquad\qquad
        {} +
        \ind{\db{B}\sigma_s = \true} \cdot
        \int \db{C_0}(\sigma)(\dd (\sigma',w')) \int \kappa(\sigma')(\dd(\sigma'',w''))\,\ind{(\sigma'',w'\cdot w'') \in A}
        \\
        & {} =
        \ind{\db{B}\sigma_s \neq \true} \cdot
        \ind{(\sigma,1) \in A}
        \\
        & \qquad\qquad
        {} +
        \ind{\db{B}\sigma_s = \true} \cdot
        \int \db{C_0}(\sigma)(\dd (\sigma',w')) \int \bigsqcup_n \kappa_n(\sigma')(\dd(\sigma'',w''))\,\ind{(\sigma'',w'\cdot w'') \in A}
        \\
        & {} =
        \ind{\db{B}\sigma_s \neq \true} \cdot
        \ind{(\sigma,1) \in A}
        \\
        & \qquad\qquad
        {} +
        \ind{\db{B}\sigma_s = \true} \cdot
        \int \db{C_0}(\sigma)(\dd (\sigma',w')) \lim_n \int \kappa_n(\sigma')(\dd(\sigma'',w''))\,\ind{(\sigma'',w'\cdot w'') \in A}
        \\
        & {} =
        \ind{\db{B}\sigma_s \neq \true} \cdot
        \ind{(\sigma,1) \in A}
        \\
        & \qquad\qquad
        {} +
        \ind{\db{B}\sigma_s = \true} \cdot
        \lim_n \int \db{C_0}(\sigma)(\dd (\sigma',w')) \int \kappa_n(\sigma')(\dd(\sigma'',w''))\,\ind{(\sigma'',w'\cdot w'') \in A}
        \\
        & {} =
        \lim_n\Big(\ind{\db{B}\sigma_s \neq \true} \cdot
        \ind{(\sigma,1) \in A}
        \\
        & \qquad\qquad\qquad
        {} +
        \ind{\db{B}\sigma_s = \true} \cdot
        \int \db{C_0}(\sigma)(\dd (\sigma',w')) \int \kappa_n(\sigma')(\dd(\sigma'',w''))\,\ind{(\sigma'',w'\cdot w'') \in A}\Big)
        \\
        & {} =
        \lim_n F(\kappa_n)(\sigma)(A)
        = \Big(\bigsqcup_n F(\kappa_n)\Big)(\sigma)(A).
\end{align*}
The third equality uses the continuity of integration with respect to the integrating measure and the non-negative integrand (Lemma~\ref{lemma:integration-continuity-measure}), and the fourth equality uses the monotone convergence theorem.

The next case is $C \equiv (x:=\csample_\cnorm(S,E_1,E_2))$. The requirement that 
\[
        \db{x:=\csample_\cnorm(S,E_1,E_2)}(\sigma)
\]
is a measure for every $\sigma$ follows immediately from the semantic equation for this sample statement, which says that $\db{x:=\csample_\cnorm(S,E_1,E_2)}(\sigma)$ is either the constant-$0$ measure or the pushforward of a normal distribution with mean $\db{E_1}\sigma_s$ and standard deviation $\db{E_2}\sigma_s$ by some measurable function. In fact, the following calculation shows that $\db{x:=\csample_\cnorm(S,E_1,E_2)}$ is a subprobability measure:
\begin{align*}
        & \db{x:=\csample_\cnorm(S,E_1,E_2)}(\sigma)(A)
        \\
        & {} =
        \ind{\db{S}\sigma_s \not\in \dom(\sigma_r)} 
        \cdot
        \ind{\db{E_2}\sigma_s \not\in (0,\infty)}
        \cdot
        \int \dd v\,\big(
              \cN(v;\db{E_1}\sigma_s,\db{E_2}\sigma_s)
              \cdot
              \ind{((\sigma_s[x\mapsto v],\,\sigma_r[\db{E_2} \sigma_s \mapsto v]), 1) \in A}\big)
        \\
        & {} \leq
        \ind{\db{S}\sigma_s \not\in \dom(\sigma_r)}
        \cdot
        \ind{\db{E_2}\sigma_s \not\in (0,\infty)}
        \cdot
        1
        \\
        & {} \leq 1.
\end{align*}
Also, since all of $\cN$, $\db{S}$, $\db{E_0}$ and $\db{E_1}$ are measurable, 
\[
        \{\sigma \,\mid\, \db{x:=\csample_\cnorm(S,E_1,E_2)}(\sigma)(A) \leq r\}
\]
is a measurable subset of $\State$ for all $A$ and $r$. This means that $\db{x:=\csample_\cnorm(S,E_1,E_2)}$ is a measurable function. From what we have proved, it follows that $\db{x:=\csample_\cnorm(S,E_1,E_2)}$ is a subprobability kernel as required.

The last case is $C \equiv \cscore_\cnorm(E_0,E_1,E_2)$. Because of the semantic equation for this score statement, $\db{\cscore_\cnorm(E_0,E_1,E_2)}(\sigma)$ is a measure for all $\sigma$. Furthermore, $\db{\cscore_\cnorm(E_0,E_1,E_2)}(\sigma)(A)$ is the product of two indicator functions,
and so it is at most $1$. That is, $\db{\cscore_\cnorm(E_0,E_1,E_2)}(\sigma)$ is a subprobability measure.
By the semantic equation for the score statement again, the measurability of the normal density and the measurability of $E_0$, $E_1$ and $E_2$,
the set $\{\sigma \,\mid\, \db{\cscore_\cnorm(E_0,E_1,E_2)}(\sigma)(A) \leq r\}$ is  measurable for all $A$ and $r$. This measurability implies that $\db{\cscore_\cnorm(E_0,E_1,E_2)}$ is a measurable function. By putting all these together, we can derive the required fact that $\db{\cscore_\cnorm(E_0,E_1,E_2)}$ is a subprobability kernel.
\end{proof}


{\sc Lemma~\ref{lemma:rdb-disj-measurable}.\ }{\it 
  For every measurable $h : \Rdb \times \Rdb \times \Rdb \to \mathbb{R}$,
  the function $(r,r') \longmapsto \ind{r \# r'} \times h(r,r',r\uplus r')$
  from $\Rdb \times \Rdb$ to $\mathbb{R}$ is measurable.}

\begin{proof}
        Let ${*} : \Rdb \times \Rdb \to \Rdb$ be the following extension of the partial function $\uplus : \Rdb \times \Rdb \to_\smath{partial} \Rdb$ to a total function:
        \[
                r * r' \defeq 
                \left\{\begin{array}{ll}
                        r \uplus r' & \text{if } r\#r',
                        \\[0.5ex]
                        {[]} & \text{otherwise}.
                \end{array}\right.
        \]
        Then,  for all $r,r'$,
        \[
                \ind{r \# r'} \cdot h(r,r',r\uplus r') 
                =
                \ind{r \# r'} \cdot h(r,r',r * r').
        \]
        We will show that $(r,r') \longmapsto \ind{r \# r'}$ and $(r,r') \longmapsto r*r'$ are measurable. 
        This is enough because the function on the RHS of the above equation 
        (namely, $(r,r') \longmapsto \ind{r \# r'} \cdot h(r,r',r * r')$) is constructed by applying pairing, composition and multiplication operators on measurable functions and so it is measurable. 

        Pick a real number $v \in \mathbb{R}$. We should show that $B = \{(r,r') \,\mid\, \ind{r \# r'} \leq v\}$
        is measurable. If $v \geq 1$, the set $B$ is $\Rdb \times \Rdb$, and so it is measurable. If $v < 0$,
        then $B$ is the empty set, so that it is measurable as well. If $0 \leq v < 1$, then
        \[
                \{(r,r') \,\mid\, \ind{r \# r'} \leq v\}
                = \bigcup_{K \subseteq_\fin \Str} \bigcup_{\substack{K' \subseteq_\fin \Str \\ K\cap K' \neq \emptyset}} [K \to \bR] \times [K' \to \bR],
        \]
        which is measurable because the RHS of the above equation is the countable union of measurable sets.

        Pick a measurable subset $A$ of $\Rdb$. We have to prove that $B = \{(r,r') \,\mid\, r * r' \in A\}$ is measurable. If $A = \{[]\}$, then $B = \{(r,r') \,\mid\, \ind{r \# r'} \leq 0.5\} \cup \{([],[])\}$, so that it is measurable. If $[] \not\in A$, then 
        \begin{align*}
                & \{(r,r') \,\mid\, r * r' \in A\} = {}
                \\
                & \qquad \bigcup_{K \subseteq_\fin \Str} \bigcup_{\substack{K' \subseteq_\fin \Str \\ K\cap K' = \emptyset}} \{ (r,r') \in [K \to \bR] \times [K' \to \bR] \,\mid\, r \uplus r' \in \{r'' \in A \,\mid\, \dom(r'') = K \cup K'\}\}.
        \end{align*}
	But $\{r'' \in A \,\mid\, \dom(r'') = K \cup K'\}$ is measurable, and the set on the RHS of the above equation is the countable union of measurable sets. Thus, $B$ is measurable in this case as well. 
        The remaining case is that $A$ contains $[]$ and some other element. We can prove this case by splitting $A$ into $\{[]\}$ and $A \setminus \{[]\}$ and using what we proved in the previous two cases. 
        %
        %
\end{proof}

{\sc Lemma~\ref{lem:4:6:cpo}.\ }{\it $(\cD,\sqsubseteq)$ is an $\omega$-complete partial order and has the least element $a \longmapsto \bot$.  Thus, every continuous function $G$ on $\cD$ has the least fixed point.}
\begin{proof}
  All constant functions from a measurable space to another are measurable. Thus, the constant function $a \longmapsto \bot$ is measurable. It is immediate that this constant function is local and smaller than any other functions in $\cD$.

  Let $g_0 \sqsubseteq g_1 \sqsubseteq g_2 \sqsubseteq \ldots$ be an $\omega$-chain.
  We show that the following function is the least upper bound of the sequence in $\cD$:
  for all $s,r$,
  \[
  \Big(\bigsqcup_i g_i\Big)( s, r ) =
  \left\{
    \begin{array}{ll}
            \bot & \text{ if } g_i( s, r ) = \bot \text{ for all $i$,}\\[0.5ex]
            c \not= \bot & \text{ if } g_i( s, r ) = c \not= \bot \text{ for some $i$.}
    \end{array}
  \right.
  \]
  By definition, this function is the least upper bound in $\cD$ that we are looking for, if it is measurable and local. To check the locality, suppose that $(\bigsqcup_i g_i)(s,r) = (s',r',w',p')$. Then, there exists some $i$ such that $g_i(s,r) = (s',r',w',p')$. By the locality of $g_i$, we have that
  \begin{align*}
        & (\exists r''.\, r'\# r'' \wedge r = r' \uplus r'' \wedge g_i(s,r'') = (s',[],w',p'))
        \\
        &
        \qquad\qquad
        {}
        \wedge
        (\forall r''.\, r\#r'' \implies g_i(s,r\uplus r'') = (s',r' \uplus r'',w',p')).
  \end{align*}
Thus, the definition of $\bigsqcup_i g_i$ implies that the above property holds when we replace $g_i$ by $\bigsqcup_i g_i$. We have just shown that $\bigsqcup_i g_i$ is local.

  It remains to show that $\bigsqcup_i g_i$ is measurable. Let $A \subseteq \{\bot\} \cup (\Store \times \Rdb \times [0,\infty) \times [0,\infty))$ be a measurable set. We prove that $(\bigsqcup_i g_i)^{-1}( A )$ is a measurable set.
  \begin{compactitem}
  \item Case $A = \{ \bot \}$: \\
    \begin{align*} 
            (s,r) \in \Big(\bigsqcup_i g_i\Big)^{-1}( \{ \bot \} ) 
            \iff g_i(s,r) = \bot \text{ for all $i$} 
            & {} \iff (s,r) \in g_i^{-1}( \{\bot\}) \text{ for all $i$}
            \\ 
            & {} \iff (s,r) \in \bigcap_i g_i^{-1}( \{ \bot \} ).
    \end{align*}
    The intersection of countably many measurable sets are measurable,
    and $g_0, g_1, \ldots$ are measurable functions.
    Hence we get the measurability of $(\bigsqcup_i g_i)^{-1}(\{ \bot \})$. 
  \item Case $\bot \not\in A$: \\
    \begin{align*} 
            (s,r) \in \Big(\bigsqcup_i g_i\Big)^{-1}( A ) 
            \iff g_i(s,r) \in A \text{ for some $i$}
            & \iff (s,r) \in g_i^{-1}( A ) \text{ for some $i$} 
            \\
            & \iff (s,r) \in \bigcup_i g_i^{-1}( A ).
    \end{align*}
    The first equivalence follows from the definition of the order $\sqsubseteq$. The union of 
    measurable sets are measurable, and $g_0, g_1, \ldots$ are measurable functions. Thus, the desired result follows.
  \item Case $A = \{ \bot \} \cup A'$ with $\bot \not\in A'$: \\
    Then $A'$ is also measurable as $\{ \bot \}$ is and the set subtraction
    preserves measurability.
    Moreover $(\bigsqcup_i g_i)^{-1}( A ) = (\bigsqcup_i g_i)^{-1}( \{ \bot \} )
    \cup (\bigsqcup_i g_i)^{-1}( A' )$, and the union of countably many
    measurable sets is measurable, and so we can apply the two cases above.
  \end{compactitem}
\end{proof} 

The $\omega$-cpo $\cD$ permits an operator for composing functions in it. We say that a function 
\[
        g : 
        \Big(\{\bot\} \cup (\Store \times \Rdb \times [0,\infty) \times [0,\infty))\Big)
        \to 
        \Big(\{\bot\} \cup (\Store \times \Rdb \times [0,\infty) \times [0,\infty))\Big)
\]
is \emph{local} if for all $(s,r,w,p), (s',r',w',p') \in \Store \times \Rdb \times [0,\infty) \times [0,\infty)$,
\begin{align*}
        g(s,r,w,p) = (s',r',w',p') 
        \implies  {}
        &
        (\exists r''.\, r'\# r'' \wedge r = r' \uplus r'' \wedge g(s,r'',w,p) = (s',[],w',p'))
        \\
        &
        {}
        \wedge
        (\forall r''.\, r\#r'' \implies g(s,r\uplus r'',w,p) = (s',r' \uplus r'',w',p')).
\end{align*}
We say that $g$ is \emph{strict} if $g(\bot) = \bot$. Let
\begin{align*}
        \cD^\ddagger
        \defeq
        \Big\{
                g :
        \Big(\{\bot\} \cup (\Store \times \Rdb \times [0,\infty) \times [0,\infty))\Big)
        \to 
        \Big(\{\bot\} \cup (\Store \times \Rdb \times [0,\infty) \times [0,\infty))\Big)
        &
        \\
        \qquad\qquad\qquad\qquad
        ~\Big|~
        \text{ $g$ is measurable, local and strict}
        \Big\}.
        &
\end{align*}
Define a partial order $\sqsubseteq$ on $\cD^\ddagger$ by
\[
        g \sqsubseteq g' \iff
        \forall a \in \{\bot\} \cup (\Store \times \Rdb \times [0,\infty) \times [0,\infty)).\,
        (g(a) = \bot \vee g(a) = g'(a)).
\]
\begin{lemma}
        $(\cD^\ddagger,\sqsubseteq)$ is an $\omega$-complete partial order and has the least element.
\end{lemma}
\begin{proof}
Let
\begin{align*}
        \cE
        \defeq
        \Big\{
                g :
        \Big(\Store \times \Rdb \times [0,\infty) \times [0,\infty)\Big)
        \to 
        \Big(\{\bot\} \cup (\Store \times \Rdb \times [0,\infty) \times [0,\infty))\Big)
        &
        \\
        \qquad\qquad\qquad\qquad
        ~\Big|~
        \text{ $g$ is measurable and local}
        \Big\}.
        &
\end{align*}
        Essentially by the same argument as the one used in the proof of Lemma~\ref{lem:4:6:cpo}, the set $\cE$ with the pointwise order $\sqsubseteq$ is an $\omega$-cpo and has the least element. But $(\cD^\ddagger,\sqsubseteq)$ and $(\cE,\sqsubseteq)$ are isomorphic as partially order sets. Thus, $(\cD^\ddagger,\sqsubseteq)$ has the property claimed by the lemma.
\end{proof}

The following lemma says that the lifting $g^\ddagger$ of $g \in \cD$ in \S\ref{subsec:posterior-density}
is a well-defined function from $\cD$ to this new domain.
\begin{lemma}
  \label{lem:4:6:dagger}
  The function $g \longmapsto g^\ddagger$ from $\cD$ to $\cD^\ddagger$ is
  well-defined and continuous.
\end{lemma}
\begin{proof} 
        We first show that the function is well-defined. Let \( g \in \cD \). Then, \( g^\ddagger( \bot ) = \bot \) by the definition of $g^\ddagger$. The measurability of $g^\ddagger$ follows from the measurability of \( g \), projection, and multiplication and the fact that both the pairing of measurable functions (into a function to a product measurable space) and the function composition preserve measurability. The locality of $g^\ddagger$ follows from that of $g$.

        Next, we show the claimed continuity. The monotonicity is an immediate consequence of the definitions of the $-^\ddagger$ operator and the orders on $\cD$ and $\cD^\ddagger$. It remains to show the limit preservation. Let $g_0 \sqsubseteq g_1 \sqsubseteq g_2 \sqsubseteq \ldots$ be an $\omega$-chain in \( \cD \). Then, by the monotonicity of $-^\ddagger$, $\{g^\ddagger_n\}_n$ is also an $\omega$-chain in $\cD^\ddagger$. We should show that
        \[
                \Big(\bigsqcup_n g_n\Big)^\ddagger(a) =
                \Big(\bigsqcup_n g_n^\ddagger\Big)(a)
        \]
        for all $a$ from $(\{\bot\} \cup \Store \times \Rdb \times [0,\infty) \times [0,\infty))$. Pick such an $a$. If $a = \bot$, the required holds because both sides become $\bot$. If $a \neq \bot$ but $g_n(a) = \bot$ for all $n$, then again both sides of the equality are $\bot$, and so the equality holds.
        If neither of these two cases hold, there exists $n$ such that
        $g_n(a) \neq \bot$ and $g_n(a) = g_m(a)$ for all $m \geq n$.
        Let $(s',r',w',p') = g_n(a)$, and $(s,r) = a$. Then,
        \[
                \Big(\bigsqcup_n g_n\Big)^\ddagger(a) = 
                \Big(s',r',w \times w', p \times p'\Big) = 
                \Big(\bigsqcup_n g_n^\ddagger\Big)(a).
        \]
\end{proof}

{\sc Lemma~\ref{lem:4:7:dsemwd}.\ }{\it For every command $C$, its semantics $\db{C}_d$ is well-defined and belongs to $\cD$.}

\begin{proof}
  The proof that \( \db{C}_d \) is well-defined and measurable proceeds by induction on the structure of commands $C$. 
  \begin{compactitem}
  \item Case of \( \db{\cskip}_d \): \\
          The semantics is clearly well-defined in this case. Note that $\db{\cskip}_d$ is essentially the pairing of the identity function and a constant function. Since both of these functions are measurable and the pairing preserves measurability, $\db{\cskip}_d$ is measurable. Also, since $\db{\cskip}_d$ neither changes nor depends on the $r$ component, it is local.

  \item Case of \( \db{x:=E}_d \): \\
    The proof is identical to the previous case except that instead of the identity function, we
    should use the function $(s,r) \longmapsto (s[x\mapsto \db{E}s],r)$, which is measurable
    because it can be constructed by composing and pairing three measurable functions, $\db{E}$,
    \( (s,v) \longmapsto s[x \mapsto v] \), and projection.
  \item Case of \( \db{\cif\,B\,\{C_0\}\,\celse\,\{C_1\}}_d \): \\
    The well-definedness and the locality are immediate consequences of the induction hypothesis on $\db{C_0}_d$ and $\db{C_1}_d$.  Let \( A \subseteq \{\bot\} \cup \Store \times \Rdb \times [0,\infty) \times [0,\infty) \) be a measurable set.
    Then,
    \begin{align*} 
            & 
            \db{\cif\,B\,\{C_0\}\,\celse\,\{C_1\}}_d^{-1}(A) 
            \\ 
            & \qquad {} = 
            \Big(\{ (s,r) \mid \db{B}s = \true \} \sinter \db{C_0}_d^{-1}( A ) \Big)
            \sunion 
            \Big( \{ (s,r) \mid \db{B}s \not= \true \} \sinter \db{C_1}_d^{-1}( A ) \Big).
    \end{align*}
    Moreover, \( \{ \true \} \) is measurable, and so is its complement.
    Since \( \db{B} \) is a measurable function, \( \{ (s,r) \mid \db{B}s
    = \true \} \) and \( \{ (s,r) \mid \db{B}s \not= \true \} \) are
    measurable.  As a result the above set is measurable.
    So, \( \db{\cif\,B\,\{C_0\}\,\celse\,\{C_1\}}_d \) is well-defined, local and
    measurable as claimed.
  \item Case of \( \db{C_0;C_1}_d \): \\
    By induction hypothesis, both $\db{C_0}_d$ and $\db{C_1}_d$ are well-defined, local and measurable.  This and Lemma \ref{lem:4:6:dagger} imply that $\db{C_1}^{\ddagger}_d$ is also local and measurable.  The function $\db{C_0;C_1}_d$ is nothing but the composition of two well-defined measurable functions $\db{C_1}^{\ddagger}_d$ and $\db{C_0}_d$, and so it satisfies the claimed well-definedness and measurability. To see the locality of $\db{C_0;C_1}_d$, suppose that
          \[
                  (\db{C_1}_d^\ddagger \circ \db{C_0}_d)(s,r) = (s'',r'',w'',p'').
          \]
          Then, there exist $s',r',w',p'$ such that
          \[
                  \db{C_0}_d(s,r) = (s',r',w',p')
                  \wedge
                  \db{C_1}^\ddagger_d(s',r',w',p') = (s'',r'',w'',p'').
          \]
          By the locality of $\db{C_0}_d$ and $\db{C_1}^\ddagger_d$, there exist
          $r'_0$ and $r''_0$ such that
          \begin{align*}
                  &
                  r_0' \# r' \wedge
                  r_0' \uplus r' = r \wedge
                  \db{C_0}_d(s,r_0') = (s',[],w',p') 
                  \\
                  &
                  {} \wedge 
                  r_0'' \# r'' \wedge
                  r_0'' \uplus r'' = r' \wedge
                  \db{C_1}^\ddagger_d(s',r_0'',w',p') = (s'',[],w'',p'').
          \end{align*}
          Note  that $r_0' \# r_0''$. Let $r_0 \defeq r_0' \uplus r_0''$. Then,
          \[
                  r_0 \# r'' \wedge
                  r_0 \uplus r'' = r \wedge
                  (\db{C_1}^\ddagger_d \circ \db{C_0}_d)(s,r_0) 
                  = \db{C_1}^\ddagger_d(s',r_0'',w',p')
                  = (s'',[],w'',p'').
          \]
          The first equality from above uses the locality of $\db{C_0}_d$. Thus, the first conjunct in the definition of locality holds if we use $r_0$ as a witness in the conjunct. To prove the second in the definition, consider $r_1$ such that $r_1  \# r$. Then, $r_1 \# r'$. Thus, by the locality of $\db{C_0}_d$ and $\db{C_1}^\ddagger_d$, we have
          \[
                  (\db{C_1}^\ddagger_d \circ \db{C_0}_d)(s,r \uplus r_1)
                  = \db{C_1}^\ddagger_d(s',r' \uplus r_1,w',p')
                  = (s'',r'' \uplus r_1,w'',p'').
          \]
  \item Case of \( \db{\cwhile\,B\,\{C\}}_d \): \\
    Since \( \cD \) is an $\omega$-cpo and has the least element, we simply need to prove that \( G \) in
    the definition of $\db{\cwhile\,B\,\{C\}}_d$ is well-defined and continuous so that Kleene's fixed point theorem 
    applies (Lemma~\ref{lem:4:6:cpo}).
    By induction hypothesis, \( \db{C}_d \) is well-defined, local, and measurable.
    Thus, \( G \) is a well-defined function from \( \cD \) to \( \cD \); the proof 
    that the image of \( G \) consists of local measurable functions is similar to the case of the if statement extended with the argument with the case of sequencing. Furthermore, \( G \) is monotone
    because both the lifting $-^\ddagger$ and the function composition are monotone. Thus,
    $G$ transforms an $\omega$-chain into an $\omega$-chain. Finally,
    \( G \) preserves the least upper bound of such an $\omega$-chain. This is because 
    both the lifting $-^\ddagger$ and the function composition preserve the least upper bound.
    We have just shown that $G$ is well-defined and continuous.

  \item Case of \( \db{x:=\csample_\cnorm(S,E_1,E_2)}_d \): \\
    The well-definedness and the locality are immediate. To prove the measurability, 
    pick a measurable subset $A$ of $\{\bot\} \cup (\Store \times \Rdb \times [0,\infty) \times [0,\infty)$. Then,
    \[
        \{(s,r) \,\mid\, \db{x:=\csample_\cnorm(S,E_1,E_2)}_d(s,r) \in A\} = A_0 \cup A_1 \cup A_2  \cap A_3
    \]
    where
    \begin{align*} 
            A_0 & = \{(s,r) \,\mid\, \bot \in A \wedge \db{S}s \not\in \dom(r)\},
            \\
            A_1 & = \{(s,r) \,\mid\, \bot \in A \wedge \db{E_2}s \not\in (0,\infty)\},
            \\
            A_2 & = \{(s,r) \,\mid\, \db{S}s \in \dom(r) \wedge \db{E_2}s \in (0,\infty)\},
            \\
            A_3 & = A_2 \cap \{(s,r) \,\mid\, (s[x \mapsto r(\db{S}s)],\, r \setminus \db{S}s,\, 1,\, \cN(r(\db{S}s);\db{E_1}s,\db{E_2}s)) \in A\}.
    \end{align*}
    We show that all of $A_0$, $A_1$, $A_2$ and $A_3$ are measurable. The measurability of the first three sets follow from the measurability of the following two sets:
    \[ 
    A'_0 \defeq \{(s,r) \,\mid\,\db{E_2}s \in (0,\infty)\}
    \quad\mbox{and}\quad
    A'_1 \defeq \{(s,r) \,\mid\, \db{S}s \in \dom(r)\}.
    \]
    The set $A'_0$ is measurable because it is the inverse image of the composition of two measurable functions (namely, $\db{E_1}$ and projection) to the measurable set $(0,\infty)$. For the measurability of $A'_1$, note that
    \[
            \{(s,r) \,\mid\, \db{S}s \in \dom(r)\}
            =
            \bigcup_{K \subseteq_\fin \Str} \{(s,r) \,\mid\,\dom(r) = K\} \cap \{(s,r) \,\mid\,\db{S}s \in K\}.
    \]
    The set on the RHS is measurable, because it is a countable union of the intersection of two measurable subsets; the measurability of $\{(s,r) \,\mid\,\dom(r) = K\}$ comes from the fact that $\Rdb$ is built by the disjoint-union construction over $[K \to \bR]$ for all finite $K$, and the measurability of $\{(s,r) \,\mid\,\db{S}s \in K\}$ follows from the measurability of the function $\db{S}$ and the set $K$. It remains to show the measurability of $A_3$. To do so, we note that $A_3$ is the inverse image of a function constructed by applying measurability-preserving operators on $\db{S}$, $\db{E_1}$, $\db{E_2}$ and $\cN$, and the application and restriction functions on random databases. Since the first four are measurable, it is sufficient to show the measurability of the last two, which we spell out below:
    \begin{align*}
            f_1 & : \Rdb \times \Str \to \bR,
            \qquad
            f_1(r,\alpha) 
            \defeq 
            \left\{\begin{array}{ll}
                    r(\alpha) & \text{if $\alpha \in \dom(r)$},
                    \\[0.5ex]
                    0 & \text{otherwise},
            \end{array}\right.
            \\
            f_2 & : \Rdb \times \Str \to \Rdb,
            \qquad
            f_2(r,\alpha)(\beta) 
            \defeq 
            \left\{\begin{array}{ll}
                    \text{undefined} & \text{if $\alpha = \beta$},
                    \\[0.5ex]
                    r(\beta) & \text{otherwise}.
            \end{array}\right.
    \end{align*}
    For any measurable subset $A$ of $\bR$, 
    \begin{align*}
            f_1^{-1}(A) 
            & = 
            \bigcup_{\alpha \in \Str}
            \bigcup_{K \subseteq_\fin \Str}
            \Big(
            \{(r,\beta) \,\mid\, \dom(r) = K \wedge \beta = \alpha \wedge \alpha\not\in K \wedge 0 \in A\}
            \\
            & \qquad\qquad\qquad\qquad\quad
            {} \cup \{(r,\beta) \,\mid\, \dom(r) = K \wedge \beta = \alpha \wedge \alpha \in K \wedge r(\alpha) \in A\}\Big).
    \end{align*}
    All the explicitly-defined sets on the RHS of this equation are measurable,
    and the RHS only uses finite or countable union. Thus, $f_1^{-1}(A)$ is measurable. Now
    consider a measurable subset $A'$ of $\Rdb$. Then,
    \begin{align*}
            f_2^{-1}(A')
            & 
            = 
              \bigcup_{\alpha \in \Str}
              \bigcup_{K \subseteq_\fin \Str} 
              \Big(
                \{(r,\beta) \,\mid\, \beta = \alpha \wedge \dom(r) = K \wedge \alpha \not\in K \wedge r \in A'\}
            \\
            & \qquad\qquad\qquad\qquad\quad
                {} \cup \{(r,\beta) \,\mid\, \beta = \alpha \wedge \dom(r) = K \wedge \alpha \in K \wedge r \setminus \alpha \in A'\}\Big).
    \end{align*}
    Again all the explicitly-defined sets on the RHS of this equation are measurable,
    and the RHS only uses finite or countable union. Thus, $f_2^{-1}(A')$ is measurable.
  \item Case of \( \db{\cscore_\cnorm(E_0,E_1,E_2)}_d \): \\
    The well-definedness and the locality are immediate in this case. Pick a measurable set 
    \[
            A \subseteq \{\bot\} \cup (\Store \times \Rdb \times [0,\infty) \times [0,\infty)).
    \]
    Then,
    \begin{align*}
            \db{\cscore_\cnorm(E_0,E_1,E_2)}_d^{-1}(A) 
            & =
            \{ (s,r) \mid (s,r,\cN(\db{E_0}s;\db{E_1}s,\db{E_2}s),1) \in A \} 
            \\
            & {} \cup \{ (s,r) \mid \bot \in A \wedge \db{E_2}s \not\in (0,\infty) \} 
    \end{align*}
    Both sets in the union here are measurable because all of $\cN$, $\db{E_0}$, $\db{E_1}$ and $\db{E_2}$ are measurable and the composition and the pairing operators on functions preserve measurability. Thus, the union itself is measurable.
  \end{compactitem}
\end{proof}

{\sc Lemma~\ref{lemma:density-get-measurability}.\ }{\it
        For all $g \in \cD$, the following functions from $\Store \times \Rdb$ to $\mathbb{R}$ and $\{\bot\} \cup \Store$ are measurable: $(s,r) \longmapsto \density(g,s)(r)$ and $(s,r) \longmapsto \get(g,s)(r)$.}

\begin{proof}
        We have to show that for every real $v \in \mathbb{R}$ and measurable subset $A \subseteq \{\bot\} \cup \Store$,
        \[ 
                B_0 \defeq \{(s,r) \,\mid\, \density(g,s)(r) \leq v\}
                \quad\text{and}\quad
                B_1 \defeq \{(s,r) \,\mid\, \get(g,s)(r) \in A\}
        \]
        are measurable. The set $B_0$ is measurable because 
        \begin{align*}
                & \{(s,r) \,\mid\, \density(g,s)(r) \leq v\} 
                \\
                & \qquad {} = 
                \{(s,r) \,\mid\, g(s,r) = \bot \wedge 0 \leq v\}
                \\
                & \qquad\qquad
                {} \cup \{(s,r) \,\mid\, \exists s',r',w',p'.\,g(s,r) = (s',r',w',p') \wedge r' \neq [] \wedge 0 \leq v\}
                \\
                & \qquad\qquad
                {} \cup \{(s,r) \,\mid\, \exists s',r',w',p'.\, g(s,r) = (s',r',w',p') \wedge r' = [] \wedge w' \cdot p' \leq v\}
        \end{align*}
        and the three sets on the RHS of the equation are all measurable. The measurability of the other $B_1$
        follows from a similar argument based on case split. That is,
        \begin{align*}
                & \{(s,r) \,\mid\, \get(g,s)(r) \in A\}
                \\
                & \qquad {} =
                \{(s,r) \,\mid\, g(s,r) = \bot \wedge \bot \in A\}
                \\
                & \qquad\qquad
                {} \cup \{(s,r) \,\mid\, \exists s',r',w',p'.\,g(s,r) = (s',r',w',p') \wedge r' \neq [] \wedge \bot \in A\}
                \\
                & \qquad\qquad
                {} \cup \{(s,r) \,\mid\, \exists s',r',w',p'.\, g(s,r) = (s',r',w',p') \wedge r' = [] \wedge s' \in A\}
        \end{align*}
        and all three sets on the RHS of the equation are measurable. Thus, $B_1$ is a measurable subset.
\end{proof}

In the next lemma, we regard $\{\bot\} \cup \Store$ as a partially-ordered set with the expected order: $a \sqsubseteq b \iff a = \bot \vee a = b$,
for all $a,b \in \{\bot\} \cup \Store$.
\begin{lemma}
  \label{lemma:density-get-continuity}
        Let $\{g_n\}_n$ be an $\omega$-chain in $\cD$.  
  Then, for all stores $s$ and random databases $r$,
  the sequences $\{\density(g_n,s)(r)\}_n$ and
  $\{\get(g_n,s)(r)\}_n$ are increasing,
  \[
  \lim_n \density(g_n,s)(r) = 
  \density\Big(\bigsqcup_n g_n,s\Big)(r),
  \quad\mbox{and}\quad
  \lim_n \get(g_n,s)(r) = 
  \get\Big(\bigsqcup_n g_n,s\Big)(r).
  \]
\end{lemma}
\begin{proof}
        Let $s$ be a store, $r$ be a random database, and $\{g_n\}_n$ be an $\omega$-chain in $\cD$. 
        
        We first show that the sequences $\{\density(g_n,s)(r)\}_n$ and
        $\{\get(g_n,s)(r)\}_n$ are increasing. We do so by showing the monotonicity of $\density(-,s)(r)$and $\get(-,s)(r)$. Consider $g,g' \in \cD$ such that $g \sqsubseteq g'$. Then, $g(s,r)=\bot$ or $g(s,r)=g'(s,r)$.  In the former case, $\density(g,s)(r)=0 \leq \density(g',s)(r)$ and $\get(g,s)(r) = \bot \sqsubseteq \get(g',s)(r)$. In the latter case, $\density(g,s)(r)= \density(g',s)(r)$ and $\get(g,s)(r) = \get(g',s')(r)$. Thus, $\density(-,r)(s)$ and $\get(-,r)(s)$ are monotone. 
       
        Next we show that $\density(-,s)(r)$ and $\get(-,s)(r)$ preserve the limit. By what we have already shown, the sequence $\{g_n(s,r)\}_{n}$ is monotone. But this sequence is ultimately stable. The stable value becomes the least upper bound $\bigsqcup_n g_n(s,r)$, which is also equal to $(\bigsqcup_n g_n)(s,r)$.  Thus, $\{\density(g_n,s)(r)\}_n$ is ultimately stable as well, and its limit is 
        \[ 
        \lim_n \density(g_n,s)(r) = \density\Big( \bigsqcup_n g_n, s \Big)(r).  
        \] 
        By the same reason, $\{\get(g_n,s)(r)\}_n$ is ultimately stable, and has the limit
        \[ 
        \lim_n \get(g_n,s)(r) = \get\Big( \bigsqcup_n g_n, s \Big)(r).  
        \] 
\end{proof}

{\sc Lemma~\ref{lemma:dens-get-sequencing}.\ }{\it
        For all non-negative bounded measurable functions $h : (\{\bot\} \cup \Store) \times \Rdb \to \mathbb{R}$, stores $s$,
        and functions $g_1,g_2 \in \cD$, we have that}
        \begin{align*}
                & \int \rho(\dd r)\, \Big(\density(g_2^\ddagger \circ g_1,s)(r) \cdot h\Big(\get(g_2^\ddagger \circ g_1, s)(r), r\Big)\Big)
                \\
                & {}
                = 
                \int \rho(\dd r_1)\,
                \Big(\density(g_1,s)(r_1)
                \cdot \ind{\get(g_1,s)(r_1) \neq \bot} 
                \\
                &
                \qquad
                \qquad
                \cdot \int \rho(\dd r_2)\,
                \Big(
                \density(g_2,\get(g_1,s)(r_1))(r_2) \cdot \ind{r_1 \# r_2} \cdot h\Big(\get(g_2,\get(g_1,s)(r_1))(r_2), r_1 \uplus r_2\Big)\Big)
                \Big).
        \end{align*}
\begin{proof}
        For $g \in \cD$ and $s \in \Store$, let 
        \begin{align*}
                R_1(g,s) & : \Rdb \to \Rdb,
                \\
                R_1(g,s)(r) & \defeq
                \left\{\begin{array}{ll}
                        r \setminus \dom(r') & \text{if } g(s,r) = (s',r',w',p') \text{ for some $(s',r',w',p')$},
                        \\[0.5ex]
                        [] & \text{otherwise},
                \end{array}\right.
                \\[2ex]
                R_2(g,s) & : \Rdb \to \Rdb,
                \\
                R_2(g,s)(r) & \defeq
                \left\{\begin{array}{ll}
                        r' & \text{if } g(s,r) = (s',r',w',p') \text{ for some $(s',r',w',p')$},
                        \\[0.5ex]
                        [] & \text{otherwise}.
                \end{array}\right.
        \end{align*}
        Here $r \setminus K$ means the restriction of $r$ to $\dom(r) \setminus K$. Both $R_1(g,s)$ and $R_2(g,s)$ are measurable. The measurability of the latter is an immediate consequence of the measurability of $g$. For the former, we note that for every measurable subset $A$ of $\Rdb$ with $[] \not\in A$,
        \begin{align*}
                R_1(g,s)^{-1}(A) 
                {} =
                \bigcup_{\substack{K_1,K_2 \subseteq_\fin \Str\\ K_1 \cap K_2 = \emptyset}}
                \begin{array}[t]{@{}l@{}}
                        \{r_1 \uplus r_2 \,\mid\, r_1 \in (A \cap [K_1 \to \bR]) {} \wedge r_2 \in [K_2 \to \bR]\}
                        \\[1ex]
                        \qquad\qquad
                        {} \cap
                        g^{-1}(s,-)(\Store \times [K_2 \to \bR] \times [0,\infty) \times [0,\infty)).
                \end{array}
        \end{align*}
        The partially applied function $g(s,-)$ is measurable, so that the RHS of this equation is the countable union of measurable subsets. If $A = \{[]\}$, then
        \begin{align*}
                R_1(g,s)^{-1}(A) 
                = 
                g^{-1}(s,-)(\{\bot\})
                \cup \Big(A \cap g^{-1}(s,-)(\Store \times \{[]\} \times [0,\infty) \times [0,\infty))\Big).
        \end{align*}
        Thus, $R_1(g,s)^{-1}(A)$ is measurable. The remaining case is that $[] \in A$. We can be handle this case by splitting $A$ into $\{[]\}$ and $A \setminus \{[]\}$, and dealing with the slit cases separately.
        
        For $g \in \cD$, $s \in \Store$, and $r \in \Rdb$, let $\consumed(g,s,r)$ be the predicate defined by
        \[
                \consumed(g,s,r) \iff \exists s',w',p'.\,(g(s,r) = (s',[],w',p')).
        \]
        Then, $r \longmapsto \ind{\consumed(g,s,r)}$ is a measurable function.

        Now we note a few useful equalities about
        the entities $h,s,g_1,g_2$ assumed in the lemma. 
        
        First, for all $r$, if 
        \[
                \consumed(g_2^\ddagger \circ g_1, s, r),
        \]
        then 
        \begin{align*}
                r & = R_1(g_1,s)(r) \uplus R_2(g_1,s)(r),
                \\[1ex]
                \get(g_2^\ddagger \circ g_1,s)(r) & =
                \get\Big(g_2,\get(g_1,s)(R_1(g_1,s)(r))\Big)(R_2(g_1,s)(r)),
                \\[1ex]
                \density(g_2^\ddagger \circ g_1,s)(r) & =
                \density(g_1,s)(R_1(g_1,s)(r)) 
                \cdot
                \density\Big(g_2,\get(g_1,s)(R_1(g_1,s)(r))\Big)(R_2(g_1,s)(r)).
        \end{align*}
        The first equality is just an immediate consequence of the definitions of $R_1$ and $R_2$, and the other two equalities hold mainly because $g_1$ and $g_2$ are local.

        Next, consider the density functions $f_1$ on $\Rdb$ and $f_2$ on $\Rdb \times \Rdb$:
        \begin{align*}
                f_1 & : \Rdb \to [0,\infty),
                &
                f_1(r) & \defeq \ind{\consumed(g_2^\ddagger \circ g_1, s, r)}, 
                \\
                f_2 & : \Rdb \times \Rdb \to [0,\infty)
                &
                f_2(r_1,r_2) & \defeq
                \ind{\consumed(g_1, s, r_1)} 
                \cdot \ind{\consumed(g_2, \get(g_1,s)(r_1), r_2)} 
                \\
                &&& 
                \phantom{{} \defeq \ind{\consumed(g_1, s, r_1)}}
                {} \cdot \ind{r_1\#r_2}.
        \end{align*}
        The densities are taken with respect to $\rho$ and $\rho \otimes \rho$, respectively. There is a measurable bijection $\beta$ between the supports of $f_1$ and $f_2$, and the pushforward of the measure of $f_1$ by this bijection is the measure of $f_2$. Concretely, the function $\beta$ is:
        \[
                \beta : \Rdb \to \Rdb \times \Rdb,
                \qquad
                \beta(r) = 
                \left\{\begin{array}{ll}
                        (R_1(g_1,s)(r),\,R_2(g_1,s)(r)) & \text{if } \consumed(g_2^\ddagger \circ g_1, s, r)
                        \\[0.5ex]
                        (r_d,\,r_d) & \text{otherwise},
                \end{array}\right.
        \]
        where $r_d \in \Rdb$ is $[``s"\mapsto \nil]$. Note that $\beta$ is not a bijection. But the restriction of $\beta$ to $A_1 = \{ r \,\mid\, f_1(r) > 0\}$ is injective, and its image is $A_2 = \{(r_1,r_2) \,\mid\, f_2(r_1,r_2) > 0 \}$. These properties follow from the definitions of $\consumed$, $R_1$ and $R_2$ and the locality of $g$. The function $\beta$ is measurable because $R_1(g_1,s)$ and $R_2(g_2,s)$ are measurable functions and $\{r \,\mid\,\consumed(g_2^\ddagger \circ g_1,s,r)\}$ is a measurable set. It remains to show the measure preservation of $\beta$. Pick a measurable subset $A'$ of $A_2$. We have to show that
        \[
                (\rho \otimes \rho)(A') = \rho(\beta^{-1}(A')).
        \]
        We calculate the equation as follows:
        \begin{align*}
                (\rho \otimes \rho)(A')
                & {} = (\rho \otimes \rho)\Big(\bigcup_{\substack{K_1,K_2 \subseteq_\fin \Str \\ K_1 \cap K_2 = \emptyset}} A' \cap \Big([K_1 \to \bR] \times [K_2 \to \bR]\Big)\Big)
                \\
                & {} = \bigcup_{\substack{K_1,K_2 \subseteq_\fin \Str \\ K_1 \cap K_2 = \emptyset}} \Big(\Big(\bigotimes_{\alpha_1 \in K_1} \rho_v\Big) \otimes \Big(\bigotimes_{\alpha_2 \in K_2} \rho_v\Big)\Big(A' \cap \Big([K_1 \to \bR] \times [K_2 \to \bR]\Big)\Big)
                \\
                & {} = \bigcup_{\substack{K_1,K_2 \subseteq_\fin \Str \\ K_1 \cap K_2 = \emptyset}} \Big(\bigotimes_{\alpha \in K_1 \cup K_2} \rho_v\Big)\Big(\beta^{-1}(A' \cap ([K_1 \to \bR] \times [K_2 \to \bR]))\Big)
                \\
                & {} = \bigcup_{\substack{K_1,K_2 \subseteq_\fin \Str \\ K_1 \cap K_2 = \emptyset}} \rho\Big(\beta^{-1}(A' \cap ([K_1 \to \bR] \times [K_2 \to \bR]))\Big)
                \\
                & {} = \rho\Big(\beta^{-1}\Big(\bigcup_{\substack{K_1,K_2 \subseteq_\fin \Str \\ K_1 \cap K_2 = \emptyset}} (A' \cap ([K_1 \to \bR] \times [K_2 \to \bR]))\Big)\Big)
                \\
                & {} = \rho(\beta^{-1}(A')).
        \end{align*}
       
        Using what we have shown so far, we calculate the claim of the lemma. In the calculation
        we often omit the $g_1$ and $s$ parameters from $R_1(g_1,s)(r)$ and $R_2(g_1,s)(r)$, and just write $R_1(r)$ and $R_2(r)$.
        \begin{align*}
                & \int \rho(\dd r)\, \Big(\density(g_2^\ddagger \circ g_1,s)(r) \cdot h\Big(\get(g_2^\ddagger \circ g_1, s)(r), r\Big)\Big)
                \\
                & {} = 
                \int \rho(\dd r)\, 
                \Big(
                \ind{\consumed(g_2^\ddagger \circ g_1,s,r)}
                \cdot \density(g_2^\ddagger \circ g_1,s)(r) 
                \cdot h(\get(g_2^\ddagger \circ g_1, s)(r), r)\Big)
                \\
                & {} = 
                \int \rho(\dd r)\, 
                \Big(
                \ind{\consumed(g_2^\ddagger \circ g_1,s,r)}
                \cdot \density(g_2^\ddagger \circ g_1,s)(r) 
                \cdot \ind{\get(g_1,s)(R_1(r_1)) \neq \bot} 
                \cdot h(\get(g_2^\ddagger \circ g_1, s)(r), r)\Big)
                \\
                & {} = 
                \int \rho(\dd r)\, 
                \Big(
                \ind{\consumed(g_2^\ddagger \circ g_1,s,r)}
                \cdot \density(g_1,s)(R_1(r))
                \cdot \ind{\get(g_1,s)(R_1(r_1)) \neq \bot} 
                \\
                &
                \qquad\qquad\qquad
                {} \cdot \density(g_2,\get(g_1,s)(R_1(r)))(R_2(r)) 
                {} \cdot h(\get(g_2, \get(g_1,s)(R_1(r)))(R_2(r)), R_1(r) \uplus R_2(r))\Big)
                \\
                & {} = 
                \int \rho(\dd r_1)\,\int \rho(\dd r_2) 
                \Big(
                \ind{\consumed(g_1,s,r_1)}
                \cdot 
                \ind{\consumed(g_2,\get(g_1,s)(r_1),r_2)}
                \cdot 
                \ind{r_1 \# r_2}
                \\
                & 
                \qquad\qquad\qquad\qquad\qquad\qquad\qquad
                {}
                \cdot 
                \density(g_1,s)(r_1)
                \cdot \ind{\get(g_1,s)(r_1) \neq \bot} 
                \cdot \density(g_2,\get(g_1,s)(r_1))(r_2) 
                \\
                & 
                \qquad\qquad\qquad\qquad\qquad\qquad\qquad
                {}
                {} \cdot h(\get(g_2, \get(g_1,s)(r_1))(r_2), r_1 \uplus r_2)\Big)
                \\
                & {} = 
                \int \rho(\dd r_1)\,\int \rho(\dd r_2) 
                \Big(
                \ind{r_1 \# r_2}
                \cdot
                \density(g_1,s)(r_1)
                \cdot 
                \ind{\get(g_1,s)(r_1) \neq \bot} 
                \cdot 
                \density(g_2,\get(g_1,s)(r_1))(r_2) 
                \\
                & 
                \qquad\qquad\qquad\qquad\qquad\qquad\qquad\qquad\qquad
                {}
                {} \cdot h(\get(g_2, \get(g_1,s)(r_1))(r_2), r_1 \uplus r_2)\Big)
                \\
                & {}
                = 
                \int \rho(\dd r_1)
                \Big(\density(g_1,s)(r_1)
                \cdot \ind{\get(g_1,s)(r_1) \neq \bot} 
                \\
                & 
                \qquad
                \qquad
                {} \cdot \int \rho(\dd r_2)\,
                \Big(\density(g_2,\get(g_1,s)(r_1))(r_2) \cdot \ind{r_1 \# r_2} \cdot h\Big(\get(g_2,\get(g_1,s)(r_1))(r_2), r_1 \uplus r_2\Big)\Big)
                \Big).
        \end{align*}
\end{proof}

{\sc Theorem~\ref{thm:correspondence:measure-density}.\ }{\it
        For all non-negative bounded measurable monotone functions $h : (\{\bot\} \cup \Store) \times \Rdb \to \mathbb{R}$ and states $\sigma$,}
        \[
                \int \db{C}(\sigma)(\dd (\sigma',w'))\, (w' \cdot h(\sigma'_s,\sigma'_r))
                = \int \rho (\dd r')\, \Big(\density(C,\sigma_s)(r') \cdot \ind{r' \# \sigma_r} \cdot h(\get(C,\sigma_s)(r'), r' \uplus \sigma_r)\Big).
        \]
\begin{proof}
        We prove the theorem by induction on the structure of $C$. 

        When $C \equiv \cskip$, we derive the claimed equality as follows:
        \begin{align*}
                & \int \db{\cskip}(\sigma)(\dd(\sigma',w'))\, (w' \cdot h(\sigma'_s,\sigma'_r))
                \\
                & {} = h(\sigma_s,\sigma_r)
                \\
                & {} = \int \rho(\dd r')\, \Big(\ind{r' = []} \cdot \ind{[] \# \sigma_r} \cdot h(\sigma_s,[] \uplus \sigma_r)\Big)
                \\
                & {} = \int \rho (\dd r')\, \Big(\ind{r' = []} \cdot \ind{r' \# \sigma_r} \cdot 
                 h(\sigma_s, r' \uplus \sigma_r)\Big)
                \\
                & {} = \int \rho (\dd r')\, \Big(\density(\cskip,\sigma_s)(r') \cdot \ind{r' \# \sigma_r} \cdot 
                 h(\get(\cskip,\sigma_s)(r'), r' \uplus \sigma_r)\Big).
        \end{align*}

        Next we handle the case that $C \equiv (x:=E)$.
        \begin{align*}
                & \int \db{x:=E}(\sigma)(\dd(\sigma',w'))\, (w' \cdot h(\sigma'_s,\sigma'_r))
                \\
                & {} = h(\sigma_s[x\mapsto \db{E}\sigma_s],\sigma_r)
                \\
                & {} = \int \rho(\dd r')\, \Big(\ind{r' = []} \cdot \ind{[] \# \sigma_r} \cdot h(\sigma_s[x\mapsto \db{E}\sigma_s],[] \uplus \sigma_r)\Big)
                \\
                & {} = \int \rho (\dd r')\, \Big(\ind{r' = []} \cdot \ind{r' \# \sigma_r} \cdot 
                 h(\sigma_s[x\mapsto \db{E}\sigma_s], r' \uplus \sigma_r)\Big)
                \\
                & {} = \int \rho (\dd r')\, \Big(\density(x:=E,\sigma_s)(r') \cdot \ind{r' \# \sigma_r} \cdot 
                 h(\get(x:=E,\sigma_s)(r'), r' \uplus \sigma_r)\Big).
        \end{align*}

        We move on to the case that $C \equiv (\cif\,B\,\{C_0\}\,\celse\,\{C_1\})$, and prove the desired equality:
        \begin{align*}
                & \int \db{\cif\,B\,\{C_0\}\,\celse\,\{C_1\}}(\sigma)(\dd(\sigma',w'))\,(w' \cdot h(\sigma'_s, \sigma'_r))
                \\
                & {} = \ind{\db{B}\sigma_s = \true} \cdot \int \db{C_0}(\sigma)(\dd(\sigma',w'))\,(w' \cdot h(\sigma'_s, \sigma'_r))
                \\
                & \qquad {} + \ind{\db{B}\sigma_s \neq \true} \cdot \int \db{C_1}(\sigma)(\dd(\sigma',w'))\,(w' \cdot h(\sigma'_s, \sigma'_r))
                \\
                & {} = \ind{\db{B}\sigma_s = \true} \cdot \int \rho (\dd r')\, \Big(\density(C_0,\sigma_s)(r') \cdot \ind{r' \# \sigma_r} \cdot h(\get(C_0,\sigma_s)(r'), r' \uplus \sigma_r)\Big) 
                \\
                & \qquad {} + \ind{\db{B}\sigma_s \neq \true} \cdot \int \rho (\dd r')\, \Big(\density(C_1,\sigma_s)(r') \cdot \ind{r' \# \sigma_r} \cdot h(\get(C_1,\sigma_s)(r'), r' \uplus \sigma_r)\Big)
                \\
                & {} = \int \rho (\dd r')\, \Big(\density(\cif\,B\,\{C_0\}\,\celse\,\{C_1\},\sigma_s)(r') \cdot \ind{r' \# \sigma_r} \cdot h(\get(\cif\,B\,\{C_0\}\,\celse\,\{C_1\},\sigma_s)(r'), r' \uplus \sigma_r)\Big).
        \end{align*}

        Now we prove the claimed equality for $C \equiv (C_0;C_1)$. 
        \begin{align*}
                & \int \db{C_0;C_1}(\sigma)(\dd(\sigma',w'))\, (w' \cdot h(\sigma'_s,\sigma'_r))
                \\
                & {} = \int \db{C_0}(\sigma)(\dd(\sigma',w'))\int \db{C_1}(\sigma')(\dd(\sigma'',w''))\,(w' \cdot w'' \cdot h(\sigma''_s,\sigma''_r))
                \\
                & {} = \int \db{C_0}(\sigma)(\dd(\sigma',w'))\Big(w' \cdot \int \db{C_1}(\sigma')(\dd(\sigma'',w''))\,(w'' \cdot h(\sigma''_s,\sigma''_r))\Big)
                \\
                & {} = \int \db{C_0}(\sigma)(\dd(\sigma',w'))\Big(w' \cdot \int \rho(\dd r_2)\,(\density(C_1,\sigma'_s)(r_2) \cdot \ind{r_2 \# \sigma'_r} \cdot h(\get(C_1,\sigma'_s)(r_2),r_2 \uplus \sigma'_r))\Big)
                \\
                & {} = \int \db{C_0}(\sigma)(\dd(\sigma',w'))
                \\
                & \qquad\qquad
                \Big(w' 
                \cdot \ind{\sigma'_s \neq \bot} 
                \cdot \int \rho(\dd r_2)\,(\density(C_1,\sigma'_s)(r_2) \cdot \ind{r_2 \# \sigma'_r} \cdot h(\get(C_1,\sigma'_s)(r_2),r_2 \uplus \sigma'_r))\Big)
                \\
                & {} = 
                \int \rho(\dd r_1)\, \Big(\density(C_0,\sigma_s)(r_1) \cdot \ind{r_1 \# \sigma_r} 
                \\
                & 
                \qquad\qquad\qquad
                  {} \cdot \ind{\get(C_0,\sigma_s)(r_1) \neq \bot} 
                  \cdot 
                  \int \rho(\dd r_2)\,\Big(\density(C_1,\get(C_0,\sigma_s)(r_1))(r_2) \cdot \ind{r_2 \# (r_1 \uplus \sigma_r)} 
                \\
                & 
                \qquad\qquad\quad\qquad\qquad\qquad\qquad\qquad\qquad\qquad
                  {} \cdot 
                  h(\get(C_1,\get(C_0,\sigma_s)(r_1))(r_2), r_2 \uplus (r_1 \uplus \sigma_r)\Big)\Big)
                \\
                & {} = 
                \int \rho(\dd r_1)\, \Big(\density(C_0,\sigma_s)(r_1) 
                  \cdot \ind{\get(C_0,\sigma_s)(r_1) \neq \bot} 
                \\
                & 
                \qquad\qquad\qquad
                  {} \cdot 
                  \int \rho(\dd r_2)\,\Big(\density(C_1,\get(C_0,\sigma_s)(r_1))(r_2) 
                  \cdot \ind{r_1 \# r_2}
                  \cdot \ind{(r_1 \uplus r_2) \# \sigma_r} 
                \\
                & 
                \qquad\qquad\quad\qquad\qquad\qquad\qquad
                  {} \cdot 
                  h(\get(C_1,\get(C_0,\sigma_s)(r_1))(r_2), (r_1 \uplus r_2) \uplus \sigma_r)\Big)\Big)
                \\
                & {} = \int \rho(\dd r)\, \Big(\density(C_0;C_1)(r) \cdot \ind{r \# \sigma_r} \cdot h\Big(\get(C_0;C_1)(r), r \uplus \sigma_r\Big)\Big).
        \end{align*}
        The third and fifth equalities follow from the induction hypothesis, and the last equality from
        Lemma~\ref{lemma:dens-get-sequencing}.

        The next case is $C \equiv (\cwhile\,B\,\{C_0\})$. 
        Let $\cR \subseteq \cK \times \cD$ be the relation defined by
        \begin{align*}
                \kappa \,[\cR]\, g \iff 
                & 
                \text{for all states $\sigma$ and non-negative bounded measurable monotone functions $h$, }
                \\
                &
                \qquad
                \int \kappa(\sigma)(\dd (\sigma',w'))\, (w' \cdot h(\sigma'_s,\sigma'_r))
                \\
                & \qquad\qquad
                {} = \int \rho (\dd r')\, \Big(\density(g,\sigma_s)(r') \cdot \ind{r' \# \sigma_r} \cdot h\Big(\get(g,\sigma_s)(r'), r' \uplus \sigma_r\Big)\Big).
        \end{align*}
        Note that the least elements from $\cK$ and $\cD$ are related by $\cR$ because they make
        both sides of the equality in the definition of $\cR$ be zero. Furthermore, for all $\omega$-chains $\{\kappa_n\}_n$ and $\{g_n\}_n$ in $\cK$ and $\cD$, if $\kappa_n \,[\cR]\, g_n$ for all $n$, then
        \[
                \bigsqcup_n \kappa_n \,[\cR]\, \bigsqcup_n g_n.
        \]
        This is because for all stores $\sigma$ and non-negative bounded measurable functions $h$,
        \begin{align*}
                & \int \Big(\bigsqcup_n \kappa_n\Big)(\sigma)(\dd (\sigma',w'))\, (w' \cdot h(\sigma'_s,\sigma'_r))
                \\
                & = \lim_n \int \kappa_n(\sigma)(\dd (\sigma',w'))\, (w' \cdot f(\sigma'_r))
                \\
                & = \lim_n \int \rho (\dd r')\, \Big(\density(g_n,\sigma_s)(r') \cdot \ind{r' \# \sigma_r} \cdot h(\get(g_n,\sigma_s)(r'), r' \uplus \sigma_r)\Big)
                \\
                & = \int \rho (\dd r')\, \lim_n \Big(\density(g_n,\sigma_s)(r') \cdot \ind{r' \# \sigma_r} \cdot h(\get(g_n,\sigma_s)(r'), r' \uplus \sigma_r)\Big)
                \\
                & = \int \rho (\dd r')\, \Big(\density\Big(\bigsqcup_n g_n,\sigma_s\Big)(r') \cdot \ind{r' \# \sigma_r} \cdot h\Big(\get\Big(\bigsqcup_n g_n,\sigma_s\Big)(r'),r' \uplus \sigma_r\Big)\Big).
        \end{align*}
        The first equation follows from Lemma~\ref{lemma:integration-continuity-measure}, 
        and the third from the monotone convergence theorem and Lemma~\ref{lemma:density-get-continuity}. The last equality holds because of Lemma~\ref{lemma:density-get-continuity} again.

        Thus, we can complete the proof of this case if we show one more thing. Let $K$ and $G$ be the functions on $\cK$ and $\cD$ used in the interpretation of the loop (namely, $\cwhile\,B\,\{C_0\}$) in the two semantics. What we have to show is:
        \[
                \kappa \,[\cR]\, g 
                \implies
                K(\kappa) \,[\cR]\, G(g).
        \]
        Pick $\kappa$ and $g$ with $\kappa \,[\cR]\, g$. Then, for all $\sigma$ and non-negative bounded measurable $h$,
        \begin{align*} 
                & \int (K(\kappa))(\sigma)(\dd (\sigma',w'))\, (w' \cdot h(\sigma'_s, \sigma'_r))
                \\
                & = \ind{\db{B}\sigma_s \neq \true} \cdot h(\sigma_s,\sigma_r) 
                \\
                & \qquad
                {} + \ind{\db{B}\sigma_s = \true} \cdot \int \db{C}(\sigma)(\dd (\sigma',w')) \int \kappa(\sigma')(\dd (\sigma'',w''))\,(w' \cdot w'' \cdot h(\sigma''_s,\sigma''_r)) 
                \\ 
                & = \ind{\db{B}\sigma_s \neq \true} \cdot h(\sigma_s,\sigma_r) 
                \\
                & \qquad
                {} + \ind{\db{B}\sigma_s = \true} \cdot \int \db{C}(\sigma)(\dd (\sigma',w')) \Big(w' \cdot \int \kappa(\sigma')(\dd (\sigma'',w''))\,(w'' \cdot h(\sigma''_s,\sigma''_r))\Big).
        \end{align*}
        But
        \begin{align*}
                & \int \db{C}(\sigma)(\dd (\sigma',w')) \Big(w' \cdot \int \kappa(\sigma')(\dd (\sigma'',w''))\,(w'' \cdot h(\sigma''_s, \sigma''_r))\Big) 
                \\
                & = \int \db{C}(\sigma)(\dd (\sigma',w')) \Big(w' \cdot \int \rho(\dd r'') \Big(\density(g,\sigma'_s)(r'') \cdot \ind{r'' \# \sigma'_r} \cdot h(\get(g,\sigma'_s)(r''), r'' \uplus \sigma'_r)\Big)\Big) 
                \\
                & = \int \db{C}(\sigma)(\dd (\sigma',w')) 
                \\
                & \qquad\qquad\qquad 
                \Big(w' 
                \cdot \ind{\sigma'_s \neq \bot} \cdot \int \rho(\dd r'') \Big(\density(g,\sigma'_s)(r'') \cdot \ind{r'' \# \sigma'_r} \cdot h(\get(g,\sigma'_s)(r''), r'' \uplus \sigma'_r)\Big)\Big) 
                \\ 
                \\ 
                & = \int \rho(\dd r')\,\Big(\density(C,\sigma_s)(r') \cdot \ind{r' \# \sigma_r} 
                \\
                & 
                \qquad\qquad\qquad
                {}
                \cdot
                \ind{\get(C,\sigma_s)(r') \neq \bot}
                \cdot 
                \int \rho(\dd r'') \Big(\density(g,\get(C,\sigma_s)(r'))(r'') \cdot \ind{r'' \# (r' \uplus \sigma_r)} 
                \\
                & 
                \qquad\qquad\qquad\qquad\qquad\qquad\qquad\qquad\qquad\qquad
                {} \cdot h(\get(g,\get(C,\sigma_s))(r''), r'' \uplus (r' \uplus \sigma_r))\Big)\Big) 
                \\
                & = \int \rho(\dd r')\,\Big(\density(C,\sigma_s)(r') 
                \cdot 
                \int \rho(\dd r'') \Big(\density(g,\get(C,\sigma_s)(r'))(r'')
                \cdot \ind{r' \# r''} \cdot \ind{(r' \uplus r') \# \sigma_r} 
                \\
                & 
                \qquad\qquad\qquad\qquad\qquad\qquad\qquad\qquad
                {} \cdot h(\get(g,\get(C,\sigma_s))(r''), (r' \uplus r'') \uplus \sigma_r))\Big)\Big) 
                \\
                & = \int \rho(\dd r)\,\Big(\density(g^\ddagger \circ \db{C}_d,\sigma_s)(r) \cdot \ind{r \# \sigma_r} \cdot h(\get(g^\ddagger \circ \db{C}_d,\sigma_s)(r), r \uplus \sigma_r)\Big). 
        \end{align*}
        The last equality uses Lemma~\ref{lemma:dens-get-sequencing}.
        Thus, if we continue our calculation that we paused momentarily, we get
        \begin{align*}
                & \ind{\db{B}\sigma_s \neq \true} \cdot h(\sigma_s,\sigma_r) 
                \\
                & \quad
                {} + \ind{\db{B}\sigma_s = \true} \cdot \int \db{C}(\sigma)(\dd (\sigma',w')) \Big(w' \cdot \int \kappa(\sigma')(\dd (\sigma'',w''))\,(w'' \cdot h(\sigma''_s,\sigma''_r))\Big) 
                \\
                & = \ind{\db{B}\sigma_s \neq \true} \cdot h(\sigma_s,\sigma_r) 
                \\
                & \quad
                {} + \ind{\db{B}\sigma_s = \true} \cdot 
                \int \rho(\dd r)\,\Big(\density(g^\ddagger \circ \db{C}_d,\sigma_s)(r) \cdot \ind{r \# \sigma_r} \cdot h(\get(g^\ddagger \circ \db{C}_d,\sigma_s)(r), r \uplus \sigma_r)\Big)
                \\
                & = \ind{\db{B}\sigma_s \neq \true} \cdot \int \rho(\dd r')\,\Big(
                \ind{r' = []} \cdot \ind{[] \# \sigma_r} \cdot h(\sigma_s,[] \uplus \sigma_r)\Big)
                \\
                & \quad
                {} + \ind{\db{B}\sigma_s = \true} \cdot 
                \int \rho(\dd r)\,\Big(\density(g^\ddagger \circ \db{C}_d,\sigma_s)(r) \cdot \ind{r \# \sigma_r} \cdot h(\get(g^\ddagger \circ \db{C}_d,\sigma_s)(r), r \uplus \sigma_r)\Big)
                \\
                & = \ind{\db{B}\sigma_s \neq \true} \cdot \int \rho(\dd r')\,\Big(
                \ind{r' = []} \cdot \ind{r' \# \sigma_r} \cdot h(\sigma_s,r' \uplus \sigma_r)\Big)
                \\
                & \quad
                {} + \ind{\db{B}\sigma_s = \true} \cdot 
                \int \rho(\dd r)\,\Big(\density(g^\ddagger \circ \db{C}_d,\sigma_s)(r) \cdot \ind{r \# \sigma_r} \cdot h(\get(g^\ddagger \circ \db{C}_d,\sigma_s)(r), r \uplus \sigma_r)\Big)
                \\
                & = \int \rho(\dd r')\,\Big(\density(G(g),\sigma_s)(r') \cdot \ind{r' \# \sigma_r} \cdot
                h(\get(G(g),\sigma_s)(r'), r'\uplus \sigma_r)\Big).
        \end{align*}

        Next we handle the case that $C \equiv (x:=\csample_\cnorm(S,E_1,E_2))$. 
        \begin{align*}
                &
                \int \db{x:=\csample_\cnorm(S,E_1,E_2)}(\sigma)(\dd(\sigma',w'))\,(w' \cdot h(\sigma'_s,\sigma'_r))
                \\
                & = 
                \ind{\db{S}\sigma_s \not\in \dom(\sigma_r)} 
                \cdot \ind{\db{E_2}\sigma_s \in (0,\infty)}
                \\
                &
                \qquad
                {} 
                \cdot \int \dd v\,\Big(\cN(v;\db{E_1}\sigma_s, \db{E_2}\sigma_s) \cdot h(\sigma_s[x\mapsto v],\sigma_r[\db{S}\sigma_s \mapsto v])\Big)
                \\
                & = 
                \ind{\db{E_2}\sigma_s \in (0,\infty)}
                \\
                &
                \qquad
                {} 
                \cdot \int \dd v\,
                \Big(\cN(v;\db{E_1}\sigma_s, \db{E_2}\sigma_s) 
                \cdot \ind{[\db{S}\sigma_s \mapsto v] \# \sigma_r} 
                \cdot h(\sigma_s[x\mapsto v], [\db{S}\sigma_s \mapsto v] \uplus \sigma_r)\Big)
                \\
                & = 
                \ind{\db{E_2}\sigma_s \in (0,\infty)}
                {} \cdot
                \int \rho(\dd r')\,
                \Big(
                \ind{\dom(r') = \{\db{S}\sigma_s\}} 
                \cdot \cN(r'(\db{S}\sigma_s); \db{E_1}\sigma_s, \db{E_2}\sigma_s)
                \\
                &
                \qquad\qquad\qquad\qquad\qquad\qquad
                {} \cdot \ind{r' \# \sigma_r} \cdot h(\sigma_s[x\mapsto r'(\db{E_0}\sigma_s)],r' \uplus \sigma_r)\Big)
                \\
                & {} = \int \rho (\dd r')\, 
                \Big(
                \ind{\db{E_2}\sigma_s \in (0,\infty)}
                \cdot \ind{\dom(r') = \{\db{S}\sigma_s\}} 
                \cdot \cN(r'(\db{S}\sigma_s); \db{E_1}\sigma_s, \db{E_2}\sigma_s)
                \\
                &
                \qquad\qquad\qquad\qquad\qquad\qquad
                {} \cdot \ind{r' \# \sigma_r} 
                \cdot h(\sigma_s[x\mapsto r'(\db{E_0}\sigma_s)], r' \uplus \sigma_r)\Big)
                \\
                & {} = \int \rho (\dd r')\, 
                \Big(\density(x := \csample_\cnorm(S,E_1,E_2),\sigma_s)(r') 
                \\
                &
                \qquad\qquad\qquad\qquad\qquad\qquad
                {} \cdot \ind{r' \# \sigma_r} 
                \cdot h(\get(x := \csample_\cnorm(S,E_1,E_2), \sigma_s)(r'), r' \uplus \sigma_r)\Big).
        \end{align*}

        It remains to prove the case that $C \equiv \cscore_\cnorm(E_0,E_1,E_2)$. Here is our proof for this last case:
        \begin{align*}
                & \int \db{\cscore_\cnorm(E_0,E_1,E_2)}(\sigma)(\dd (\sigma',w'))\,(w' \cdot h(\sigma'_s,\sigma'_r))
                \\
                & {} = 
                \ind{\db{E_2}\sigma_s \in (0,\infty)} 
                \cdot \cN(\db{E_0}\sigma_s;\db{E_1}\sigma_s,\db{E_2}\sigma_s) 
                \cdot h(\sigma_s, \sigma_r)
                \\
                & {} = 
                \ind{\db{E_2}\sigma_s \in (0,\infty)} 
                \cdot \int \rho (\dd r')\, 
                \Big(\ind{r' = []} 
                \cdot \cN(\db{E_0}\sigma_s;\db{E_1}\sigma_s,\db{E_2}\sigma_s) 
                \cdot \ind{[] \# \sigma_r} 
                \cdot h(\sigma_s, [] \uplus \sigma_r)\Big)
                \\
                & {} = \ind{\db{E_2}\sigma_s \in (0,\infty)}
                \cdot \int \rho (\dd r')\, 
                \Big(\ind{r' = []} 
                \cdot \cN(\db{E_0}\sigma_s;\db{E_1}\sigma_s,\db{E_2}\sigma_s) 
                \cdot \ind{r' \# \sigma_r} 
                \cdot h(\sigma_s, r' \uplus \sigma_r)\Big)
                \\
                & {} = \int \rho (\dd r')\, 
                \Big(\density(\cscore_\cnorm(E_0,E_1,E_2),\sigma_s)(r') 
                \\
                &
                \qquad\qquad\qquad\qquad\qquad\qquad\qquad
                {}
                \cdot \ind{r' \# \sigma_r} 
                \cdot h(\get(\cscore_\cnorm(E_0,E_1,E_2), \sigma_s)(r'), r' \uplus \sigma_r)\Big).
        \end{align*}
\end{proof}


\section{Proof of Theorem in \S\ref{sec:inference}}
$\,$

{\sc Theorem~\ref{thm:score-unbiasedness}.\ }{\it 
Let $C$ be a model, $D_\theta$ be a guide, and $N \neq 0 \in \bN$. 
Define $\KL_{(-)} : \bR^p \to \bR_{\geq 0}$ as
$\KL_\theta$ $\defeq$ $\KL(\density(D_\theta,s_I) {||} \density(C,s_I)/Z_C)$.
Then, $\KL_{(-)}$ is well-defined and continuously differentiable with
\begin{align*}
        &
        \nabla_\theta \KL_\theta
        =
        \mathbb{E}_{\prod_i\density(D_\theta,s_I)(r_i)}
        \left[ 
        \frac{1}{N}\sum_{i=1}^N
                \Big(\nabla_\theta \log \density(D_\theta,s_I)(r_i)\Big)
                \log \frac{\density(D_\theta,s_I)(r_i)}{\density(C,s_I)(r_i)}
        \right]
        \tag{\ref{eqn:score-estimator}}
\end{align*}
if
\begin{enumerate}
\item[(\ref{req:1})]
  $\density(C,s_I)(r) = 0 \implies \density(D_\theta,s_I)(r) = 0$,
        for all $r \in \Rdb$ and $\theta \in \mathbb{R}^p$;
\item[(\ref{req:2})]
  for all $(r,\theta,j) \in \Rdb \times \mathbb{R}^p \times [p]$, the function $v \longmapsto \density(D_{\theta[j:v]},s_I)(r)$ on $\bR$ is differentiable;
\item[(\ref{req:3})]
  for all $\theta \in \mathbb{R}^p$,
\[
        \int \rho(\dd r)\,
                \left(\density(D_\theta,s_I)(r) \cdot
                \log\frac{\density(D_\theta,s_I)(r)}{\density(C,s_I)(r)}\right)
                < \infty;
\]
\item[(\ref{req:4})]
  for all $(\theta,j) \in \mathbb{R}^p \times [p]$, the function
\[
        v \longmapsto 
        \int \rho(\dd r)\,
                \left(\density(D_{\theta[j:v]},s_I)(r) \cdot
                \log\frac{\density(D_{\theta[j:v]},s_I)(r)}{\density(C,s_I)(r)}\right)
\]
on $\mathbb{R}$ is continuously differentiable;
\item[(\ref{req:5})]
  for all $\theta \in \mathbb{R}^p$,
\begin{align*}
        & \nabla_\theta
        \int \rho(\dd r)
                \left(\density(D_\theta,s_I)(r) \cdot
                \log \frac{\density(D_\theta,s_I)(r)}{\density(C,s_I)(r)}\right)
        \\
        & \qquad\qquad {} =
        \int \rho(\dd r)\,
        \nabla_\theta
        \left(\density(D_\theta,s_I)(r) \cdot
                \log \frac{\density(D_\theta,s_I)(r)}{\density(C,s_I)(r)}\right);
\end{align*}
\item[(\ref{req:6})]
  for all $\theta \in \mathbb{R}^p$,
\[
        \int \rho(\dd r)\,
        \nabla_\theta
        \density(D_\theta,s_I)(r)
        =
        \nabla_\theta
        \int \rho(\dd r)\,
        \density(D_\theta,s_I)(r).
\]
\end{enumerate}
}
As we explained in the main text already,
the theorem (except the continuous differentiability of $\KL_\theta$)
is well-known with a well-known proof.
We include its proof for completeness and also to help the reader to see why each of the requirements in the theorem is needed.
\begin{proof}
  $\KL_\theta$ is well-defined because of the following derivation
  with \ref{req:1} and~\ref{req:3}:
  \begin{align*}
    \KL_\theta
    & = \KL\left(\density(D_\theta,s_I) {\Big|\Big|} \frac{\density(C,s_I)}{Z_C}\right)
    \\
    & =
    \int \rho(\dd r)\,
    \left(\density(D_\theta,s_I)(r) \cdot
    \log\left(\frac{\density(D_\theta,s_I)(r)}{\density(C,s_I)(r)/Z_C}\right)\right)
    \\
    & =
    \int \rho(\dd r)\,
    \left(\density(D_\theta,s_I)(r) \cdot
    \left(\log Z_C + \log\frac{\density(D_\theta,s_I)(r)}{\density(C,s_I)(r)}\right)
    \right)
    \\
    & =
    \log Z_C +
    \int \rho(\dd r)\,
    \left(\density(D_\theta,s_I)(r) \cdot
    \log\frac{\density(D_\theta,s_I)(r)}{\density(C,s_I)(r)}\right).
  \end{align*}

  It remains to show the continuous differentiability of $\KL_\theta$
  and Equation~\eqref{eqn:score-estimator}.
  Recall the fact that for any function $f : \bR^N \to \bR^M$,
  if all partial derivatives of $f$ exist and are continuous,
  $f$ is continuously differentiable.
  Because of the fact and the linearity of expectation,
  it suffices to show that
  all partial derivatives of $\KL_\theta$ are well-defined and continuous, and
  \begin{equation}
        \nabla_\theta \KL_\theta
        =
        \mathbb{E}_{\density(D_\theta,s_I)(r)}
        \left[\Big(\nabla_\theta \log \density(D_\theta,s_I)(r)\Big)
             \log \frac{\density(D_\theta,s_I)(r)}{\density(C,s_I)(r)}
        \right].
        \label{eqn:(2):used}
  \end{equation}
  Here is a well-known derivation of~\eqref{eqn:(2):used}:
\begin{align}
  & \nabla_\theta \KL_\theta
  \nonumber
  \\ 
        & \quad
        {} = \nabla_\theta
        \left(\log Z_C +
        \int \rho(\dd r)\,
                \left(\density(D_\theta,s_I)(r) \cdot
                \log\frac{\density(D_\theta,s_I)(r)}{\density(C,s_I)(r)}\right)\right)
        \nonumber
        \\
        & \quad
        {} = \nabla_\theta
        \int \rho(\dd r)
                \left(\density(D_\theta,s_I)(r) \cdot
                \log \frac{\density(D_\theta,s_I)(r)}{\density(C,s_I)(r)}\right)
        \label{eqn:(4):used}
        \\
        & \quad
        {} =
        \int \rho(\dd r)\,
        \nabla_\theta
        \left(\density(D_\theta,s_I)(r) \cdot
                \log \frac{\density(D_\theta,s_I)(r)}{\density(C,s_I)(r)}\right)
        \label{eqn:(5):used}
        \\
        & \quad
        {} =
        \int \rho(\dd r)
        \left(
        \Big(\nabla_\theta \density(D_\theta,s_I)(r)\Big) \cdot
                \log\frac{\density(D_\theta,s_I)(r)}{\density(C,s_I)(r)}\right)
                +
        \int \rho(\dd r)\,
        \nabla_\theta \density(D_\theta,s_I)(r)
        \nonumber
        \\
        & \quad
        {} =
        \int \rho(\dd r)
        \left(
        \Big(\nabla_\theta \density(D_\theta,s_I)(r)\Big)
        \cdot 
        \log \frac{\density(D_\theta,s_I)(r)}{\density(C,s_I)(r)} 
        \right)
        +
        \nabla_\theta
        \int \rho(\dd r)\,
        \density(D_\theta,s_I)(r)
        \label{eqn:(6):used}
        \\
        & \quad
        {} =
        \int \rho(\dd r) 
                \left(\density(D_\theta,s_I)(r) 
                \cdot \Big(\nabla_\theta \log \density(D_\theta,s_I)(r)\Big) 
                \cdot \log\frac{\density(D_\theta,s_I)(r)}{\density(C,s_I)(r)}\right)
        +
        \nabla_\theta 1
        \nonumber
        \\
        & \quad
        {} = \mathbb{E}_{\density(D_\theta,s_I)(r)}
                \left[
                        \Big(\nabla_\theta \log \density(D_\theta,s_I)(r)\Big)
                        \cdot
                        \log \frac{\density(D_\theta,s_I)(r)}{\density(C,s_I)(r)}
                \right].
        \nonumber
\end{align}
The well-definedness and the continuity of $(\nabla_\theta \KL_\theta)_j$ for all $j \in [p]$
follow from \eqref{eqn:(4):used} and~\ref{req:4}, which concludes the proof.

Note that~\ref{req:2} is used to guarantee the well-definedness of
$\nabla_\theta \log \density(D_\theta,s_I)(r)$ in~\eqref{eqn:(2):used};
\ref{req:5} and \ref{req:6} are used in~\eqref{eqn:(5):used} and \eqref{eqn:(6):used},
respectively.
\end{proof}


\section{Proofs of Examples, Lemma, and Theorems in \S\ref{sec:conditions}}\label{appendix:proof-condition}
$\,$

{\sc Theorem~\ref{thm:condition-2a}.\ }{\it
	Under our assumption in \S\ref{subsec:requirement-assumption},}
        \[
                \EE{\density(D_\theta,s_I)(r)}{|\log \density(D_\theta,s_I)(r)|} < \infty.
        \]
\begin{proof}
For $i \in [M]$ and $r \in A_i$, define
\[
        g_{(i,\theta)}(r) \defeq \prod_{\alpha \in K_i}
        \cN(r(\alpha);\mu_{(i,\alpha)}(\theta),\sigma_{(i,\alpha)}(\theta)).
\]
Then,
\begin{align*} 
        & \EE{\density(D_\theta,s_I)(r)}{|\log \density(D_\theta,s_I)(r)|} 
        \\
        & \qquad {} = \int \rho(\dd r)
        \left(\Big(\sum_{i=1}^M \ind{r \in A_i} \cdot g_{(i,\theta)}(r)\Big)
        \cdot
        \Big|\log \sum_{i=1}^M \big(\ind{r \in A_i}  \cdot g_{(i,\theta)}(r)\big)\Big|\right)
        \\
        & \qquad {} = \int \rho(\dd r)
        \left(\sum_{i=1}^M \ind{r \in A_i} 
        \cdot
        g_{(i,\theta)}(r)
        \cdot
        \Big|\log g_{(i,\theta)}(r)\Big|\right)
        \\
        & \qquad {} = \sum_{i=1}^M \int \rho(\dd r)
        \left( \ind{r \in A_i} 
        \cdot
        g_{(i,\theta)}(r)
        \cdot
        \Big|\log g_{(i,\theta)}(r)\Big|\right)
        \\
        & \qquad {} \leq \sum_{i=1}^M \int \rho(\dd r)
        \left( \ind{r \in [K_i \to \bR]} \cdot
        g_{(i,\theta)}(r)
        \cdot
        \Big|\log g_{(i,\theta)}(r)\Big|\right)
        \\
        & \qquad {} = \sum_{i=1}^M
        \EE{g_{(i,\theta)}(r)}{|\log g_{(i,\theta)}(r)|}
        \\
        & \qquad {} < \infty.
\end{align*}
The last inequality uses a well-known result that the differential entropy
of any multivariate normal distribution is finite.
\end{proof}

{\sc Example~\ref{eg:condition-2a}.\ }{\it
  Consider guides $D_{(i, \theta)}$ defined as follows ($i =1,2$):
  \[
  D_{(i,\theta)} \equiv
  (x_1 := \csample_\cnorm(``a_1", \theta_1,1);\; x_2 := \csample_\cnorm(``a_2", \theta_2,E_i[x_1]))
  \]
  where for some $n \geq 1$ and $c \neq 0 \in \mathbb{R}$,
  \begin{align*}
    E_1[x_1] & \equiv \code{if}\,(x_1{=}0)\,\code{then}\, 1\,\code{else}\,{\exp(-1/|x_1|^n)},
    &
    E_2[x_1] & \equiv \exp(\exp(c \cdot x_1^3)).
  \end{align*}
  Then, the entropies of $\density(D_{(i,\theta)},s_I)$'s are all undefined.}
\begin{proof}
  We prove that 
  \[\bE_{\density(D_{(i,\theta)},s_I)(r)}{[|\log \density(D_{(i,\theta)},s_I)(r)|]} = \infty
  \quad\text{for all $i=1,2$}\]
  by bounding the the expectation as follows.
  \begin{align*}
    & \bE_{\density(D_{(i,\theta)},s_I)(r)}{[|\log \density(D_{(i,\theta)},s_I)(r)|]}
    \\
    & \quad =
    \EE{\cN(x_1; \theta_1,1) \cdot \cN(x_2; \theta_2,E_i[x_1])}
       {| \log \cN(x_1; \theta_1,1) + \log \cN(x_2; \theta_2,E_i[x_1] )|}
    \\
    & \quad \geq
    -\EE{\cN(x_1; \theta_1,1) \cdot \cN(x_2; \theta_2,E_i[x_1])}{|\log \cN(x_1; \theta_1,1)|}
    \\
    & \quad\qquad
    + \EE{\cN(x_1; \theta_1,1) \cdot \cN(x_2; \theta_2,E_i[x_1])}{|\log \cN(x_2; \theta_2,E_i[x_1])|}
  \end{align*}
  Here the two terms in the last equation are bounded as follows.
  \begin{align*}
    & {-}\EE{\cN(x_1; \theta_1,1) \cdot \cN(x_2; \theta_2,E_i[x_1])}{|\log \cN(x_1; \theta_1,1)|}
    \\
    & \quad =
    -\EE{\cN(x_1; \theta_1,1) }
    {\left|\log \sqrt{2\pi} -\frac{1}{2}(x_1-\theta_1)^2\right|}
    \\
    & \quad \geq
    -\EE{\cN(x_1; \theta_1,1)}{\log \sqrt{2\pi}}
    -\frac{1}{2}\,\EE{\cN(x_1; \theta_1,1)}{(x_1-\theta_1)^2}
    \\
    & \quad =
    -\log \sqrt{2\pi} -\frac{1}{2}\cdot 1^2
    \\
    & \EE{\cN(x_1; \theta_1,1) \cdot \cN(x_2; \theta_2,E_i[x_1])}{|\log \cN(x_2; \theta_2,E_i[x_1])|}
    \\
    & \quad =
    \EE{\cN(x_1; \theta_1,1) \cdot \cN(x_2; \theta_2,E_i[x_1] )}{\left|-\log\sqrt{2\pi} 
      - \frac{(x_2-\theta_2)^2}{2 E_i[x_1]^2}-\log E_i[x_1]\right|}
    \\
    & \quad \geq
    -\EE{\cN(x_1; \theta_1,1)}{\log\sqrt{2\pi}}
    -\frac{1}{2} \, \bE_{\cN(x_1; \theta_1,1)}
    {\bigg[\underbrace{\EE{\cN(x_2; \theta_2,E_i[x_1])}{(x_2-\theta_2)^2}}_{E_i[x_1]^2} \cdot \frac{1}{E_i[x_1]^2}\bigg]}
    \\
    & \quad\qquad
    +\EE{\cN(x_1; \theta_1,1)}{|\log E_i[x_1]|}
    \\
    & \quad =
    -\log \sqrt{2\pi} -\frac{1}{2}
    +\EE{\cN(x_1; \theta_1,1)}{|\log E_i[x_1]|}
  \end{align*}
  Note that
  \[
    \EE{\cN(x_1; \theta_1,1)}{|\log E_i[x_1]|}
    =
    \left\{
      \begin{array}{@{\,\,}ll@{\,\,}}
        \bE_{\cN(x_1; \theta_1,1)}{[\ind{x_1 \neq 0} \cdot 1/|x_1|^n]}  & \text{for $i=1$}
        \\
        \bE_{\cN(x_1; \theta_1,1)}{[\exp(c \cdot x_1^3)]}  & \text{for $i=2$}
      \end{array}
    \right\}
    = \infty
  \]
  because $\int_{[0,\epsilon]} 1/v^n \, \dd v= \infty$ for any $\epsilon>0$ and $n \geq 1$,
  and $\bE_{\cN(v;\mu,\sigma)}[\exp(c\cdot v^3)] = \infty$ for any $\mu \in \bR$ and $\sigma \in \bR_{>0}$.
  Hence, we obtain the desired result.
\end{proof}

{\sc Theorem~\ref{thm:condition-2}.\ }{\it
        Under our assumption in \S\ref{subsec:requirement-assumption}, 
	the condition \eqref{eqn:condition-2b} implies
        \[ \EE{\density(D_\theta,s_I)(r)}{|\log \density(C,s_I)(r)|} < \infty, \]
        and thus the requirement~\ref{req:3} (i.e., the objective in \eqref{eqn:SVI:objective} is well-defined).}
\begin{proof}
For $i \in [M]$, $j \in [N_i]$, $\alpha \in K_i$, 
and $r \in A_i$, let
\begin{align*}
  g_{(i,\theta)}(r) & \defeq \prod_{\beta \in K_i}
  \cN(r(\beta);\mu_{(i,\beta)}(\theta),\sigma_{(i,\beta)}(\theta)),
        \\
        h_{(i,\alpha)}(r) & \defeq \cN(r(\alpha);\mu'_{(i,\alpha)}(r),\sigma'_{(i,\alpha)}(r)),
        \quad
        k_{(i,j)}(r) \defeq \cN(c_{(i,j)};\mu''_{(i,j)}(r), \sigma''_{(i,j)}(r)).
\end{align*}
We note that
\begin{align*}
  \EE{g_{(i,\theta)}(r)}{\left|\log h_{(i,\alpha)}(r)\right|}
  &
  = \EE{g_{(i,\theta)}(r)}{\left| 
        - \log{\sqrt{2\pi}} 
        - \log \sigma'_{(i,\alpha)}(r) 
        - \frac{\left(r(\alpha)-\mu'_{(i,\alpha)}(r)\right)^2}{2\left(\sigma'_{(i,\alpha)}(r)\right)^2}
        \right|}
  \\
  &
  \leq 
  \log{\sqrt{2\pi}}
  +\EE{g_{(i,\theta)}(r)}{\left|
        \log \sigma'_{(i,\alpha)}(r)
        \right|}
  +\EE{g_{(i,\theta)}(r)}{
        \frac{\left(r(\alpha)-\mu'_{(i,\alpha)}(r)\right)^2}{2\left(\sigma'_{(i,\alpha)}(r)\right)^2}
	}.
\end{align*}
Thus, we can have 
\begin{equation}
\label{eqn:req2-proof0}
\EE{g_{(i,\theta)}(r)}{\left|\log h_{(i,\alpha)}(r)\right|} < \infty 
\end{equation}
if the following two inequalities hold:
\begin{equation}
\label{eqn:req2-proof1}
\EE{g_{(i,\theta)}(r)}{\left| \log \sigma'_{(i,\alpha)}(r)\right|} < \infty
\text{\quad and \quad}
\EE{g_{(i,\theta)}(r)}{\frac{\left(r(\alpha) - \mu'_{(i,\alpha)}(r)\right)^2}{\left(\sigma'_{(i,\alpha)}(r)\right)^2}}
< \infty.
\end{equation}
We show that the inequalities are indeed true. By the condition \eqref{eqn:condition-2b} and the parts (a) and (b) of Lemma~\ref{lem:lin-exp} (full version), there exist functions $f_i,u_i \in \cA_i$ such that
for every $r \in A_i$,
\begin{equation}
        \label{eqn:req2-proof2}
        \left| \log \sigma'_{(i,\alpha)}(r)\right|  \leq \exp(f_i(r))
        \text{\quad and \quad}
        \frac{\left(r(\alpha) - \mu'_{(i,\alpha)}(r)\right)^2}{\left(\sigma'_{(i,\alpha)}(r)\right)^2}  
        \leq \exp(u_i(r)).
\end{equation}
For every $f \in \cA_i$,
by the part (e) of Lemma~\ref{lem:lin-exp} (full version), we have that
\begin{equation}
        \label{eqn:req2-proof3}
        \EE{g_{(i,\theta)}(r)}{\exp(f(r))} < \infty.
\end{equation}
The inequalities in \eqref{eqn:req2-proof1} follow from
the properties in \eqref{eqn:req2-proof2} and \eqref{eqn:req2-proof3}.

By an argument similar to what we have just given, we also have that
\begin{equation}
\label{eqn:req2-proof4}
\EE{g_{(i,\theta)}(r)}{\left|\log k_{(i,j)}(r)\right|} < \infty.
\end{equation}

Using \eqref{eqn:req2-proof0} and \eqref{eqn:req2-proof4}, we calculate the claim of the theorem:
\begin{align*}
        & \EE{\density(D_\theta,s_I)(r)}{\left|\log \density(C,s_I)(r)\right|} 
        \\
        & {} = \int \rho(\dd r)
        \left(\left(\sum_{i=1}^M \ind{r \in A_i} \cdot g_{(i,\theta)}(r)\right)
        \cdot
        \left|\log 
        \sum_{i=1}^M  \Big(
                \ind{r \in A_i}  
                \cdot \prod_{\beta \in K_i} h_{(i,\beta)}(r) 
                \cdot \prod_{j \in [N_i]} k_{(i, j)}(r) \Big)
        \right|\right)
        \\
        & {} = \int \rho(\dd r)
        \left(\sum_{i=1}^M 
        \ind{r \in A_i} 
        \cdot g_{(i,\theta)}(r)
        \cdot \left|\log \Big(
                \prod_{\beta \in K_i} h_{(i,\beta)}(r) 
                \cdot \prod_{j \in [N_i]} k_{(i,j)}(r) \Big) \right|\right)
        \\
        & {} = 
        \sum_{i=1}^M 
        \int \rho(\dd r)
        \left(
        \ind{r \in A_i} 
        \cdot g_{(i,\theta)}(r)
        \cdot \left|\sum_{\beta \in K_i} \log h_{(i,\beta)}(r) 
                   + \sum_{j \in [N_i]} \log k_{(i,j)}(r)\right|\right)
        \\
        & {} \leq 
        \sum_{i=1}^M \sum_{\beta \in K_i} 
        \int \rho(\dd r)
        \left(
        \ind{r \in A_i} 
        \cdot g_{(i,\theta)}(r)
        \cdot \left|\log h_{(i,\beta)}(r)\right|\right)
        \\
        & 
        \qquad
        {} +
        \sum_{i=1}^M \sum_{j \in [N_i]}
        \int \rho(\dd r)
        \left(
        \ind{r \in A_i} 
        \cdot g_{(i,\theta)}(r)
        \cdot \left|\log k_{(i,j)}(r)\right|\right)
        \\
        & {} \leq 
        \sum_{i=1}^M \sum_{\beta \in K_i} 
        \int \rho(\dd r)
        \left( \ind{r \in [K_i \to \bR]} \cdot
        g_{(i,\theta)}(r)
        \cdot \left|\log h_{(i,\beta)}(r)\right|\right)
        \\
        & \qquad {} +
        \sum_{i=1}^M \sum_{j \in [N_i]}
        \int \rho(\dd r)
        \left( \ind{r \in [K_i \to \bR]} \cdot
        g_{(i,\theta)}(r)
        \cdot \left|\log k_{(i,j)}(r)\right|\right)
        \\
        & {} =
        \sum_{i=1}^M \sum_{\beta \in K_i} 
        \EE{g_{(i,\theta)}(r)}{\left|\log h_{(i,\beta)}(r)\right|}
        {} +
        \sum_{i=1}^M \sum_{j \in [N_i]}
        \EE{g_{(i,\theta)}(r)}{\left|\log k_{(i,j)}(r)\right|}
        \\
        & {} < \infty.
\end{align*}
The last inequality follows from \eqref{eqn:req2-proof0} and \eqref{eqn:req2-proof4}.
Finally, the requirement~\ref{req:3} follows from the above and Theorem~\ref{thm:condition-2a}:
\begin{align*}
  & \int \rho(\dd r)\,
  \left|\density(D_\theta,s_I)(r) \cdot
  \log\frac{\density(D_\theta,s_I)(r)}{\density(C,s_I)(r)}\right|
  \\
  & \qquad{} \leq
  \EE{\density(D_\theta,s_I)(r)}{\big|\log \density(D_\theta,s_I)(r)\big|}
  + \EE{\density(D_\theta,s_I)(r)}{\big|\log \density(C,s_I)(r)\big|}
  \\
  & \qquad{} < \infty.
\end{align*}
\end{proof}

{\sc Example~\ref{eg:condition-2b}.\ }{\it
  Consider models $C_1, \ldots, C_4$ and a guide $D_\theta$ defined
  as follows:
  \[\begin{array}{rll}
          C_i \equiv {} \!\!\!\!\!
          & (x_1 := \csample_\cnorm(``a_1", 0,1);\; x_2 := \csample_\cnorm(``a_2", E_i[x_1],1)) 
          & \text{for $i=1,2$}
          \\[.5ex]
          C_i \equiv {} \!\!\!\!\!
          & (x_1 := \csample_\cnorm(``a_1", 0,1);\; x_2 := \csample_\cnorm(``a_2",  0, E_i[x_1]))
          &\text{for $i=3,4$}
          \\[.5ex]
          D_\theta \equiv {} \!\!\!\!\!
          & (x_1 := \csample_\cnorm(``a_1", \theta_1, 1);\; x_2 := \csample_\cnorm(``a_2", \theta_2, 1))
  \end{array}\]
  where for some $n \geq 1$ and $c \neq 0 \in \mathbb{R}$,
  \begin{align*} 
          E_1[x_1] & \equiv \code{if}\,(x_1{=}0)\,\code{then}\,0\,\code{else}\,1/x_1^n,
          &
          E_2[x_1] & \equiv E_4[x_1] \equiv \exp(c \cdot x_1^3),
          \\
          E_3[x_1] & \equiv \code{if}\,(x_1{=}0)\,\code{then}\, 1\,\code{else}\,{|x_1|^n}.
  \end{align*}
  Then, the requirement~\ref{req:3} does not hold
  (i.e., the objective in~\eqref{eqn:SVI:objective} is undefined)
  for all $i =1,\ldots,4$.}
\begin{proof}
  Since we have
  \begin{align*}
    \int \rho(\dd r)\,
    \left|\density(D_\theta,s_I)(r) \cdot
    \log\frac{\density(D_\theta,s_I)(r)}{\density(C_i,s_I)(r)}\right|
    & \geq
    - \bE_{\density(D_{\theta},s_I)(r)}{[|\log \density(D_{\theta},s_I)(r)|]}
    \\ & \qquad
    + \bE_{\density(D_{\theta},s_I)(r)}{[|\log \density(C_{i},s_I)(r)|]}
    \\
    \bE_{\density(D_{\theta},s_I)(r)}{[|\log \density(D_{\theta},s_I)(r)|]}
    & < \infty,
  \end{align*}
  it suffices to show that
  \begin{equation}
    \label{eqn:eg:condition-2b}
    \bE_{\density(D_{\theta},s_I)(r)}{[|\log \density(C_{i},s_I)(r)|]} = \infty
    \quad\text{for all $i= 1, \ldots, 4$.}
  \end{equation}

  {\bf Case of $i=1,2$:}
  The quantity in~\eqref{eqn:eg:condition-2b} is bounded as follows.
  \begin{align*}
    & \bE_{\density(D_\theta,s_I)(r)}{[|\log \density(C_i,s_I)(r)|]}
    \\
    & \quad =
    \EE{\cN(x_1; \theta_1,1) \cdot \cN(x_2; \theta_2,1)}
       {| \log \cN(x_1; 0,1) + \log \cN(x_2; E_i[x_1],1 )|}
    \\
    & \quad \geq
    -\EE{\cN(x_1; \theta_1,1) \cdot \cN(x_2; \theta_2,1)}{|\log \cN(x_1; 0,1)|}
    \\
    & \quad \qquad
    + \EE{\cN(x_1; \theta_1,1) \cdot \cN(x_2; \theta_2,1)}{|\log \cN(x_2; E_i[x_1],1)|}
  \end{align*}
  Here the two terms in the last equation are bounded as follows.
  \begin{align*}
    & -\EE{\cN(x_1; \theta_1,1) \cdot \cN(x_2; \theta_2,1)}{|\log \cN(x_1; 0,1)|}
    \\
    & \quad =
    -\EE{\cN(x_1; \theta,1) }
    {\left|\log \sqrt{2\pi} -\frac{1}{2}x_1^2\right|}
    \\
    & \quad \geq
    -\EE{\cN(x_1; \theta,1)}{\log \sqrt{2\pi}}
    -\frac{1}{2}\EE{\cN(x_1; \theta,1)}{x_1^2}
    \\
    & \quad =
    -\log \sqrt{2\pi} -\frac{1}{2}(1+\theta^2)
    \\
    & \EE{\cN(x_1; \theta_1,1) \cdot \cN(x_2; \theta_2,1)}{|\log \cN(x_2; E_i[x_1],1)|}
    \\
    & \quad =
    \EE{\cN(x_1; \theta_1,1) \cdot \cN(x_2; \theta_2,1)}{\left|-\log\sqrt{2\pi}
      - \frac{1}{2}(x_2-E_i[x_1])^2\right|}
    \\
    & \quad \geq
    -\EE{\cN(x_1; \theta_1,1)}{\log\sqrt{2\pi}}
    + \frac{1}{2} \, \EE{\cN(x_1; \theta_1,1) \cdot \cN(x_2; \theta_2,1)}{\left|
      x_2^2 - 2x_2 \cdot E_i[x_1] + E_i[x_1]^2\right|}
    \\
    & \quad \geq
    -\log \sqrt{2\pi}
    -\frac{1}{2} \, \EE{\cN(x_2; \theta_2,1)}{x_2^2}
    + \frac{1}{2} \, \EE{\cN(x_1; \theta_1,1) \cdot \cN(x_2; \theta_2,1)}{\left|
      E_i[x_1]^2 -2x_2 \cdot E_i[x_1]\right|}
    \\
    & \quad \geq
    -\log \sqrt{2\pi}
    -\frac{1}{2}(1+\theta_2^2)
    + \frac{1}{2} \, \EE{\cN(x_1; \theta_1,1) \cdot \cN(x_2; \theta_2,1)}{\left|
      E_i[x_1]^2 -2x_2 \cdot E_i[x_1]\right|}
  \end{align*}
  We bound the last term in the last equation as follows, assuming $c>0$.
  \begin{align*}
    & \EE{\cN(x_1; \theta_1,1) \cdot \cN(x_2; \theta_2,1)}{\left|
      E_i[x_1]^2 -2x_2 \cdot E_i[x_1]\right|}
    \\
    & \quad \geq
    \bE_{\cN(x_1; \theta_1,1) \cdot \cN(x_2; \theta_2,1)}{\big[
      \ind{x_1 \geq 0 \land x_2 \leq 0} \cdot 
      \big|
      E_i[x_1]^2 + \underbrace{(-2x_2)}_{\geq 0} \cdot \underbrace{E_i[x_1]}_{\geq 0}\big|\big]}
    \\
    & \quad \geq
    \EE{\cN(x_1; \theta_1,1) \cdot \cN(x_2; \theta_2,1)}{
      \ind{x_1 \geq 0 \land x_2 \leq 0} \cdot 
      \left(
      E_i[x_1]^2 + (-2x_2) \cdot E_i[x_1]\right)}
    \\
    & \quad \geq
    \EE{\cN(x_1; \theta_1,1)}{\ind{x_1 \geq 0} \cdot E_i[x_1]^2}
    \cdot
    \EE{\cN(x_2; \theta_2,1)}{\ind{x_2 \leq 0}}
    \\
    & \quad\qquad
    +
    \EE{\cN(x_1; \theta_1,1)}{\ind{x_1 \geq 0} \cdot E_i[x_1]}
    \cdot
    \EE{\cN(x_2; \theta_2,1)}{\ind{x_2 \leq 0} \cdot (-2x_2)}
    \\
    & \quad \geq \infty \qquad\text{for all $i=1,2$}
  \end{align*}
  Here the last inequality comes from that
  the two expectations over $x_2$ are finite
  while the other two over $x_1$ are both $\infty$,
  because $\int_{[0,\epsilon]} 1/v^n \, \dd v= \infty$ for any $\epsilon>0$ and $n \geq 1$,
  and $\bE_{\cN(v;\mu,\sigma)}[\exp(c\cdot v^3)] = \infty$ for any $\mu \in \bR$ and $\sigma \in \bR_{>0}$.
  Note that we can use a similar calculation for the case of $c < 0$.
  This proves the equality in~\eqref{eqn:eg:condition-2b}.

  {\bf Case of $i=3,4$:}
  The proof is similar to the above case, so we omit it.
  Remark that in the proof, we additionally use the following fact:
  $\int_{[0,1]} |\log v| \,\dd v < \infty$.
\end{proof}

{\sc Lemma~\ref{lem:lin-exp} (Full Version).\ }{\it
  Pick $i \in [M]$.
  Let $(\alpha_1,\ldots,\alpha_J)$ be an enumeration of the elements in $K_i$,
  and $\overline{r} \defeq (r(\alpha_1),\ldots,r(\alpha_J)) \in \bR^J$
  for $r \in A_i$.
  \begin{itemize}
  \item[(a)]
    For every $l_1, l_2 \in \cA_i$ and every $\odot \in \{ +, -, \times, /, \max\}$,
    there exists $l \in \cA_i$ such that
    \[
            \exp(l_1(r)) \odot \exp(l_2(r)) \leq \exp(l(r))
            \text{ for all $r \in A_i$}.
    \]
  \item[(b)]
    For every polynomial $\mathit{poly} : \bR^J \to \bR$,
    there exists $l \in \cA_i$ such that
    \[
            |\mathit{poly}(\overline{r})| \leq \exp(l(r))
            \text{ for all $r \in A_i$}.
    \]
    In particular, for every $l \in \cA_i$, there exists $l' \in \cA_i$ such that
    \[
            |l(r)| \leq \exp(l'(r))
            \text{ for all $r \in A_i$}.
    \]
  \item[(c)]
    For every affine-bounded neural network $\nn: \bR^J \to \bR$, there exists $l \in \cA_i$ such that
    \[ 
           |\nn(\overline{r})| \leq l(r) 
           \text{ for all $r \in A_i$.}
    \]
  \item[(d)]
    There exist affine functions $l_1,l_2 : \bR \to \bR$ such that
    \[ \exp(l_1(|v|)) \leq \mathrm{softplus}(v) \leq \exp(l_2(|v|))
    \text{ for all $v \in \bR$.} \]
  \item[(e)]
    For every $l \in \cA_i$ and every $J$-dimensional normal distribution $q$,
    \[
            \EE{q(r(\alpha_1),\ldots,r(\alpha_J))}{\exp(l(r))} < \infty.
    \]
  \end{itemize}
}
\begin{proof}[\bf Proof of (a)]
  We split the cases.
  \begin{compactitem}
  \item Case $\odot \in \{\times, /\}$: Take $l = l_1 \pm l_2$.
  \item Case $\odot =\max$:
    For $m=1,2$, let 
    \[ 
          l_m(r) = c_{m,0} + \sum_{j=1}^J c_{m,j} \cdot |r(\alpha_j)|
     \]
    where $c_{m,j} \in \bR$. For all $r \in A_i$, we have
    \begin{align*}
      \max(l_1(r),l_2(r))
      & \leq \max\Big(c_{1,0} + \sum_{j=1}^J |c_{1,j}| \cdot |r(\alpha_j)|,\,
                      c_{2,0} + \sum_{j=1}^J |c_{2,j}| \cdot |r(\alpha_j)|\Big)
        \leq l(r),
    \end{align*}
    where 
    \[
       l(r) \defeq \max(c_{1,0},c_{2,0}) + \sum_{j=1}^J \max(|c_{1,j}|,|c_{2,j}|) \cdot |r(\alpha_j)| \in \cA_i.
    \]
    Hence, for every $r \in A_i$,
    \[
        \max(\exp(l_1(r)),\exp(l_2(r))) = \exp(\max(l_1(r),l_2(r))) \leq \exp(l(r)).
    \]

  \item Case $\odot \in \{+, -\}$: Using the case $\odot=\max$, we have the following
    for some $l \in \cA_i$:
    \begin{align*}
       \exp(l_1(r)) + \exp(l_2(r))
       & {} \leq 2 \cdot \max(\exp(l_1(r)), \exp(l_2(r)))
       \\
       & {} \leq \exp(\log 2) \cdot \exp(l(r)) 
       \\
       & {} = \exp(l(r) + \log 2).
    \end{align*}
  \end{compactitem}

\vspace{1mm}
\noindent{\sc\bf Proof of (b).} Let $n$ be an index in $\{1,\ldots,J\}$ such that
  \[
          \mathit{poly}(r(\alpha_1),\ldots,r(\alpha_J)) = \mathit{poly}'(r(\alpha_1),\ldots,r(\alpha_n))
  \]
  for some polynomial $\mathit{poly}' : \mathbb{R}^n \to \mathbb{R}$. We show that
  there exists $l \in \cA_i$ satisfying
  \[
          \mathit{poly}'(r(\alpha_1),\ldots,r(\alpha_n))
          \leq \exp(l(r)).
  \]
  Note that the claim of the lemma follows from this. Our proof is by induction on $n$.
  \begin{compactitem}
  \item Case $n=1$:
          Let $v \in \bR$ and $\mathit{poly}'(v) = \sum_{d=0}^D c_d \cdot v^d$
          where $c_d \in \bR$. We use the following fact.
    \begin{quote}
       {\sc Fact:} For every $d \in \bN$, there exists an affine function $l_d : \bR \to \bR$ such that 
        \[
                |v|^d \leq \exp(l_d(|v|))\ \text{ for all $v \in \mathbb{R}$.}
        \]
      \begin{proof}
        Since $v^d = \cO(\exp(v))$ for positive $v \in \mathbb{R}$,
        there exist $C_1, C_2 >0$ such that
        $|v| > C_1$ implies $|v|^d < C_2\exp(|v|)$.
        Since $|v|^d$ is continuous,
        $C_3 \defeq \sup\{ |v|^d \,\mid\, |v| \leq C_1\}$
        satisfies $0 \leq C_3 < \infty$.
        Putting these together, for every $v \in \bR$,
        \begin{align*}
                |v|^d 
                & {} \leq \max(C_2\exp(|v|), C_3) 
                \\
                & {} \leq \max(\max(C_2, 1) \cdot \exp(|v|), C_3+1) 
                \\
                & {} \leq C_4  \exp(|v|),
        \end{align*}
        where $C_4 \defeq \max(C_2, 1) \cdot (C_3+1) \geq 1$. Thus, $l_d(v) \defeq (v + \log C_4)$ is a desired affine function.
      \end{proof}
    \end{quote}
    Using the fact, we have
    \begin{align*}
      |\mathit{poly}'(r(\alpha_1))|
      & \leq \sum_{d=0}^D |c_d| \cdot |r(\alpha_1)|^d
      \\
      & \leq \sum_{d=0}^D |c_d| \cdot {\exp}(l_d(|r(\alpha_1)|))
      = \sum_{d=0, c_d \neq 0}^D {\exp}(l'_d(r))
      \\
      & \leq \exp(l(r))
    \end{align*}
    for some $l'_d,l \in \cA_i$. The existence of $l$ satisfying the last inequality follows from~(a).
  \item Case $n>1$:
    Let $v \in \bR^n$, and define $v_{1:n-1}$ to be the projection of $v$ to its first $n-1$ components.
    Also, let
    \[
            \mathit{poly}'(v) = \sum_{d=0}^D \mathit{poly}'_d(v_{1:n-1}) \cdot v_n^d
    \]
    where $\mathit{poly}'_d : \bR^{n-1} \to \bR$ is a polynomial. Using the induction hypothesis, we have
    \begin{align*}
      |\mathit{poly}'(r(\alpha_1),\ldots,r(\alpha_n))| 
      &
      \leq \sum_{d=0}^D |\mathit{poly}'_d(r(\alpha_1),\ldots,r(\alpha_n-1))| \cdot |r(\alpha_n)|^d
      \\
      &
      \leq \sum_{d=0}^D \exp(l'_d(r)) 
           \cdot \exp(l_d(|r(\alpha_n)|))
      =  \sum_{d=0}^D \exp(l''_d(r))
      \\
      &
      \leq \exp(l(r))      
    \end{align*}
    for some affine functions $l_d : \bR \to \bR$,
    and functions $l'_d,l''_d,l \in \cA_i$. The last inequality follows from (a).
  \end{compactitem}

  \vspace{1mm}
  \noindent{\sc\bf Proof of (c).} By the definition of affine-bounded neural network, there exist functions 
  \[ 
        f_1 : \bR^{n_1} \to \bR^{n_2},\quad 
        f_2 : \bR^{n_2} \to \bR^{n_3},\quad
        \ldots,\quad 
        f_d : \bR^{n_d} \to \bR^{n_{d+1}}
  \]
  and affine functions 
  \[
        l'_j : \bR^{n_j} \to \bR\quad
        \text{ for all $1 \leq j \leq d$}
  \]
  such that (i) $n_1 = J$ and $n_{d+1} = 1$; (ii) $\nn = f_d \circ \cdots \circ f_1$; and (iii) for all $1 \leq j \leq d$ and $v \in \bR^{n_j}$,
  \[ 
        \sum_{k=1}^{n_{j+1}} |f_j(v)_k| \leq l'_j(|v_1|,\ldots,|v_{n_j}|).
  \]
  We prove the claim of the lemma by induction on the depth $d$.
  For $d=1$, the claim follows immediately from the defining property (iii) of the affine-bounded neural network $\mathit{nn}$. Assume $d>1$. By induction hypothesis, there exist affine functions $l_1,\ldots,l_{n_{d}} : \bR^J \to \bR$ such that for all $k \in \{1,\ldots,n_{d}\}$ and $v \in \mathbb{R}^J$,
  \[
          |(f_{d-1} \circ \cdots \circ f_1)(v)_k| \leq l_k(|v_1|,\ldots,|v_J|)
  \]
  This is because the $(f_{d-1} \circ \cdots \circ f_1)(v)_k$ are affine-bounded neural networks of depth $d-1$. Let 
  \[
          l'_d(v_1,\ldots,v_{n_{d}})
          = c_0 + \sum_{k=1}^{n_{d}} c_k \cdot |v_k|.
  \]
  Using what we have proved so far, we calculate:
  \begin{align*}
    |\nn(v)| 
    & = \big|f_d\big((f_{d-1} \circ \cdots \circ f_1)(v)\big)\big|
    \\
    & \leq 
    l'_d\big(|(f_{d-1} \circ \cdots \circ f_1)(v)_1|,
         \ldots,
         |(f_{d-1} \circ \cdots \circ f_1)(v)_{n_{d}}|\big)
    \\
    & = c_0 + \sum_{k=1}^{n_{d}} c_k \cdot |(f_{d-1} \circ \cdots \circ f_1)(v)_k|
    \\
    & \leq c_0 + \sum_{k=1}^{n_{d}} |c_k| \cdot |(f_{d-1} \circ \cdots \circ f_1)(v)_k|
    \\
    & \leq c_0 + \sum_{k=1}^{n_{d}} |c_k| \cdot l_k\big(|v_1|,\ldots,|v_J|\big).
  \end{align*}
  Let $l(r) = c_0 + \sum_{k=1}^{n_{d}} |c_k| \cdot l_k(|r(\alpha_1)|,\ldots,|r(\alpha_J)|)$. The function $l$ is essentially the composition of two affine functions, and so it belongs to $\cA_i$. This and the conclusion of the above calculation show that $l$ is the function in $\cA_i$ that we look for.

  \vspace{1mm}
  \noindent{\sc\bf Proof of (d).}
  Since $\log(1+v) \geq v/2$ for all $v \in [0,1]$,
  we have
  \begin{align*}
    \log(1+\exp(v)) &\geq \exp(v)/2 = \exp(-|v| - \log2)
    &&\text{for all $v \in (-\infty, 0]$, \quad and}
    \\
    \log(1+\exp(v)) &\leq \log(2 \cdot \exp(v)) = |v| + \log 2
    &&\text{for all $v \in [0, \infty)$.}
  \end{align*}
  Let $l_1(v) = -v-\log2$ and $l_2(v) = v+\log2$.
  As $\mathrm{softplus}(-)$ is increasing,
  we have
  \[\exp(l_1(|v|)) \leq \mathrm{softplus}(v) \leq \exp(l_2(|v|))
  \quad\text{for all $v \in \bR$.}\]

  \vspace{1mm}
  \noindent{\sc\bf Proof of (e).} Let $l(r) = c_0 + \sum_{j=1}^J c_j \cdot |r(\alpha_j)|$,
  where $c_j \in \bR$. The moment generating function $G_p : \bR^J \to \bR$ of
  a random variable of density $p(v)$ on $\bR^J$
  is defined by $G_p(t) \defeq \EE{p(v)}{\exp(t\cdot v)}$ where $t \cdot v$ is the inner product of $t$ and $v$.
  It is well-known that the moment generating functions are well defined for multivariate normal distributions 
  $q(v)$ on $\bR^J$ in the lemma,
  i.e., $G_q(t) < \infty$ for all $t \in \bR^J$.
  From this fact and that $\exp(|v|) \leq \exp(v)+\exp(-v)$, it follows that
  \begin{align*}
          \EE{q(r(\alpha_1),\ldots,r(\alpha_J))}{\exp(l(r))}
          & = \exp(c_0) \cdot \EE{q(r(\alpha_1),\ldots,r(\alpha_J))}{\exp\left(\sum_{j=1}^J c_j \cdot |r(\alpha_j)|\right)}
          \\
          & \leq \exp(c_0) \cdot \sum_{k_1,\ldots,k_J \in \{0,1\}}
          G_q\left((-1)^{k_1}  c_1,\ldots,(-1)^{k_J} c_J\right) 
          \\
          & < \infty.
  \end{align*}
\end{proof}


The proof of Theorem~\ref{thm:condition-456}
uses two more ingredients:
Theorem~\ref{thm:diff-under-int-cont}
which, compared to Theorem~\ref{thm:diff-under-int},
additionally guarantees the continuity of $\nabla_v \!\int\! f_v \,\dd\mu$
by assuming the continuity of $\nabla_v f_v$;
and Lemma~\ref{lem:bound-moving-normal}
which bounds the densities of multiple normal distributions
using the density of a~single, fixed normal distribution.

\begin{theorem}
  \label{thm:diff-under-int-cont}
  Let $V \subset \bR$ be an open interval, and $(X,\Sigma,\mu)$ be a measure space.
  Suppose that a measurable function $f: V \times X \to \bR$ satisfies the following conditions:
  \begin{itemize}
  \item[(a)]
    The function $v \in V \longmapsto \int \mu(\dd x)\, f_v(x) \in \bR$
    is well-defined.  
  \item[(b)]
    For almost all $x \in X$ (w.r.t.~$\mu$),
    the function $v \in V \longmapsto \nabla_v f_v(x) \in \bR$
    is well-defined, continuous.
  \item[(c)]
    There is a measurable function $h : X \to \bR$ such that $\int \mu(\dd x)\, h(x)$ is well-defined
    and $\left|\nabla_v f_v(x)\right| \leq h(x)$ for all $v \in V$ and almost all $x \in X$ (w.r.t.~$\mu$).
  \end{itemize}
  Then, for all $v \in V$,
  both sides of the below equation are well-defined,
  and the equality holds:
  \begin{equation}
    \label{eqn:diff-under-int}
    \nabla_v \int \mu(\dd x)\, f_v(x) 
    = 
    \int \mu(\dd x)\,\nabla_v f_v(x).
  \end{equation}
  Moreover, the function $v \in V \longmapsto \nabla_v \int \mu(\dd x)\, f_v(x)$
  is continuous.
\end{theorem}
\begin{proof}
  Theorem~\ref{thm:diff-under-int} implies the former part of the conclusion:
  for all $v \in V$, each side of \eqref{eqn:diff-under-int} is well-defined
  and the equation~\eqref{eqn:diff-under-int} holds.
  To prove the latter part of the conclusion,
  define $G : V \to \bR$ and $g:V \times X \to \bR$ by
  \[
  G(v) \defeq \nabla_v \int \mu(\dd x)\, f_v(x) = \int \mu(\dd x)\, g(v,x),
  \qquad
  g(v,x) \defeq \nabla_v f_v(x),
  \]
  and let $(v_n)_{n \in \bN}$ be any sequence in $V$ such that
  $\lim_{n \to \infty} v_n = v^* \in V$.
  Then, it suffices to show that
  \[\lim_{n \to \infty} G(v_n) = G(v^*).\]
  To apply the dominated convergence theorem,
  define $g_n,\, g^* : X \to \bR$ by $g_n(x) \defeq g(v_n, x)$ and $g^*(x) \defeq g(v^*, x)$.
  By (b), the function $g(\cdot, x) : V \to \bR$ is continuous for almost all $x \in X$ (w.r.t.~$\mu$),
  so \[\lim_{n \to \infty} g_n(x) = \lim_{n \to \infty} g(v_n,x)
  = g(v^*, x) = g^*(x) \qquad\text{for almost all $x \in X$ (w.r.t.~$\mu$)}.\]
  By (c), for all $n \in \bN$,
  $|g_n(x)| \leq h(x)$ for almost all $x \in X$ (w.r.t.~$\mu$),
  and $h(x)$ is integrable (w.r.t.~$\mu$).
  Therefore, we can apply the dominated convergence theorem to
  $(g_n)_{n \in \bN}$ and $g^*$ with $h$, which gives:
  \[\lim_{n \to \infty} G(v_n) = \lim_{n \to \infty} \int \mu(\dd x)\, g_n(x)
  = \int \mu(\dd x)\, g^*(x) = G(v^*) .\]
\end{proof}

\begin{lemma}
  \label{lem:bound-moving-normal}
  Suppose $[\mu_1, \mu_2]\subset \bR$ and $[\sigma_1, \sigma_2] \subset (0,\infty)$
  Then, there exists $c \in \bR$ such that 
  \[ \cN(v;\mu, \sigma) \leq c \cdot \cN(v; 0, 2\sigma_2)\]
  for all $v \in \bR$, $\mu \in [\mu_1, \mu_2]$, and $\sigma \in [\sigma_1, \sigma_2]$.
\end{lemma}
\begin{proof}
  We claim that $c = \frac{2\sigma_2}{\sigma_1} \exp(\frac{1}{6\sigma_1^2} \max(\mu_1^2, \mu_2^2))$ is the desired coefficient.
  For the constant $c$, the claim we need show is that
  \begin{align*}
    \frac{1}{\sqrt{2\pi}\sigma} \exp\left(-\frac{(v-\mu)^2}{2\sigma^2} \right)
    \leq
    \frac{1}{\sqrt{2\pi}\sigma_1} \exp\left(-\frac{v^2}{8\sigma_2^2}
    + \frac{1}{6\sigma_1^2} \max(\mu_1^2, \mu_2^2) \right)
  \end{align*}
  for any $v \in \bR$, $\mu \in [\mu_1, \mu_2]$, and $\sigma \in [\sigma_1, \sigma_2]$.
  Since $\frac{1}{\sqrt{2\pi}\sigma_1} \geq \frac{1}{\sqrt{2\pi}\sigma}$,
  it suffices to show
  \begin{align*}
    & & -\frac{v^2}{8\sigma_2^2}
    + \frac{1}{6\sigma_1^2} \max(\mu_1^2, \mu_2^2)
    &\geq
    -\frac{(v-\mu)^2}{2\sigma^2}
    \\
    &\iff
    & \frac{1}{2\sigma^2} \left( av^2 - 2\mu v + \mu^2
    + \frac{\sigma^2}{3\sigma_1^2} \max(\mu_1^2, \mu_2^2) \right)
    &\geq 0,
    \quad\text{where $a=1-\frac{\sigma^2}{4\sigma_2^2}$}
    \\
    &\iff
    &a\left(v-\frac{\mu}{a}\right)^2
    + \left(1-\frac{1}{a}\right) \mu^2
    + \frac{\sigma^2}{3\sigma_1^2} \max(\mu_1^2, \mu_2^2)
    & \geq 0
    \\
    &\impliedby
    & \left(1-\frac{1}{a}\right) \mu^2
    + \frac{\sigma^2}{3\sigma_1^2} \max(\mu_1^2, \mu_2^2)
    & \geq 0,
    \quad\text{since }
    a = 1-\frac{\sigma^2}{4\sigma_2^2} \geq \frac{3}{4}.
  \end{align*}
  The last inequality holds because
  \begin{align*}
    \left(1-\frac{1}{a}\right) \mu^2
    + \frac{\sigma^2}{3\sigma_1^2} \max(\mu_1^2, \mu_2^2)
    \geq
    -\frac{1}{3}\mu^2 + \frac{1}{3} \max(\mu_1^2, \mu_2^2) \geq 0.
  \end{align*}
\end{proof}

{\sc Theorem~\ref{thm:condition-456}.\ }{\it
        If both our assumption in \S\ref{subsec:requirement-assumption}
        and the condition \eqref{eqn:condition-2b} hold, then
        the condition \eqref{eqn:condition-456} implies the requirements~\ref{req:4}-\ref{req:6}.}

\begin{proof}
Before staring the proof, we note that 
by the condition~\eqref{eqn:condition-456},
there is an open interval $V_{(i,\alpha,j,\theta)} \subseteq \bR$
for all $i \in [M]$, $\alpha \in K_i$, $j \in [p]$ and $\theta \in \mathbb{R}^p$,
such that
\begin{compactitem}
  \item $V_{(i,\alpha,j,\theta)}$ contains $\theta_j$;
  \item the function $v\in\bR \longmapsto \mu_{(i,\alpha)}(\theta[j : v])$
    is continuously differentiable in $V_{(i,\alpha,j,\theta)}$;
  \item the function $v\in\bR \longmapsto \sigma_{(i,\alpha)}(\theta[j : v])$
    is continuously differentiable in $V_{(i,\alpha,j,\theta)}$,
    and $\inf \{\sigma_{(i,\alpha)}(\theta[j : v]) \,|\,
    v \in V_{(i,\alpha,j,\theta)}\} > 0$.
\end{compactitem}

Let $F_1$ and $F_2$ be functions from $\bR^p \times \Rdb$ to $\bR$ defined as follows: for all $\theta \in \bR^p$ and $r \in \Rdb$,
\begin{align*}
        F_1(\theta,r) & \defeq \density(D_\theta,s_I)(r) \cdot \log\frac{\density(D_\theta,s_I)(r)}{\density(C,s_I)(r)},
        & F_2(\theta,r) & \defeq \density(D_\theta,s_I)(r).
\end{align*}
We have to show that for every $\theta \in \mathbb{R}^p$ and $j \in [p]$,
\begin{itemize} 
\item the partial derivatives
        \[
                \nabla_{\theta_j} \int \rho(\dd r)\,F_1(\theta,r)
                \quad\text{ and }\quad
                \nabla_{\theta_j} \int \rho(\dd r)\,F_2(\theta,r)
        \]
      are well-defined and are continuous functions of $\theta_j \in \bR$;
\item these derivatives commute with integration:
        \[
                \nabla_{\theta_j} \int \rho(\dd r)\,F_1(\theta,r)
                =
                \int \rho(\dd r)\, \nabla_{\theta_j} F_1(\theta,r)
                \quad\text{ and }\quad
                \nabla_{\theta_j} \int \rho(\dd r)\,F_2(\theta,r)
                =
                \int \rho(\dd r)\, \nabla_{\theta_j} F_2(\theta,r).
        \]
\end{itemize} 

We discharge these proof obligations using Theorem~\ref{thm:diff-under-int-cont}. Pick $\theta^* \in \bR^p$ and $j \in [p]$. We instantiate Theorem~\ref{thm:diff-under-int-cont} twice for $F_1$ and $F_2$ as follows. 
\begin{itemize} 
        \item For both cases of $F_1$ and $F_2$, we set $V$ to be an open interval $(a,b)$ such that both $a$ and $b$ are finite, the closure $[a,b]$ is contained in
          $V_{(j,\theta^*)} \defeq \bigcap_{i \in [M]} \bigcap_{\alpha \in K_i} V_{(i,\alpha,j,\theta^*)}$, and $(a,b)$ contains $\theta^*_j$.
          Note that such an open interval exists because
          $V_{(j,\theta^*)}$
          is open and contains $\theta^*_j$. 
        \item The measurable space $(X,\Sigma)$ in the theorem is $\Rdb$ with its $\sigma$-algebra in both cases.  
        \item The measure $\mu$ is the reference measure $\rho$ on $\Rdb$ in both cases.  
        \item The measurable function $f$ is instantiated twice for $(v,r) \longmapsto F_i(\theta^*[j:v],r)$ with $i = 1,2$. 
\end{itemize}
Under this instantiation, the conclusion of Theorem~\ref{thm:diff-under-int-cont} implies all of the requirements~\ref{req:4}-\ref{req:6}.

It remains to show that both instantiations satisfy the three provisos of Theorem~\ref{thm:diff-under-int-cont} marked (a), (b) and (c). We tackle the provisos one-by-one. We use the notations in \S\ref{subsec:requirement-assumption} without recalling them explicitly. For $i \in [M]$, $l \in [N_i]$, $\alpha \in K_i$, $v \in \bR$, $r \in A_i$ and $\theta \in \bR^p$, let
\begin{align*}
        g_{(i,\alpha,\theta)}(v) & \defeq \cN(v;\mu_{(i,\alpha)}(\theta),\sigma_{(i,\alpha)}(\theta)),
        &
        g_{(i,\theta)}(r) & \defeq \prod_{\beta \in K_i} g_{(i,\beta,\theta)}(r(\beta)),
        \\
        h_{(i,\alpha)}(r) & \defeq \cN(r(\alpha);\mu'_{(i,\alpha)}(r),\sigma'_{(i,\alpha)}(r)),
        &
        h_i(r) & \defeq \prod_{\beta \in K_i} h_{(i,\beta)}(r),
        \\
        k_{(i,l)}(r) & \defeq \cN(c_{(i,l)};\mu''_{(i, l)}(r), \sigma''_{(i, l)}(r)),
        &
        k_i(r) & \defeq \prod_{m=1}^{N_i} k_{(i,m)}(r).
\end{align*}

\vspace{2mm}
\noindent{\bf Proof of (a).} Since we assume the condition~\ref{eqn:condition-2b}, Theorem~\ref{thm:condition-2} ensures that the integral $\int \mu(\dd r)\, F_1(\theta^*[j:v],r)$ is well-defined for all $v \in V$. The other integral involving $F_2$ is the integral of the density of $D_{\theta^*[j:v]}$. Since $D_{\theta^*[j:v]}$ is a guide, the integral is well-defined and has the value $1$.

\vspace{2mm}
\noindent{\bf Proof of (b).} Note that 
\begin{align*}
        F_1(\theta^*[j:v], r) 
        & {} = \density(D_{\theta^*[j:v]},s_I)(r) \cdot \log\frac{\density(D_{\theta^*[j:v]},s_I)(r)}{\density(C,s_I)(r)}
        \\
        & {} = 
        \sum_{i=1}^M 
        \ind{r \in A_i}
        \cdot
        g_{(i,\theta^*[j:v])}(r) 
        \cdot
        \log 
        \frac{ g_{(i,\theta^*[j:v])}(r)}{h_i(r) \cdot k_i(r)}
        \\
        & {} = 
        \sum_{i=1}^M 
        \ind{r \in A_i}
        \cdot
        g_{(i,\theta^*[j:v])}(r) 
        \cdot
        \left(- \log(h_i(r) \cdot k_i(r)) + \log(g_{(i,\theta^*[j:v])}(r))\right)
        \\
        & {} = 
        \sum_{i=1}^M 
        \ind{r \in A_i}
        \cdot
        \prod_{\alpha \in K_i} \cN(r(\alpha);\mu_{(i,\alpha)}(\theta^*[j:v]),\sigma_{(i,\alpha)}(\theta^*[j:v]))
        \\
        & \qquad
        {}
        \cdot
        \left(
        - \log(h_i(r) \cdot k_i(r))
        + \sum_{\alpha \in K_i} \log \cN(r(\alpha);\mu_{(i,\alpha)}(\theta^*[j:v]),\sigma_{(i,\alpha)}(\theta^*[j:v]))
        \right),
        \\[2ex]
        F_2(\theta^*[j:v], r) 
        & {} = \density(D_{\theta^*[j:v]},s_I)(r) 
        \\
        & {} = 
        \sum_{i=1}^M 
        \ind{r \in A_i}
        \cdot
        \prod_{\alpha \in K_i} \cN(r(\alpha);\mu_{(i,\alpha)}(\theta^*[j:v]),\sigma_{(i,\alpha)}(\theta^*[j:v])).
\end{align*}
Recall that every $r \in A_i$ is a map from $K_i$ to $\bR$. The normal density $\cN(v'; \mu',\sigma')$ and its log are smooth functions on $(v',\mu',\sigma') \in \bR \times \bR \times (0,\infty)$. Also, by the condition \eqref{eqn:condition-456}, the functions 
\[
        v \in V_{(i,\alpha,j,\theta^*)} \longmapsto \mu_{(i,\alpha)}(\theta^*[j:v])
        \quad\text{ and }\quad
        v \in V_{(i,\alpha,j,\theta^*)} \longmapsto \sigma_{(i,\alpha)}(\theta^*[j:v])
\]
are continuously differentiable for all $i \in [M]$ and $\alpha \in K_i$. Thus, they are also continuously differentiable when we restrict their domains to $V$. One more thing to notice is that neither $\ind{r \in A_i}$ nor $- \log(h_i(r) \cdot k_i(r))$ depends on $v$, so that they can be viewed as constant functions on $v$, which are obviously smooth. Now note that both $v \in V \longmapsto F_1(\theta^*[j:v], r)$ and $v \in V \longmapsto F_2(\theta^*[j:v], r)$ are obtained from these smooth or continuously differentiable functions by function composition, addition and multiplication. Thus, they are continuously differentiable.

\vspace{2mm}
\noindent{\bf Proof of (c).} For all $i \in [M]$, we will construct measurable function $H_i,H'_i : A_i \to [0,\infty)$ such that
\begin{itemize} 
        \item both $\int \rho(\dd r)\,(\ind{r\in A_i} \cdot H_i(r))$ and $\int \rho(\dd r)\,(\ind{r\in A_i} \cdot H'_i(r))$ are well-defined;
        \item for all $r \in A_i$ and $v \in V$,
        \[
                \left|\nabla_v\left(g_{(i,\theta^*[j:v])}(r)\cdot \log\frac{g_{(i,\theta^*[j:v])}(r)}{h_i(r) \cdot k_i(r)}\right)\right|
                \leq H_i(r);
        \]
        \item for all $r \in A_i$ and $v \in V$,
        \[
        \left|\nabla_v g_{(i,\theta^*[j:v])}(r)\right| \leq H'_i(r).  
        \]
\end{itemize} 
Once we have such $H_i$'s, we can use them to define desired measurable functions $H,H' : \Rdb \to \mathbb{R}$ as follows:
\[
        H(r) \defeq
        \sum_{i = 1}^M \ind{r \in A_i} \cdot H_i(r)
        \quad\text{ and }\quad
        H'(r) \defeq
        \sum_{i = 1}^M \ind{r \in A_i} \cdot H'_i(r).
\]
Since both $H_i$ and $H'_i$ are integrable for all $i$, so are $H$ and $H'$. We prove this only for $H$ below. The case for $H'$ is similar.
\begin{align*}
        \int \rho(\dd r)\, |H(r)|
        \leq
        \int \rho(\dd r)\, \sum_{i = 1}^M \left|\ind{r \in A_i} \cdot H_i(r)\right|
        = 
        \sum_{i=1}^M \int \rho(\dd r)\, \left|\ind{r \in A_i} \cdot H_i(r)\right|
        < \infty.
\end{align*}
Both $H_i$ and $H'_i$ bound the magnitudes of the partial derivatives involving $g_{(i,\theta^*[j:v])}$ for every $i$. Using this, we show that $H$ and $H'$ meet the condition about bounding the magnitude of a derivative:
\begin{align*}
        \left|\nabla_v F_1(\theta^*[j:v],r)\right|
        & {} = 
        \left|\nabla_v \sum_{i=1}^M \ind{r \in A_i} \cdot \Big(g_{(i,\theta^*[j:v])}(r)\cdot \Big(- \log(h_i(r) \cdot k_i(r)) + \log(g_{(i,\theta^*[j:v])}(r))\Big)\Big)\right|
        \\
        & {} \leq
        \sum_{i=1}^M \ind{r \in A_i} \cdot \left|\nabla_v \Big(g_{(i,\theta^*[j:v])}(r)\cdot \Big(- \log(h_i(r) \cdot k_i(r)) + \log(g_{(i,\theta^*[j:v])}(r))\Big)\Big)\right|
        \\
        & {} \leq 
        \sum_{i=1}^M \ind{r \in A_i} \cdot H_i(r) 
        = H(r),
        \\
        \left|\nabla_v F_2(\theta^*[j:v],r)\right|
        & {} = 
        \left|\nabla_v \sum_{i=1}^M \ind{r \in A_i} \cdot g_{(i,\theta^*[j:v])}(r)\right|
        \\
        & {} \leq 
        \sum_{i=1}^M \ind{r \in A_i} \cdot \left|\nabla_v g_{(i,\theta^*[j:v])}(r)\right|
        \\
        & {} \leq 
        \sum_{i=1}^M \ind{r \in A_i} \cdot H'_i(r)
        = H'(r).
\end{align*}

It remains to construct measurable $H_i,H'_i : A_i \to \mathbb{R}$ with the required properties for every $i \in [M]$.
Pick $i \in [M]$. Note that
\begin{align*}
        & \nabla_v\left(g_{(i,\theta^*[j:v])}(r)\cdot \log \frac{g_{(i,\theta^*[j:v])}(r)}{h_i(r) \cdot k_i(r)}\right)
        \\
        & \qquad{} = 
        \left(\nabla_v g_{(i,\theta^*[j:v])}(r)\right) \cdot \log \frac{g_{(i,\theta^*[j:v])}(r)}{h_i(r) \cdot k_i(r)}
        +
        \left(g_{(i,\theta^*[j:v])}(r)\cdot \nabla_v \log(g_{(i,\theta^*[j:v])}(r))\right)
        \\
        & \qquad{} = 
        \left(\nabla_v g_{(i,\theta^*[j:v])}(r)\right) \cdot \left(\log \frac{g_{(i,\theta^*[j:v])}(r)}{h_i(r) \cdot k_i(r)} + 1\right).
\end{align*}
We will build $H_i$ and $H'_i$ by finding good bounds for $|\nabla_v g_{(i,\theta^*[j:v])}(r)|$,
$|\log g_{(i,\theta^*[j:v])}(r)|$ and $|\log h_i(r)\cdot k_i(r)|$ and then combining these bounds. We do so in three steps.

First, we build functions that bound $|\nabla_v g_{(i,\theta^*[j:v])}(r)|$ and 
$|\log g_{(i,\theta^*[j:v])}(r)|$ over $v \in V$.
For every $\alpha \in K_i$, 
\begin{align*} 
        \nabla_v \log g_{(i,\alpha,\theta^*[j:v])}(w)
        & {} = 
        \nabla_v 
        \left(-\log \sqrt{2\pi} - \log \sigma_{(i,\alpha)}(\theta^*[j:v]) 
        -\frac{\left(w - \mu_{(i,\alpha)}(\theta^*[j:v])\right)^2}{2\sigma_{(i,\alpha)}(\theta^*[j:v])^2} \right)
        \\
        & {} = 
        - \frac{\nabla_v \sigma_{(i,\alpha)}(\theta^*[j:v])}{\sigma_{(i,\alpha)}(\theta^*[j:v])}
        - \frac{(\mu_{(i,\alpha)}(\theta^*[j:v]) - w)\nabla_v \mu_{(i,\alpha)}(\theta^*[j:v])}{\sigma_{(i,\alpha)}(\theta^*[j:v])^2}
        \\
        & \phantom{{} = 
        - \frac{\nabla_v \sigma_{(i,\alpha)}(\theta^*[j:v])}{\sigma_{(i,\alpha)}(\theta^*[j:v])}}
        + \frac{\left(w - \mu_{(i,\alpha)}(\theta^*[j:v])\right)^2 \nabla_v \sigma_{(i,\alpha)}(\theta^*[j:v])}{\sigma_{(i,\alpha)}(\theta^*[j:v])^3}
        \\
        & {} = 
        \nabla_v \sigma_{(i,\alpha)}(\theta^*[j:v])
        \cdot
        \frac{\left(w - \mu_{(i,\alpha)}(\theta^*[j:v])\right)^2 - \sigma_{(i,\alpha)}(\theta^*[j:v])^2}{\sigma_{(i,\alpha)}(\theta^*[j:v])^3}
        \\
        & {} \qquad+
         \nabla_v \mu_{(i,\alpha)}(\theta^*[j:v]) \cdot
         \frac{w - \mu_{(i,\alpha)}(\theta^*[j:v])}{\sigma_{(i,\alpha)}(\theta^*[j:v])^2}.
\end{align*}
Thus, we have
\begin{align}
        \nabla_v g_{(i,\theta^*[j:v])}(r)
        & = \nabla_v \prod_{\alpha \in K_i} g_{(i,\alpha,\theta^*[j:v])}(r(\alpha))
        \nonumber \\
        & = \sum_{\alpha \in K_i} 
                \nabla_v g_{(i,\alpha,\theta^*[j:v])}(r(\alpha))
                \cdot \prod_{\substack{\beta \in K_i\\ \beta \neq \alpha}} g_{(i,\beta,\theta^*[j:v])}(r(\beta))
        \nonumber \\
        & = \sum_{\alpha \in K_i} 
                g_{(i,\alpha,\theta^*[j:v])}(r(\alpha))
                \cdot \nabla_v \log g_{(i,\alpha,\theta^*[j:v])}(r(\alpha))
                \cdot \prod_{\substack{\beta \in K_i\\ \beta \neq \alpha}} g_{(i,\beta,\theta^*[j:v])}(r(\beta))
        \nonumber \\
        & = g_{(i,\theta^*[j:v])}(r) \cdot \sum_{\alpha \in K_i} \nabla_v \log g_{(i,\alpha,\theta^*[j:v])}(r(\alpha))
        \nonumber \\
        & = g_{(i,\theta^*[j:v])}(r) \cdot \sum_{\alpha \in K_i} 
        \Bigg(
        \nabla_v \sigma_{(i,\alpha)}(\theta^*[j:v])
        \cdot
        \frac{\left(r(\alpha) - \mu_{(i,\alpha)}(\theta^*[j:v])\right)^2 - \sigma_{(i,\alpha)}(\theta^*[j:v])^2}{\sigma_{(i,\alpha)}(\theta^*[j:v])^3}
        \nonumber \\
        &
        \qquad\qquad\qquad\qquad\qquad\qquad
        {} +
        \nabla_v \mu_{(i,\alpha)}(\theta^*[j:v]) \cdot
        \frac{r(\alpha) - \mu_{(i,\alpha)}(\theta^*[j:v])}{\sigma_{(i,\alpha)}(\theta^*[j:v])^2}\Bigg)
        \nonumber \\
        & = g_{(i,\theta^*[j:v])}(r) \cdot \sum_{\alpha \in K_i}\sum_{n=0}^2 G_{(1,\alpha,n)}(v) \cdot r(\alpha)^n
        \label{eqn:nabla-v-g}
\end{align}
for some continuous functions $G_{(1,\alpha,n)} : V_{(j,\theta^*)} \to \bR$
that do not depend on $r$.
The continuity of $G_{(1,\alpha,n)}$ follows from our condition \eqref{eqn:condition-456}. Also,
\begin{align*}
        \log g_{(i,\theta^*[j:v])}(r) 
        & {} = \sum_{\alpha \in K_i} \log g_{(i,\alpha,\theta^*[j:v])}(r(\alpha)) 
        \\
        & {} = \sum_{\alpha \in K_i}
        \left(
        - \log \sqrt{2\pi} 
        - \log \sigma_{(i,\alpha)}(\theta^*[j:v]) 
        - \frac{\left(r(\alpha) - \mu_{(i,\alpha)}(\theta^*[j:v])\right)^2}{2\sigma_{(i,\alpha)}(\theta^*[j:v])^2} 
        \right)
        \\
        & = \sum_{\alpha \in K_i}\sum_{n=0}^2 G_{(2,\alpha,n)}(v) \cdot r(\alpha)^n
\end{align*}
for some continuous functions $G_{(2,\alpha,n)} : V_{(j,\theta^*)} \to \bR$
that do not depend on $r$.
The continuity of $G_{(2,\alpha,n)}$ follows from the condition \eqref{eqn:condition-456}.
Since $\overline{V} \subset V_{(j,\theta^*)}$,
the functions $G_{(m,\alpha,n)}$ are all continuous
on the closed (bounded) interval $\overline{V}$,
where $\overline{V}$ is the closure of $V$.
Thus, by the extreme value theorem, 
\[
        C_\mathit{max} \defeq \max_{(m,\alpha,n)} \sup_{v \in \overline{V}} |G_{(m,\alpha,n)}(v)| < \infty.
\]

Since the term $g_{(i,\theta^*[j:v])}(r)$ in~\eqref{eqn:nabla-v-g} 
still depends on $v$, we bound it over $v \in V$ using Lemma~\ref{lem:bound-moving-normal}
which is instantiated with
\begin{align*} 
	\mu_1 
	& \defeq \min_{\alpha \in K_i} \inf_{v \in \overline{V}} \mu_{(i,\alpha)}(\theta^*[j:v])
	& \mu_2 
	& \defeq \max_{\alpha \in K_i} \sup_{v \in \overline{V}} \mu_{(i,\alpha)}(\theta^*[j:v]),
	\\
	\sigma_1 
	& \defeq  \min_{\alpha \in K_i} \inf_{v \in \overline{V}} \sigma_{(i,\alpha)}(\theta^*[j:v])
	& \sigma_2 
	& \defeq \max_{\alpha \in K_i} \sup_{v \in \overline{V}} \sigma_{(i,\alpha)}(\theta^*[j:v]).
\end{align*}
We check that the assumptions of the lemma are satisfied:
since $\overline{V} \subset V_{(j,\theta^*)}$ and the functions
\[v \in V_{(j,\theta^*)} \mapsto \mu_{(i,\alpha)}(\theta^*[j:v])
\quad\text{and}\quad
v \in V_{(j,\theta^*)} \mapsto \sigma_{(i,\alpha)}(\theta^*[j:v])\]
are continuous by the condition \eqref{eqn:condition-456},
$\mu_m$ and $\sigma_m$ ($m=1,2$) are all finite by the extreme value theorem;
moreover, $\sigma_1$ is positive by the choice of $V_{(i,\alpha,j,\theta^*)}$.
Thus, by Lemma \ref{lem:bound-moving-normal}, there exists $c \in \bR$
such that
\begin{align*}
	g_{(i,\theta^*[j:v])}(r) 
        & = 
	\prod_{\alpha \in K_i} \cN(r(\alpha); \mu_{(i,\alpha)}(\theta^*[j:v]), \sigma_{(i,\alpha)}(\theta^*[j:v]))
	\\
        & \leq \prod_{\alpha \in K_i} c \cdot \cN(r(\alpha); 0, 2\sigma_2)
	= c' \cdot g'(r)
\end{align*}
for all $v \in V$,
where $c' \defeq c^{|K_i|}$ and $g'(r) \defeq \prod_{\alpha \in K_i} \cN(r(\alpha); 0, 2\sigma_2)$.

Combining what we have proved, we obtain the following bounds:
for all $v \in V$,
\begin{align*}
        \left|\nabla_v g_{(i,\theta^*[j:v])}(r)\right| 
	& = 
	g_{(i,\theta^*[j:v])}(r)
	\cdot
	\left| \sum_{\alpha \in K_i}\sum_{n=0}^2 G_{(1,\alpha,n)}(v) \cdot r(\alpha)^n \right|       
	\\
	& {} \leq
	c' \cdot g'(r) \cdot 
	\sum_{\alpha \in K_i} \sum_{n=0}^2 C_\mathit{max} \cdot |r(\alpha)^n|
        \\ 
        & {} \leq
	c' \cdot g'(r) \cdot \exp(l_1(r)) 
	\qquad \text{for some $l_1 \in \cA_i$ (Lemma \ref{lem:lin-exp} (a) and (b))},
        \\
	\left|\log g_{(i,\theta^*[j:v])}(r)\right| 
	& =
	\left| \sum_{\alpha \in K_i}\sum_{n=0}^2 G_{(2,\alpha,n)}(v) \cdot r(\alpha)^n \right|
        \\
        & \leq
        \sum_{\alpha \in K_i} \sum_{n=0}^2 C_\mathit{max} \cdot |r(\alpha)^n|
	\leq 
	\exp(l_1(r)).
\end{align*}

Second, we bound $|\log h_i(r)\cdot k_i(r)|$ by $\exp(l_3(r))$ for some $l_2 \in \cA_i$.
\begin{align*}
        |\log h_i(r)\cdot k_i(r)|
        & {} \leq  
        \sum_{\alpha \in K_i} |\log h_{(i,\alpha)}(r)| + \sum_{l = 1}^{N_i} |\log k_{(i,l)}(r)|
        \\
        & {} =
        \sum_{\alpha \in K_i} \left|
        -\log\sqrt{2\pi}-\log \sigma'_{(i,\alpha)}(r) - \frac{\left(r(\alpha) - \mu'_{(i,\alpha)}(r)\right)^2}{2 \sigma'_{(i,\alpha)}(r)^2} \right| 
        \\
        & \qquad\qquad\qquad
        {} + 
        \sum_{l = 1}^{N_i} \left|
        -\log\sqrt{2\pi}-\log \sigma''_{(i,l)}(r) - \frac{\left(c_{(i,l)} - \mu''_{(i,l)}(r)\right)^2}{2 \sigma''_{(i,l)}(r)^2}
        \right|
        \\
        & {} \leq
        (|K_i|+N_i) \cdot \left|\log\sqrt{2\pi}\right|
        +
        \sum_{\alpha \in K_i} 
        \left(
        \left|\log \sigma'_{(i,\alpha)}(r)\right| + \frac{\left(r(\alpha) - \mu'_{(i,\alpha)}(r)\right)^2}{2 \sigma'_{(i,\alpha)}(r)^2}\right)
        \\
        & \phantom{{} \leq
        (|K_i|+N_i) \cdot \left|\log\sqrt{2\pi}\right|}
        {} + 
        \sum_{l = 1}^{N_i} \left(\left|\log \sigma''_{(i, l)}(r)\right|
        + \frac{\left(c_{(i,l)} - \mu''_{(i,l)}(r)\right)^2}{2 \sigma''_{(i,l)}(r)^2}\right)
        \\
        & {} \leq
        \exp(l_2(r)) \qquad \text{ for some $l_2 \in \cA_i$ (Lemma~\ref{lem:lin-exp} and Condition~\ref{eqn:condition-2b})}.
\end{align*}

Finally, we define $H_i$ and $H'_i$ by combining the bounds that we got so far. Here are our definitions for them:
\begin{align*}
        H_i(r) & 
        \defeq c' \cdot g'(r) \cdot \exp(l_1(r)) \cdot (\exp(l_1(r)) + \exp(l_2(r)) + 1),
        \\
        H'_i(r) & 
        \defeq c' \cdot g'(r) \cdot \exp(l_1(r)).
\end{align*}
Then, by the choice of $l_1$ and $l_2$, for all $r \in A_i$ and $v \in V$,
\[
        \left|\nabla_v \left(g_{(i,\theta^*[j:v])}(r)\cdot \log\frac{g_{(i,\theta^*[j:v])}(r)}{h_i(r) \cdot k_i(r)}\right)\right| \leq H_i(r)
        \quad\text{ and }\quad
        \left|\nabla_v g_{(i,\theta^*[j:v])}(r)\right| \leq H'_i(r).
\]
Also, we have
\begin{align*}
        & \int \rho(\dd r)\, \left| \ind{r \in A_i} \cdot H_i(r) \right|
        \\
        & \qquad\qquad {} =
	\int \rho(\dd r)\, \Big(\ind{r \in A_i} \cdot g'(r) \cdot c' \cdot \exp(l_1(r)) \cdot (\exp(l_1(r)) + \exp(l_2(r)) + 1)\Big)
        \\
        & \qquad\qquad {} \leq
	\int \rho(\dd r)\, \Big(\ind{r \in [K_i \to \bR]} \cdot g'(r) \cdot \exp(l_3(r))\Big)
        \qquad\text{for some $l_3 \in \cA_i$ (Lemma~\ref{lem:lin-exp} (a))}
        \\
        & \qquad\qquad {} = 
	\EE{\prod_{\alpha \in K_i} \cN(r(\alpha); 0, 2\sigma_2)
        }{
          \exp(l_3(r))
        }
        < \infty \qquad\text{(Lemma~\ref{lem:lin-exp} (e))},
\end{align*}
and
\begin{align*}
        & \int \rho(\dd r)\, \left|\ind{r \in A_i} \cdot H'_i(r)\right|
        \\
        & \qquad\qquad {} =
	\int \rho(\dd r)\, \Big(\ind{r \in A_i} \cdot g'(r) \cdot c' \cdot \exp(l_1(r))\Big)
        \\
        & \qquad\qquad {} \leq
	\int \rho(\dd r)\, \Big(\ind{r \in [K_i \to A]} \cdot g'(r) \cdot \exp(l_4(r))\Big)
        \qquad\text{for some $l_4 \in \cA_i$ (Lemma~\ref{lem:lin-exp} (a))}
        \\
        & \qquad\qquad {} = 
	\EE{\prod_{\alpha \in K_i} \cN(r(\alpha); 0, 2\sigma_2)
        }{
          \exp(l_4(r))
        }
        < \infty \qquad\text{(Lemma~\ref{lem:lin-exp} (e))},
\end{align*}
We have just shown that $H_i$ and $H'_i$ are the functions that we are looking for.
\end{proof}


\section{Proof of Theorem in \S\ref{sec:analysis}}

{\sc Theorem~\ref{thm:analysis-soundness}\ (Soundness).\ }{\it For all commands $C$, we have $\db{C}_d \in \gamma(\adb{C})$.}
\begin{proof}
        We prove the theorem by structural induction on $C$. Every case except the one for the loop follows from the assumption made for an abstract element in $\cT^\sharp$ or an operator on it that corresponds to $C$. The proof of the case of $C \equiv (\cwhile\ E\ \{C_0\})$ goes as follows. Let $t_\mathit{fix} \defeq (\wfix\ T)$. Then, by the assumption on the concretisation, $\gamma(t_\mathit{fix})$ is an admissible subset of $\cD$. Thus, it suffices to prove that when $G$ is the function in the density semantics of $\db{\cwhile\ E\ \{C_0\})}_d$, the image of $G$ on $\gamma(t_\mathit{fix})$ is included in $\gamma(t_\mathit{fix})$. To do so, pick $g \in \gamma(t_\mathit{fix})$. We have to show that $G(g) \in \gamma(t_\mathit{fix})$.  By the induction hypothesis and the assumption on the abstract composition, 
        \[ 
        (g^\ddagger \circ \db{C_0}_d) \in \gamma(t_\mathit{fix} \circ^\sharp \adb{C_0}).  
        \] 
        Also, $\db{\cskip}_d \in \gamma(\sskip^\sharp)$.
        Thus,
        \begin{align*} 
                G(g) 
                & = \lambda (s,r).\, 
                        \textbf{if}\ (\db{E}s = \true)\ 
                        \textbf{then}\ \db{\cskip}_d(s,r)\ 
                        \textbf{else}\ (g^\ddagger \circ \db{C_0}_d)(s,r) 
                 \\ 
                 & {} \in \gamma(\cond(E)^\sharp(t_\mathit{fix} \circ^\sharp \adb{C_0},\, \sskip^\sharp))
                 = \gamma(T(t_\mathit{fix})) = \gamma(t_\mathit{fix}).
       \end{align*} 
        The set membership holds because of the assumption on the abstract conditional operator.
\end{proof}

\end{document}